\documentclass[nofootinbib,aps,11pt]{revtex4}
\usepackage{graphicx}
\usepackage{cancel}
\usepackage{amssymb}
\usepackage{textcomp}
\usepackage{amsmath}
\usepackage{bm}
\usepackage{times}
\usepackage{epsfig}
\usepackage{color}
\def\beq{\begin{equation}}
\def\eeq{\end{equation}}

\def\ev{\,{\rm eV}}
\def\be{\begin{equation}}
\def\ee{\end{equation}}
\def\bea{\begin{eqnarray}}
\def\eea{\end{eqnarray}}
\begin{document}
\title{\LARGE Neutrino Masses and Heavy Triplet Leptons at the LHC: Testability of Type III Seesaw}
\bigskip
\author{Tong Li$^{1}$, Xiao-Gang He$^{1,2}$}
\address{
$^1$Center for High Energy Physics, Peking University, Beijing,
100871\\
$^2$Department of Physics and Center for Theoretical Sciences,
National Taiwan University, Taipei}
\date{\today}

\begin{abstract}
We study LHC signatures of Type III seesaw in which $SU(2)_L$
triplet leptons are introduced to supply the heavy seesaw masses. To
detect the signals of these heavy triplet leptons, one needs to
understand their decays to standard model particles  which depend on
how light and heavy leptons mix with each other. We concentrate on
the usual solutions with small light and heavy lepton mixing of
order the square root of the ratio of light and heavy masses,
$(m_\nu/M_{\nu_R})^{1/2}$. This class of solutions can lead to a visible displaced
vertex detectable at the LHC which can be used to
distinguish small mixing and large mixing between light and heavy
leptons. We show that, in this case, the couplings of light and
heavy triplet leptons to gauge and Higgs bosons, which determine the
decay widths and branching ratios, can be expressed in terms of
light neutrino masses and their mixing. Using these relations, we
study heavy triplet lepton decay patterns and production cross
section at the LHC. If these heavy triplet leptons are below a TeV
or so, they can be easily produced at the LHC due to their gauge
interactions from being non-trivial representations of $SU(2)_L$. We
consider two ideal production channels, 1) $E^+E^- \to \ell^+\ell^+
\ell^-\ell^- jj \ (\ell=e,\mu,\tau)$ and 2) $E^\pm N \to \ell^\pm
\ell^\pm jjjj$ in detail. For case 1), we find that with one or two
of the light leptons being $\tau$ it can also be effectively studied. With
judicious cuts at the LHC, the discovery of the heavy triplet
leptons as high as a TeV can be achieved with 100$fb^{-1}$
integrated luminosity.
\end{abstract}
 \maketitle

\section{Introduction}

Neutrino oscillation experiments involving neutrinos and
antineutrinos coming from astrophysical and terrestrial sources have
found compelling evidence that neutrinos have finite but small
masses. To accommodate this observation, the minimal standard model
(SM) must be extended. Generating neutrino masses through the seesaw
mechanism~\cite{seesawI,seesawII,seesawIII,review} is among the most attractive ones. It
explains the smallness of neutrino mass by supplying a suppression
factor of the ratio of electroweak scale to a new physics scale. There are
different ways to realize seesaw mechanism. They can be categorized
as Type I, Type II and Type III seesaw mechanisms. The main
ingredients of these models are as the followings.

{\bf Type I}~\cite{seesawI}: Introducing singlet right-handed
neutrinos $\nu_R$ which transform as: $(1,1,0)$ under SM
$SU(3)_C\times SU(2)_L \times U(1)_Y$ gauge group. It is clear that
$\nu_R$ does not have SM gauge interactions. The neutrino masses
$m_\nu$ are given by  $m_\nu \sim y^2_\nu v^2/ M_{\nu_R}$, where $v$
is the vacuum expectation value (vev) of the Higgs doublet in the
SM, $y_\nu^{}$ is the Yukawa coupling and $M_{\nu_R}$ is the
right-handed neutrino mass, which sets the new physics scale
$\Lambda$. If $y_\nu \simeq 1$, to obtain the light neutrino mass of
order an eV or smaller,  $M_{\nu_R}$ is required to be of order
$10^{14}\sim 10^{15}$ GeV. This makes it impossible to directly
detect $\nu_R$ at laboratory experiment. However, the Yukawa
coupling $y_\nu$ does not need to be of order one. If it turns out
to be similar to or smaller than the Yukawa coupling for electron,
$M_{\nu_R}$ can be as low as a TeV.

{\bf Type II}~\cite{seesawII}: Introducing a triplet Higgs
representation $\Delta$ transforming as: $(1,3,2)$. In this type of models, the
neutrino masses are given by: $m_\nu \approx Y_\nu  v_{\Delta}$, where
$v_{\Delta}$ is the vev of the neutral component of the triplet and
$Y_\nu$ is the Yukawa coupling. With a doublet and triplet mixing
via a dimensionful parameter $\mu$, the electroweak symmetry
breaking (EWSB) leads to a relation $v_{\Delta} \sim \mu
v^2/M_{\Delta}^2$, where $M_{\Delta}$ is the mass of the triplet. In
this case the scale $\Lambda$ is replaced by $M_{\Delta}^2/\mu$.
With $Y_\nu \approx 1$ and $\mu \sim M_{\Delta}$, the scale
$\Lambda$ is also  $10^{14}\sim 10^{15}$ GeV. Again a lower value of
order a TeV for $M_\Delta$ is possible.

{\bf Type III}~\cite{seesawIII}: Introducing triplet lepton
representations $\Sigma_L$ with $(1,3,0)$ SM quantum numbers. The resulting
mass matrix for neutrinos has the same form as that in
Type I seesaw. The high scale $\Lambda$ is replaced by the mass of
the leptons in the $SU(2)_L$ triplet representation which can also
be as low as a TeV.

In the absence of more experimental data, it is impossible to tell which, if any, of the mechanisms
is actually correct. Different models should be studied using available data or future ones.
The most direct way of verifying the seesaw mechanism is, of course, to
produce the heavy degrees of freedom in the models if they
are light enough, and study their properties. The Large Hadron Collider (LHC) at CERN with the unprecedented high energy and luminosity
is the best place to carry out such a test.

Major discoveries of exciting new physics at the Terascale at the
LHC are highly anticipated. Test of seesaw mechanism at LHC has
received a lot of attentions
recently~\cite{goran,Han:2006ip,more,processing,TypeIILHC,TypeII,production1,production2,Aguila}.
However, it is believed that any signal of $\nu_R$ would indicate a
more subtle mechanism beyond the simple Type I seesaw due to the
otherwise naturally small mixing $V_{l\nu_R} \sim
\sqrt{m_\nu/M_{\nu_R}}$ between the heavy neutrinos and the SM
leptons. Some of the ways to evade such a situation are to have some
new gauge interactions~\cite{Han:2006ip,processing} or to find
solutions where $V_{l\nu_R}$ are large which can happen in inverse
seesaw models~\cite{inverse,large,largeIII}.

The possibility of testing the Type II seesaw mechanism at the LHC
has been considered by several groups~\cite{TypeIILHC,TypeII}. Recently one group including one of us
systematically explored the parameter space in this
model~\cite{TypeII}. Using preferred parameters from experimental data, they found
that in the optimistic scenarios, by identifying the flavor
structure of the lepton number violating decays of the charged Higgs
bosons, one can establish the neutrino mass pattern of the normal
hierarchy (NH), inverted hierarchy (IH) or quasi-degenerate (QD).
Many other signatures of Type II seesaw at the LHC have been
studied~\cite{TypeIILHC,TypeII}.

There have also been studies to test Type III seesaw at the
LHC~\cite{production1,production2,Aguila}. Due to the fact that the
$SU(2)_L$ triplet $\Sigma$ has gauge interactions, the production of
the heavy triplet particles can have a much larger cross section
compared with that in Type I seesaw. The Type III seesaw can be
tested in a more comprehensive way up to the TeV range. In this
paper we further study some features of the Type III seesaw at LHC.
To detect the signals of the heavy triplet leptons, one needs to
understand their decays to SM particles which depend on
how light and heavy leptons mix with each other. Similar to Type-I
seesaw, in this model it is also possible to have small and large
mixing $V_{l\nu_R}$ between light and heavy
leptons~\cite{inverse,large,largeIII}.

The usual solutions with
light and heavy lepton mixing of order the square root of the ratio
of light and heavy masses, $(m_\nu/M_{\nu_R})^{1/2}$ could lead to a
visible displaced vertex in the detector at the
LHC~\cite{production1}. This fact can be used to distinguish small
mixing and large mixing between light and heavy leptons. The latter
does not lead to a displaced vertex. It has long been realized that it
is possible to have large light and heavy neutrino mixing originated
from the so-called inverse seesaw~\cite{inverse}. This possibility
has also received a lot of attentions
recently~\cite{large,largeIII}. With a large mixing between light and
heavy leptons, one can also study single heavy lepton
production~\cite{largeIII}. This can also be used to distinguish
model parameter spaces. We will concentrate on the usual small light and
heavy mixing solutions.

The analysis carried out in this work, in many ways,
is similar to that in Ref.~\cite{processing} since in both cases the
productions of heavy lepton pairs are through gauge boson mediation,
and also the light and heavy lepton mixing comes from seesaw
mechanism. The main differences are that in this model the heavy
leptons have electroweak interactions and the mediating gauge bosons
in productions are $W$ and $Z$, while in the model discussed in
Ref.~\cite{processing}, the heavy neutrinos do not have electroweak
interactions and the mediating particle is the new neutral gauge boson $Z'$. Our analysis also
has overlaps with that in Ref.~\cite{arhrib} where Type I+III seesaw
was studied, but detailed correlations are different since the model
in Ref.~\cite{arhrib} has both heavy neutrinos from Type I which do
not have electroweak interactions and also the triplet heavy leptons
from Type III we are considering. We have checked that when
applicable, our results agree with those obtained in
Ref.~\cite{processing,arhrib}.

We find that there is a relation
between the low energy neutrino oscillation and mass parameters, and
the heavy triplet lepton decay parameters which has not been
considered before in this model. We first derive this relation, and
then make concrete predictions of the heavy triplet lepton signals
using this relation for the small mixing solutions. We consider
two ideal production channels, 1) $E^+E^- \to \ell^+\ell^+
\ell^-\ell^- jj \ (\ell=e,\mu,\tau)$ and 2) $E^\pm N \to \ell^\pm
\ell^\pm jjjj$ in detail. We also include $\tau$ events
reconstruction in the analysis which turns out to give some
interesting additional information. With judicious cuts at the LHC,
the discovery of the heavy triplet leptons as high as a TeV can be
achieved with 100$fb^{-1}$ integrated luminosity. With 300$fb^{-1}$
integrated luminosity, the reach of the scale for heavy triplet
leptons can be higher.

The paper is arranged as the following. In Sec. II we summarize some
basic features of Type III seesaw model, paying particular attention to  the heavy triplet lepton
couplings to SM bosons and light leptons, and display relations
between the low energy neutrino oscillation and mass parameters. In Sec. III we study
constraints on the relevant parameters in the model, taking full
advantage of the relations obtained in Sec. II. In Sec. IV we study
the heavy triplet lepton decays. In Sec. V we study production of heavy triplet
leptons and the detection signals at the LHC. Finally in Sec. VI we
summarize our main results. We also include two appendices, Appendix A and Appendix B, to provide
more details on the derivation of the relation displayed in Sec. II and the general expressions for the
heavy triplet lepton decay parameters.

\section{The Type III seesaw Model}
The Type III seesaw model consists, in addition to the SM particles,
left-handed triplet leptons with zero hypercharge,
$\Sigma_L\sim (1,3,0)$ under $SU(3)_C\times SU(2)_L\times
U(1)_Y$~\cite{seesawIII}. We write the component fields as
\begin{eqnarray}
&&\Sigma_L=
\left(
  \begin{array}{cc}
    \Sigma_L^0/\sqrt{2} & \Sigma^+_L \\
    \Sigma_L^- & -\Sigma_L^0/\sqrt{2} \\
  \end{array}
\right).
\end{eqnarray}
The charge conjugated form is
\begin{eqnarray}
\ \ \Sigma^c_L= \left(
  \begin{array}{cc}
    \Sigma_L^{0c}/\sqrt{2} & \Sigma^{-c}_L \\
    \Sigma_L^{+c} & -\Sigma_L^{0c}/\sqrt{2} \\
  \end{array}
\right).
\end{eqnarray}
Note that $\Sigma^c_L $ is right-handed.

The renormalizable Lagrangian involving $\Sigma_L(\Sigma_L^c)$ is
given by
\begin{eqnarray}
\mathcal{L}&=&{\rm
Tr}[\overline{\Sigma_L}i\cancel{D}\Sigma_L]-{1\over 2}{\rm
Tr}[\overline{\Sigma_L^c}M_\Sigma
\Sigma_L+\overline{\Sigma_L}M_\Sigma^\ast
\Sigma_L^c]-\overline{L_L}\sqrt{2}Y_\Sigma^\dagger \Sigma_L^c
\tilde{H}-\tilde{H}^\dagger \overline{\Sigma_L^c}\sqrt{2}Y_\Sigma
L_L\;.
\end{eqnarray}
Here we have defined that  $\Psi\equiv \Sigma_L^-+\Sigma_L^{+c}$ with
$\Psi_L=\Sigma_L^-, \Psi_R=\Sigma_L^{+c}$. In the above $L_L \sim (1,2,-1)$ is the left-handed doublet lepton field, and
$\tilde H =i\sigma_2 H^* \sim (1,2, -1)$ is the Higgs doublet filed.

With a non-zero vacuum expectation value $\langle H\rangle =
v/\sqrt{2}$ for the Higgs field, the doublet leptons receive masses, and also mix the doublet and triplet leptons. The relevant terms in the
Lagrangian for mass matrices are given by
\begin{eqnarray}
\mathcal{L}_m&=&-
\left(
  \begin{array}{cc}
    \overline{l_R} & \overline{\Psi_R} \\
  \end{array}
\right)
\left(
  \begin{array}{cc}
    m_l & 0 \\
    Y_\Sigma v & M_\Sigma \\
  \end{array}
\right)
\left(
  \begin{array}{c}
    l_L \\
    \Psi_L \\
  \end{array}
\right) + h.c. \nonumber\\
&-& \left(
  \begin{array}{cc}
    \overline{\nu_L^c} & \overline{\Sigma_L^{0c}} \\
  \end{array}
\right) \left(
  \begin{array}{cc}
    0 & Y_\Sigma^Tv/2\sqrt{2} \\
    Y_\Sigma v/2\sqrt{2} & M_\Sigma/2 \\
  \end{array}
\right) \left(
  \begin{array}{c}
    \nu_L \\
    \Sigma^0_L \\
  \end{array}
\right) + h.c. \label{mass-matrix}
\end{eqnarray}
The second line above gives the seesaw mass matrix for neutrinos.

There are many different features for Type III seesaw compared with
the other types. Unlike Type I seesaw model, in this model the
doublet charged leptons mix with the triplet charged leptons leading
to tree level flavor changing neutral current involving changed
leptons~\cite{fcnc}. The fact that the heavy triplet leptons in Type
III seesaw have gauge interaction also leads to other different
phenomenology~\cite{flavor,pheno}. Different extensions of the
simplest model can also achieve different goals~\cite{models}.

For detailed studies, one needs to understand the mass matrices in Eq.~\ref{mass-matrix} and their diagonalization further.
The diagonalization of the mass matrices can be achieved by making unitary transformations on the triplet, the charged and neutral, leptons defined in the following
\begin{eqnarray}
&&\left(
    \begin{array}{c}
      l_{L,R} \\
      \Psi_{L,R} \\
    \end{array}
  \right)
  =U_{L,R}
  \left(
    \begin{array}{c}
      l_{mL,R} \\
      \Psi_{mL,R} \\
    \end{array}
  \right), \ \ \
\left(
    \begin{array}{c}
      \nu_{L} \\
      \Sigma^0_{L} \\
    \end{array}
  \right)
  =U_0
  \left(
    \begin{array}{c}
      \nu_{mL} \\
      \Sigma^0_{mL} \\
    \end{array}
  \right),
\end{eqnarray}
where $U_{L,R}$ and $U_0$ are $6\times6$ unitary matrices, for 3 light doublet and 3 heavy triplet lepton fields, which we
decompose into $3\times 3$ block matrices as
\begin{eqnarray}
&&U_L\equiv
\left(
  \begin{array}{cc}
    U_{Lll} & U_{Ll\Psi} \\
    U_{L\Psi l} & U_{L\Psi\Psi} \\
  \end{array}
\right), \ \ U_R\equiv \left(
  \begin{array}{cc}
    U_{Rll} & U_{Rl\Psi} \\
    U_{R\Psi l} & U_{R\Psi\Psi} \\
  \end{array}
\right), \ \ U_0\equiv \left(
  \begin{array}{cc}
    U_{0\nu\nu} & U_{0\nu\Sigma} \\
    U_{0\Sigma \nu} & U_{0\Sigma\Sigma} \\
  \end{array}
\right).
\end{eqnarray}

For our studies we need to know gauge and Higgs boson couplings to
leptonic fields. In the weak interaction basis, they can be written
as
\begin{eqnarray}
\mathcal{L}_{gauge}&=&
+e (\overline{\Psi}\gamma^\mu \Psi+\overline{l }\gamma^\mu l)A_\mu + g c_W (\overline{\Psi}\gamma^\mu \Psi+\overline{l }\gamma^\mu l)Z_\mu\nonumber \\
&-& {g \over c_W} ({1\over 2} \bar \nu_L \gamma^\mu \nu_L + {1\over 2} \bar l_L \gamma^\mu l_L + \bar l_R \gamma_\mu l_R) Z_\mu\nonumber\\
&-&g(\overline{\Psi_L}\gamma^\mu
\Sigma_L^0W^-_\mu+\overline{\Psi_R}\gamma^\mu
\Sigma_L^{0c}W^-_\mu)-{g\over
\sqrt{2}}\overline{l_L}\gamma^\mu \nu_LW_\mu^- + h.c. \;,\nonumber \\
\mathcal{L}_{Yukawa}&=&-(\overline{\nu_L}Y_\Sigma^\dagger
\Sigma_L^{0c}+\sqrt{2}\overline{l_L}Y_\Sigma^\dagger
\Psi_R){H^0\over \sqrt{2}}-\overline{l_R} m_l l_L{H^0\over
v}+ h.c.\;,
\end{eqnarray}
where $c_W = \cos\theta_W$.

In the mass eigen-state basis, the photon couplings to fermions are diagonal, but $Z$ couplings are more complicated. We have
\begin{eqnarray}
\mathcal{L}_{NCZ}
&\equiv&
\mathcal{L}_{NCZ}^A+\mathcal{L}_{NCZ}^B+(\mathcal{L}_{NCZ}^C+h.c.)+(\mathcal{L}_{NCZ}^D+h.c.)+\mathcal{L}_{NCZ}^E+\mathcal{L}_{NCZ}^F
\end{eqnarray}
where
\begin{eqnarray}
\mathcal{L}_{NCZ}^A&=&
gc_W[\overline{\Psi_m}V_{Z\Psi\Psi}^L\gamma^\mu P_L
\Psi_{m'}Z_\mu^0+\overline{\Psi_m}V_{Z\Psi\Psi}^R\gamma^\mu P_R
\Psi_{m'}Z_\mu^0],\nonumber\\
\mathcal{L}_{NCZ}^B&=&-{g\over
2c_W}\overline{\Sigma_{mL}^0}V^L_{Z\Sigma\Sigma}\gamma^\mu P_L
\Sigma_{m'L}^0Z^0_\mu\;,\nonumber\\
\mathcal{L}_{NCZ}^C&=&{g\over
2c_W}\overline{\nu_m}V^L_{Z\nu\Sigma}\gamma^\mu P_L \Sigma_{m'L}^0
Z_\mu^0,\nonumber\\
\mathcal{L}_{NCZ}^D&=&{g\over
\sqrt{2}c_W}[\overline{l_m}V_{Zl\Psi}^L\gamma^\mu P_L
\Psi_{m'}Z_\mu^0+\overline{l_m}V_{Zl\Psi}^R\gamma^\mu P_R
\Psi_{m'}Z_\mu^0],\nonumber\\
\mathcal{L}_{NCZ}^E&=&-{g\over
2c_W}\overline{\nu_m}V^L_{Z\nu\nu}\gamma^\mu P_L \nu_{m'}Z^0_\mu,\\
\mathcal{L}_{NCZ}^F&=& -{g\over
c_W}[\overline{l_m}V_{Zll}^L\gamma^\mu P_L
l_{m'}Z_\mu^0+\overline{l_m}V_{Zll}^R\gamma^\mu P_R l_{m'}Z_\mu^0],\nonumber
\end{eqnarray}
and
\begin{eqnarray}
&&V_{Z\Psi\Psi}^L=I-{1\over
2 c_W^2}U^\dagger_{Ll\Psi}U_{Ll\Psi}, \ \
V_{Z\Psi\Psi}^R=I - {1\over
c_W^2}U^\dagger_{Rl\Psi}U_{Rl\Psi},\ \ V_{Z\Sigma\Sigma}^L=U^\dagger_{0\nu\Sigma}U_{0\nu\Sigma}\;,\nonumber\\
&&V_{Z\nu\Sigma}^L=-U^\dagger_{0\nu\nu}U_{0\nu\Sigma},\ \
V_{Zl\Psi}^L=-{1\over
\sqrt{2}}U^\dagger_{Lll}U_{Ll\Psi}, \ \
V_{Zl\Psi}^R=-\sqrt{2} U^\dagger_{Rll}U_{Rl\Psi},\nonumber\\
&&V_{Z\nu\nu}^L=U^\dagger_{0\nu\nu}U_{0\nu\nu},\ \
V_{Zll}^L=-c_W^2I +{1\over
2}U^\dagger_{Lll}U_{Lll}, \ \
V_{Zll}^R=-c_W^2I +U^\dagger_{Rll}U_{Rll}.
.\nonumber
\end{eqnarray}

For the charged current interactions, we have
\begin{eqnarray}
\mathcal{L}_{CC}
&\equiv&
(\mathcal{L}_{CC}^A+\mathcal{L}_{CC}^B+\mathcal{L}_{CC}^C+\mathcal{L}_{CC}^D+h.c.)
\end{eqnarray}
where
\begin{eqnarray}
\mathcal{L}_{CC}^A&=&-{g\over \sqrt{2}}[\overline{\Psi_m}V_{\Psi
\Sigma}^{L}\gamma^\mu P_L
\Sigma_{m'L}^0W_\mu^-+\overline{\Psi_m}V_{\Psi \Sigma}^{R}\gamma^\mu
P_R \Sigma_{m'L}^{0c}W_\mu^-],\nonumber\\
\mathcal{L}_{CC}^B&=&-{g\over
\sqrt{2}}[\overline{l_m}V^L_{l\Sigma}\gamma^\mu P_L \Sigma_{m'L}^0
W^-_\mu+\overline{l_m}V^R_{l\Sigma}\gamma^\mu P_R
\Sigma_{m'L}^{0c}W^-_\mu],\nonumber\\
\mathcal{L}_{CC}^C&=&-{g\over \sqrt{2}}[\overline{\Psi_m}V_{\Psi
\nu}^{L}\gamma^\mu P_L \nu_{m'L}W_\mu^-+\overline{\Psi_m}V_{\Psi
\nu}^{R}\gamma^\mu
P_R \nu_{m'L}^{c}W_\mu^-],\\
\mathcal{L}_{CC}^D&=&-{g\over
\sqrt{2}}[\overline{l_m}V^L_{l\nu}\gamma^\mu P_L \nu_{m'L}
W^-_\mu+\overline{l_m}V^R_{l\nu}\gamma^\mu P_R \nu_{m'L}^{c}W^-_\mu]\;,\nonumber
\end{eqnarray}
and
\begin{eqnarray}
&&V_{\Psi\Sigma}^L=U^\dagger_{Ll\Psi}U_{0\nu\Sigma}+\sqrt{2}U^\dagger_{L\Psi\Psi}U_{0\Sigma\Sigma}, \ \
V_{\Psi\Sigma}^R=\sqrt{2}U_{R\Psi \Psi}^\dagger U_{0\Sigma\Sigma}^\ast,\nonumber\\
&&V_{l\Sigma}^L=U_{Lll}^\dagger U_{0\nu\Sigma}+\sqrt{2}U_{L\Psi
l}^\dagger U_{0\Sigma\Sigma}, \ \
V_{l\Sigma}^R=\sqrt{2}U_{R\Psi l}^\dagger U_{0\Sigma\Sigma}^\ast,\nonumber\\
&&V_{\Psi\nu}^L=U^\dagger_{Ll\Psi}U_{0\nu\nu}+\sqrt{2}U^\dagger_{L\Psi\Psi}U_{0\Sigma\nu},
\ \
V_{\Psi\nu}^R=\sqrt{2}U_{R\Psi \Psi}^\dagger U_{0\Sigma\nu}^\ast,\\
&&V^L_{l\nu}\equiv V_{PMNS}=U_{Lll}^\dagger
U_{0\nu\nu}+\sqrt{2}U_{L\Psi l}^\dagger U_{0\Sigma\nu}, \ \
V^R_{l\nu}=\sqrt{2}U^\dagger_{R\Psi l}U^\ast_{0\Sigma \nu}.\nonumber
\end{eqnarray}
In the above we have made the approximation $V^L_{l\nu} = V_{PMNS}$.
Strictly speaking $V^L_{l\nu}$ is not unitary as the usual
definition of the unitary $3\times 3$ $V_{PMNS}$ matrix.  The
correction is at the order of $\mathcal{O}(m_\nu/M_\Sigma)$. It is a
good approximation since we are working with the small light and
heavy neutrino mixing scenario.

One finds an interesting relation
\begin{eqnarray}
V^{L\ast}_{l\Sigma}M_{\Sigma N}^{diag}V_{l\Sigma}^{L\dagger}&=&
-V_{PMNS}^\ast m_\nu^{diag}V_{PMNS}^\dagger+m_l^{diag}U^T_{R\Psi
l}U_{L\Psi l}+U_{L\Psi l}^TU_{R\Psi l}m_l^{diag}\;. \label{VVL}
\end{eqnarray}
The detailed derivation is given in Appendix A. A similar relation
without the last two terms on the right in the above equation for
Type I seesaw has been derived in Ref.~\cite{CI,processing}.

The physical Higgs $H^0$ interactions with leptonic fields, in the mass eigen-state basis, are given by
\begin{eqnarray}
\mathcal{L}_{S}
&\equiv&\mathcal{L}_{S}^A +\mathcal{L}_{S}^B+h.c.\;,
\end{eqnarray}
where
\begin{eqnarray}
\mathcal{L}_{S}^A&=&-\overline{\nu_m}V_{S\nu\Sigma}^RP_R\Sigma_{m'L}^{0c}H^0,\nonumber\\
\mathcal{L}_{S}^B&=&-[\overline{l_m}V_{Sl\Psi}^LP_L\Psi_{m'}+\overline{l_m}V_{Sl\Psi}^RP_R\Psi_{m'}]H^0\;,
\end{eqnarray}
and
\begin{eqnarray}
&&V_{S\nu\Sigma}^R=[U^\dagger_{0\nu\nu}Y_\Sigma^\dagger U_{0\Sigma\Sigma}^\ast+U^\dagger_{0\Sigma\nu}Y_\Sigma^\ast U_{0\nu\Sigma}^\ast]/\sqrt{2},\nonumber\\
&&V_{Sl\Psi}^L=U^\dagger_{R\Psi l}Y_\Sigma U_{Ll\Psi} + {1\over
\sqrt{2} v} U^\dagger_{Lll} m_l U_{Rl\Psi}, \ \
V_{Sl\Psi}^R=U^\dagger_{Lll}Y_\Sigma^\dagger U_{R\Psi\Psi} + {1\over
\sqrt{2} v} U^\dagger_{Rll} m_l U_{Ll\Psi}\;.
\end{eqnarray}

In principle, the matrices $U_{L,R}$ and $U_{0}$ can be expressed in
terms of $Y_\Sigma$, $m_l$ and $M_\Sigma$. Since for seesaw
mechanism to work, $Y_\Sigma v M^{-1}_\Sigma$ should be small, one
can expand $U_{L,R}$ and $U_{0}$ in powers of $Y_\Sigma v
M^{-1}_\Sigma$ to keep track of the leading order contributions. For
this purpose, it is convenient to write the leading order
expressions up to $Y_\Sigma^2v^2M^{-2}_\Sigma$ in the basis where
$m_l$ and $M_\Sigma$ are already diagonalized, without loss of
generality. The following results have been obtained in the literature~\cite{fcnc}
\begin{eqnarray}
&&U_{Lll} = 1- \epsilon\;,\;\;U_{Ll\Psi} = Y^\dagger_\Sigma
M^{-1}_\Sigma v\;,\;\;\;\;\;\;\;
U_{L\Psi l} = - M^{-1}_\Sigma Y_\Sigma v\;,\;\;\;\;\;\;U_{L\Psi\Psi} = 1-\epsilon'\;,\nonumber\\
&&U_{Rll} = 1\;,\;\;\;\;\;\;\;\;U_{Rl\Psi} = m_l Y^\dagger_\Sigma
M^{-2}_\Sigma v\;,\;\;\;
U_{R\Psi l} = - M^{-2}_\Sigma Y_\Sigma m_l v\;,\;\;U_{R\Psi\Psi} = 1\;,\nonumber\\
&&U_{0\nu\nu} = (1- \epsilon/2)V_{PMNS}\;,\;\; U_{0\nu \Sigma} =
Y^\dagger_\Sigma M^{-1}_\Sigma v/\sqrt{2}\;,\;\;
U_{0\Sigma \nu} = - M^{-1}_\Sigma Y_\Sigma U_{0\nu\nu} v/\sqrt{2}\;,\nonumber\\
&&U_{0\Sigma\Sigma } =
1-\epsilon'/2\;,\;\;\;\;\;\;\;\;\;\;\;\;\;\;\; \epsilon =
Y^\dagger_\Sigma M^{-2}_\Sigma Y_\Sigma v^2/2\;,\;\;\;\; \epsilon'
= M^{-1}_\Sigma Y_\Sigma Y^\dagger_\Sigma M^{-1}_\Sigma
v^2/2\;.\nonumber \label{approx}
\end{eqnarray}

To leading order in $Y_\Sigma v M^{-1}_\Sigma$, we have
interaction terms involving heavy triplet leptons as
\begin{eqnarray}
&&\mathcal{L}_{NC(A+Z)}=e\overline{E}\gamma^\mu E A_\mu + gc_W\overline{E}\gamma^\mu E Z^0_\mu \;,\nonumber\\
&&\mathcal{L}_{NCZ}={g\over
2c_W}[\overline{\nu}(V_{PMNS}^\dagger V_{lN}\gamma^\mu
P_L-V_{PMNS}^TV^{\ast}_{lN}\gamma^\mu P_R)N_{}
+\sqrt{2}\overline{l}V_{lN}\gamma^\mu
P_LE_{}+ h.c.]Z_\mu^0\;,\nonumber\\
&&\mathcal{L}_{CC}=-g [ \overline{E}\gamma^\mu N
+{1\over \sqrt{2}}\overline{l}V_{lN}\gamma^\mu
P_LN_{} + \overline{E}V_{lN}^TV_{PMNS}^\ast\gamma^\mu
P_R\nu_{}]W^-_\mu+h.c.\;,\label{SSS}\\
&&\mathcal{L}_{S}={g\over 2M_W}[\overline{\nu}(V_{PMNS}^\dagger
V_{lN}M_{N}^{diag}P_R+V_{PMNS}^TV^{\ast}_{lN}M_{N}^{diag}P_L)N_{}+
\sqrt{2}\overline{l}V_{lN}M_{E}^{diag}P_RE_{}]H^0+h.c.\;,\nonumber
\end{eqnarray}
with $V_{lN}\equiv V_{l\Sigma}^L = -Y^\dagger_\Sigma v
M^{-1}_\Sigma/\sqrt{2}$. In the above, all fields are in mass
eigen-states. The $E$, $N$ and $M^{diag}_{E},M^{diag}_{N}$ are mass
eigen-states of $\Psi$, $\Sigma$, and the eigen-mass matrices,
respectively. Note that the interactions involving
light neutrinos in the above have the additional $V_{PMNS}$ factor compared with those involving light charged leptons.

To the same order, we also have
\begin{eqnarray}
V^{\ast}_{lN}M_{N}^{diag}V_{lN}^{\dagger}&=& -V_{PMNS}^\ast
m_\nu^{diag}V_{PMNS}^\dagger\;.\label{VL}
\end{eqnarray}
This equation plays an important role in constraining the elements in the coupling matrix  $V_{lN}$.

\section{Constraints on The Physical Parameters}

In the study of decay of $E$ and $N$ into SM particles, the
interaction matrix $V_{lN}$ plays an important role. Knowledge about
it is crucial. In this section we study constraints on $V_{lN}$ and
the decay branching ratios of $E$ and $N$. Eq.~\ref{VL} provides
very important constraints on $V_{lN}$. As have been mentioned
before that there are two classes of solutions, the cases with small and large
mixing between light and heavy leptons. The small mixing case is
characterized by the fact that in the limit, $m_\nu^{diag}$ goes
to zero, the elements in $V_{l N}$ also go to zero, and the
elements in $V_{l N}$ are of order $(m_\nu/M_{\nu_R})^{1/2}$. But
with more than one generations it is possible to
have non-trivial solutions for Eq.~\ref{VL} which have large mixing between light and heavy leptons, as have been shown
in Refs.~\cite{large} and \cite{largeIII}. The cases with small and large mixing have
very different experimental signatures.  The small mixing solution
case will lead to a visible displaced vertex in the detector at the
LHC. While for the large mixing case, one can also study
single heavy lepton production~\cite{largeIII}.  The aim of this paper
is to study the correlations of heavy lepton productions and decays
with low energy neutrino oscillation parameters and masses.
Therefore in this section we will discuss constraints on the physical
parameters for small mixing solutions.

\subsection{Neutrino Masses, Mixing and the Coupling Matrix $V_{lN}$}

On the right-handed side of Eq.~\ref{VL}, the parameters involved
are in principle measurable parameters, the neutrino masses and
mixing angles. Therefore in order to understand the constraints we
need to know as much as these parameters. As has been mentioned
before that in our case the $V_{PMNS}$ is, in general, not unitary.
However, since the deviation is of order $Y_\Sigma v/M_\Sigma$, to a
good approximation, we can neglect these corrections and use a
unitary matrix to represent it which can be written as
\begin{equation}
V_{PMNS}= \left(
\begin{array}{lll}
 c_{12} c_{13} & c_{13} s_{12} & e^{-\text{i$\delta $}} s_{13}
   \\
 -c_{12} s_{13} s_{23} e^{\text{i$\delta $}}-c_{23} s_{12} &
   c_{12} c_{23}-e^{\text{i$\delta $}} s_{12} s_{13} s_{23} &
   c_{13} s_{23} \\
 s_{12} s_{23}-e^{\text{i$\delta $}} c_{12} c_{23} s_{13} &
   -c_{23} s_{12} s_{13} e^{\text{i$\delta $}}-c_{12} s_{23} &
   c_{13} c_{23}
\end{array}
\right)\times \text{diag} (e^{i \Phi_1/2}, 1, e^{i
\Phi_2/2})\;,\label{pmns}
\end{equation}
where $s_{ij}=\sin{\theta_{ij}}$, $c_{ij}=\cos{\theta_{ij}}$, $0 \le
\theta_{ij} \le \pi/2$ and $0 \le \delta \le 2\pi$. The phase
$\delta$ is the Dirac CP phase, and $\Phi_i$ are the Majorana
phases. The experimental constraints on the neutrino masses and
mixing parameters, at $2\sigma$ level~\cite{Schwetz}, are
\begin{eqnarray}
&&7.25 \times 10^{-5} \ev^2 \  <  \Delta m_{21}^2  < \  8.11 \times 10^{-5} \ev^2,\;\;2.18 \times 10^{-3} \ev^2 \
<  |\Delta m_{31}^2| < \  2.64 \times 10^{-3} \ev^2, \nonumber\\
                  && 0.27 \  <  \sin^2{\theta_{12}}  < \  0.35,\;\;0.39 \  <  \sin^2{\theta_{23}}  <\  0.63, \;\;
                           \sin^2{\theta_{13}}  <\  0.040,
\end{eqnarray}
and no constraints on the phases. The neutrino masses are bounded by
$\sum_{i} m_{i} < \ 1.2 \ \ev$~\cite{CMB}.

For a complete discussion of these constraints see
reference~\cite{review}. Following the convention, we denote the
case $\Delta m_{31}^2 > 0$ as the normal hierarchy (NH) and otherwise the inverted hierarchy
(IH). In our later discussions unless specified for the input values of relevant parameters, when scanning the parameters space we will always
allow $s_{12,13,23}$ to run the above allowed ranges, and the lightest neutrino mass for
NH and IH cases to run the range $10^{-4 } \sim 0.4$ eV.

Eq.~\ref{VL} relates $V_{lN}$ to low energy measurable quantities,
but the elements in $V_{lN}$ can not be fully determined. Certain assumptions or new
parameters need to be introduced to describe the ranges for the elements in
$V_{lN}$. In the following we consider in details for the size of
$V_{lN}$ with the Majorana phases set to zero first, and then
comment on the effects of non-zero Majorana phases.

\subsubsection{Case I: Degenerate Heavy Triplet Leptons}

We start with a simple but interesting case where the heavy triplet
leptons are degenerate. In this case Eq.~\ref{VL} becomes simple on
the left hand side with $V^*_{lN}M_N^{diag} V^\dagger_{lN} = M_N
\sum_{j=1,2,3} V^{ij*2}_{lN}$. Here the superscript ``$i$'' runs
over the three light generation leptons and ``$j$'' runs over the
three heavy triplet leptons. $M_N = M_1 = M_2 =M_3$ is the heavy
triplet mass. We have a simple expression from Eq.~\ref{VL}
\begin{eqnarray}
M_N \sum_{j=1,2,3} (iV_{lN}^{ij*})^{2}&=&
{(V^\ast_{PMNS}m_\nu^{diag}V^\dagger_{PMNS})_{ii} }\equiv
{M_\nu^{ii}}, \ \ i=e,\mu,\tau
\end{eqnarray}
where $M_\nu = V^\ast_{PMNS}m_\nu^{diag}V^\dagger_{PMNS}$.

Explicitly we have
\begin{eqnarray}
M_N\sum_j (iV_{lN}^{ej*})^{2}&=&c_{13}^2s_{12}^2m_2+c_{12}^2c_{13}^2e^{-i\Phi_1}m_1+s_{13}^2e^{i(2\delta-\Phi_2)}m_3,\nonumber\\
M_N\sum_j (iV_{lN}^{\mu
j*})^{2}&=&(c_{12}c_{23}-s_{12}s_{13}s_{23}e^{-i\delta})^2m_2+(c_{23}s_{12}+
c_{12}s_{13}s_{23}e^{-i\delta})^2e^{-i\Phi_1}m_1\nonumber \\
&+&c_{13}^2s_{23}^2e^{-i\Phi_2}m_3,\nonumber\\
M_N \sum_j (iV_{lN}^{\tau
j*})^{2}&=&(c_{12}s_{23}+c_{23}s_{12}s_{13}e^{-i\delta})^2m_2+(s_{12}s_{23}-
c_{12}c_{23}s_{13}e^{-i\delta})^2e^{-i\Phi_1}m_1\nonumber \\
&+&c_{13}^2c_{23}^2e^{-i\Phi_2}m_3.\label{casei}
\end{eqnarray}

If the phases in $V_{l N}$ are all zero, the right-handed sides in
the above equations are all real. We can formally write
\begin{eqnarray}
M_N \sum_{j=1,2,3} (V_{lN}^{ij*})^{2} = M_N \sum_{j=1,2,3}
|V_{lN}^{ij}|^{2}\;.
\end{eqnarray}
Later when we will refer this particular case as  Case I.

If indeed the three
heavy triplet leptons are degenerate or almost degenerate,
experimentally when they are produced, one would not be able to
distinguish them and therefore must sum over the heavy ones. The above
equation allows one to fix the couplings completely in terms of low
energy parameters. We emphases that this is true only for the case
that all phases in $V_{lN}$ are zero.

\begin{figure}[tb]
\begin{center}
\begin{tabular}{cc}
\includegraphics[scale=1,width=8cm]{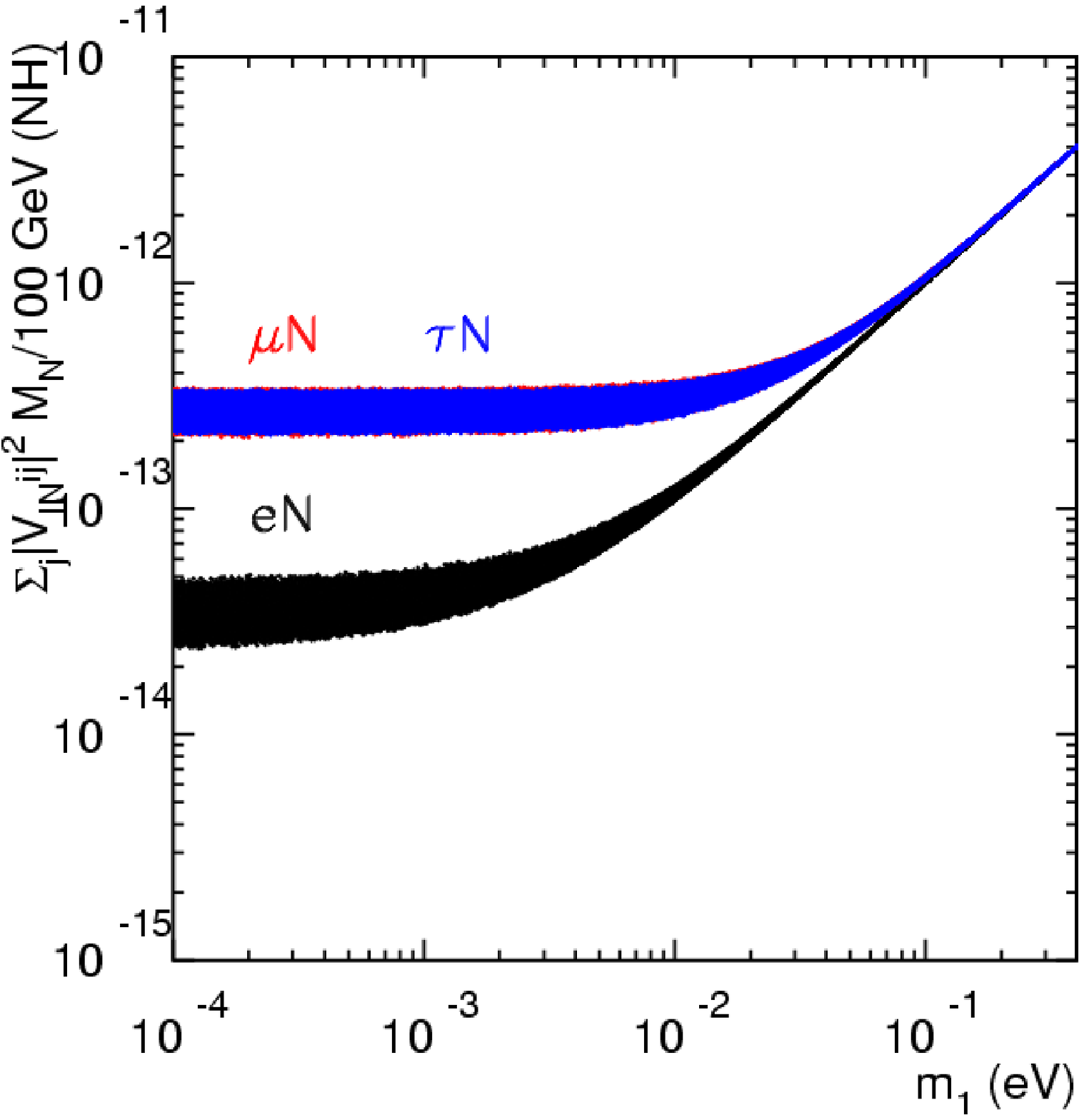}
\includegraphics[scale=1,width=8cm]{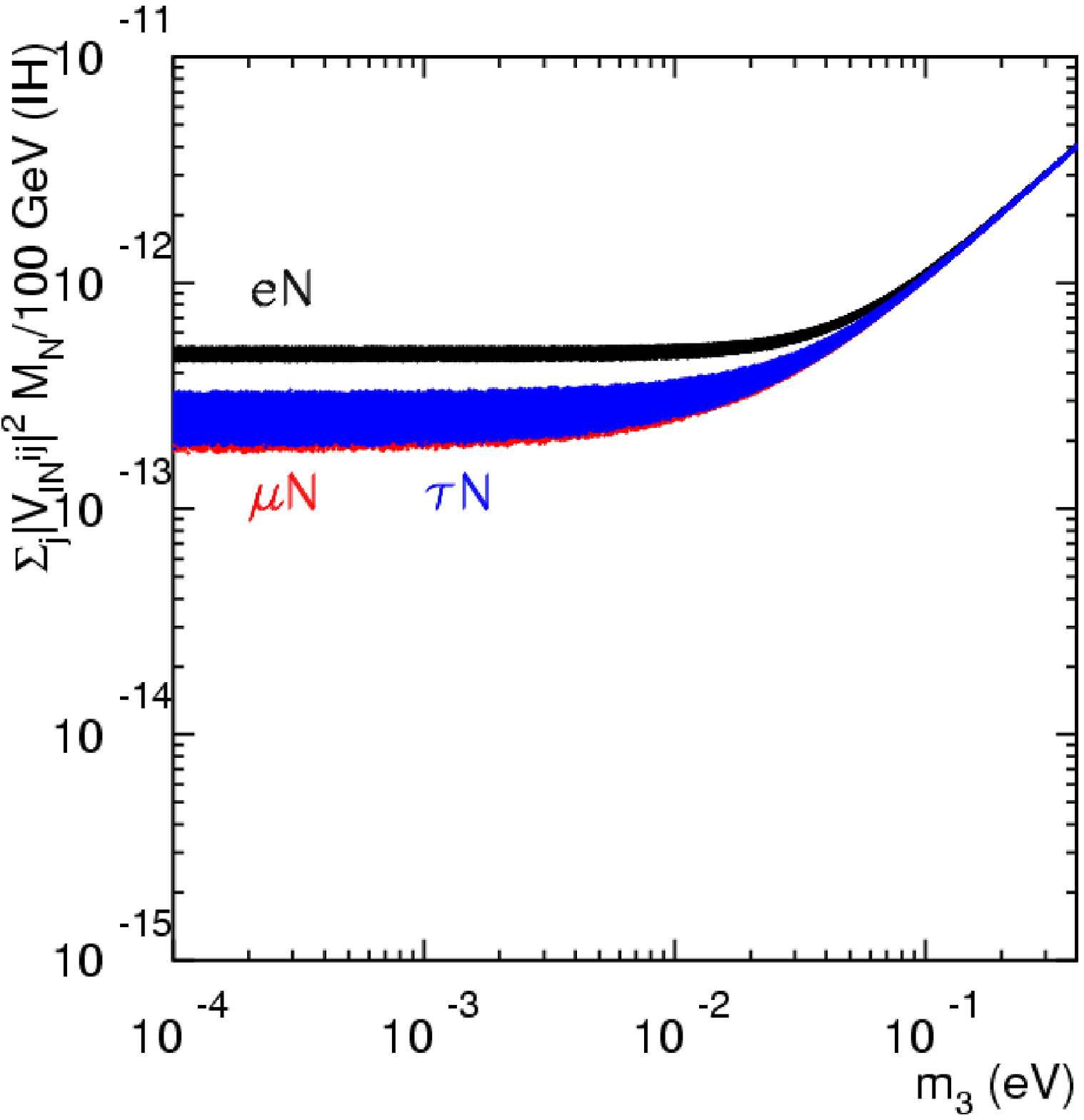}
\end{tabular}
\end{center}
\caption{$\sum_{j=1,2,3}|V_{lN}^{ij}|^2 M_N/100~{\rm GeV}$ vs. the
lightest neutrino mass for NH and IH for Case I without any phases,
where areas for $\mu$ and $\tau$ flavors overlap (same for other
figures). } \label{VLlN}
\end{figure}

The experimental information on neutrino masses and mixing indicates that the
neutrino mass matrix $M_\nu$ presents the following patterns
\begin{eqnarray}
&&M_\nu^{ee}\ll M_\nu^{\mu\mu}, M_\nu^{\tau\tau} \ \ \ {\rm for \ \ NH},\nonumber \\
&&M_\nu^{ee}> M_\nu^{\mu\mu}, M_\nu^{\tau\tau} \ \ \ {\rm for \ \
IH}.
\end{eqnarray}
More detailed discussions can be found in Ref.~\cite{TypeII}. We
plot the allowed values for the normalized couplings, $\sum_j
|V_{lN}^{ij}|^2M_N/100$ GeV, of each lepton flavor for this case
in Fig.~\ref{VLlN},
as a function of the lightest neutrino mass for
both the NH (left panels) and the IH
(right panels) cases. We see two distinctive regions in terms of the
lightest neutrino mass. In the case $m_{1(3)}<10^{-2}$ eV,
$\sum_j|V_{lN}^{ej}|^2\ll \sum_j|V_{l N}^{\mu j}|^2, \sum_j|V_{l
N}^{\tau j}|^2$ for NH and $\sum_j|V_{lN}^{ej}|^2> \sum_j|V_{l
N}^{\mu j}|^2, \sum_j|V_{l N}^{\tau j}|^2$ for IH. On the other
hand, for $m_{1(3)}>10^{-2}$ eV, we have the quasi-degenerate
spectrum $\sum_j|V_{lN}^{ej}|^2\approx \sum_j|V_{l N}^{\mu
j}|^2\approx \sum_j|V_{l N}^{\tau j}|^2$ as expected.

\subsubsection{Case II: A Class of Solutions for Small Mixing between Light and Heavy Neutrinos}

As already mentioned before the relation in Eq.\ref{VL} we can not
completely fix the form for $V_{lN}$, but we find that $V_{lN}$ can
be written in the following form using the Casas-Ibarra
parametrization~\cite{CI}
\begin{eqnarray}
V_{lN}=iV_{PMNS}(m_\nu^{diag})^{1/2} \Omega(w_{21},w_{31},w_{32}) (M_N^{diag})^{-1/2},\label{vlno}
\end{eqnarray}
where $(m_\nu^{diag})^{1/2} = diag(m_1^{1/2},m_2^{1/2},m_3^{1/2})$,
$(M^{diag}_N)^{-1/2}= diag(M^{-1/2}_1, M^{-1/2}_2, M^{-1/2}_3)$ and
$\Omega(w_{21},w_{31},w_{32})$ satisfies $\Omega \Omega^T = 1$. In
general the elements in $\Omega$ can be unbounded if complex
variables $w_{ij}$ are used. Since we are interested in
small mixing between light and heavy neutrinos, we will limit
ourselves to the case where the angles $w_{ij}$ defined in Appendix
B are real and allow them to vary in the ranges $0 < w_{21}, w_{31},
w_{32}< 2\pi$. More details about $\Omega$ are given in Appendix B
where explicit expressions for $V_{lN}^{ij} (i=e,\mu,\tau, j=1,2,3)$
are collected.

If $\Omega= 1$, the expressions for $|V_{lN}|^2$ are simple. We have
\begin{eqnarray}
&&M_1(|V_{lN}^{e1}|^2,\; |V_{lN}^{\mu 1}|^2, \;|V_{lN}^{\tau 1}|^2)
=m_1(c_{12}^2c_{13}^2,\;
|s_{12}c_{23}+c_{12}s_{13}s_{23}e^{i\delta}|^2,\;
|s_{12}s_{23}-c_{12}s_{13}c_{23}e^{i\delta}|^2),\nonumber\\
&&M_2(|V_{lN}^{e2}|^2,\;|V_{lN}^{\mu 2}|^2,\;|V_{lN}^{\tau 2}|^2)
=m_2(c_{13}^2s_{12}^2,\;
|c_{12}c_{23}-s_{12}s_{13}s_{23}e^{i\delta}|^2,\;
|c_{12}s_{23} + s_{12}s_{13}c_{23}e^{i\delta}|^2),\nonumber\\
&&M_3(|V_{lN}^{e3}|^2,\;|V_{lN}^{\mu 3}|^2,\;|V_{lN}^{\tau 3}|^2) =m_3(s_{13}^2,\;
c_{13}^2s_{23}^2,\;
c_{13}^2c_{23}^2).
\end{eqnarray}
Using known data on the mixing parameters, it is easy to see that $|V_{lN}^{e1}|^2>|V_{lN}^{\mu 1}|^2,|V_{lN}^{\tau 1}|^2$,
$|V_{lN}^{e2}|^2\approx |V_{lN}^{\mu 2}|^2\approx |V_{lN}^{\tau 2}|^2$ , and $|V_{lN}^{\mu 3}|^2,|V_{lN}^{\tau 3}|^2>|V_{lN}^{e3}|^2$ both for
NH and IH cases.

To have a better understanding about the size of $|V_{lN}|^2$, we
scan all $w_{21}$, $w_{31}$ and $w_{32}$ in the range $0< w_{ij}<
2\pi$. $\Omega = 1$ case is obtained by setting $\sin(w_{ij}) = 0$.
Case I discussed earlier, is also a special case of Case II.

\begin{figure}[tb]
\begin{center}
\begin{tabular}{cc}
\includegraphics[scale=1,width=8cm]{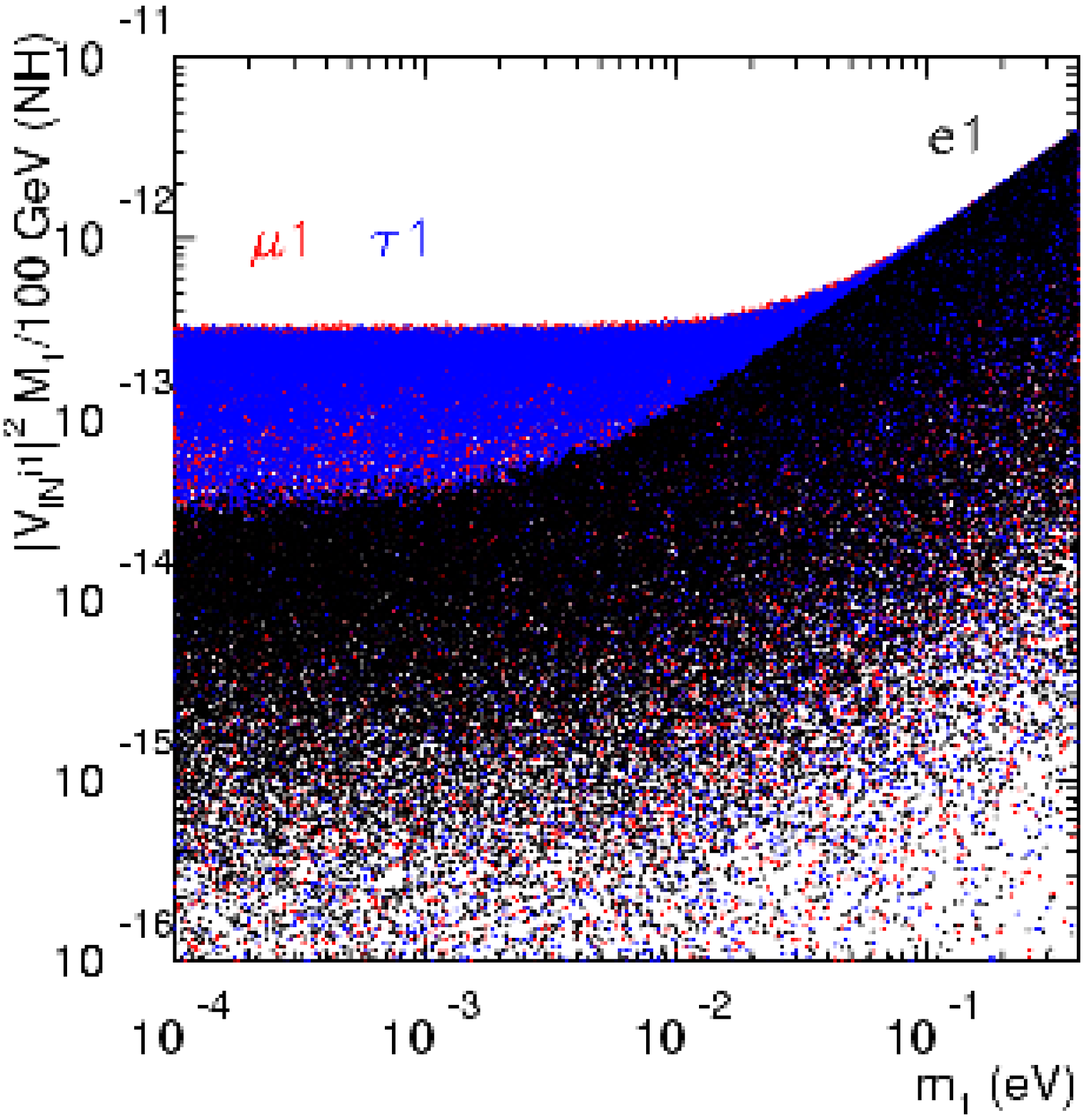}
\includegraphics[scale=1,width=8cm]{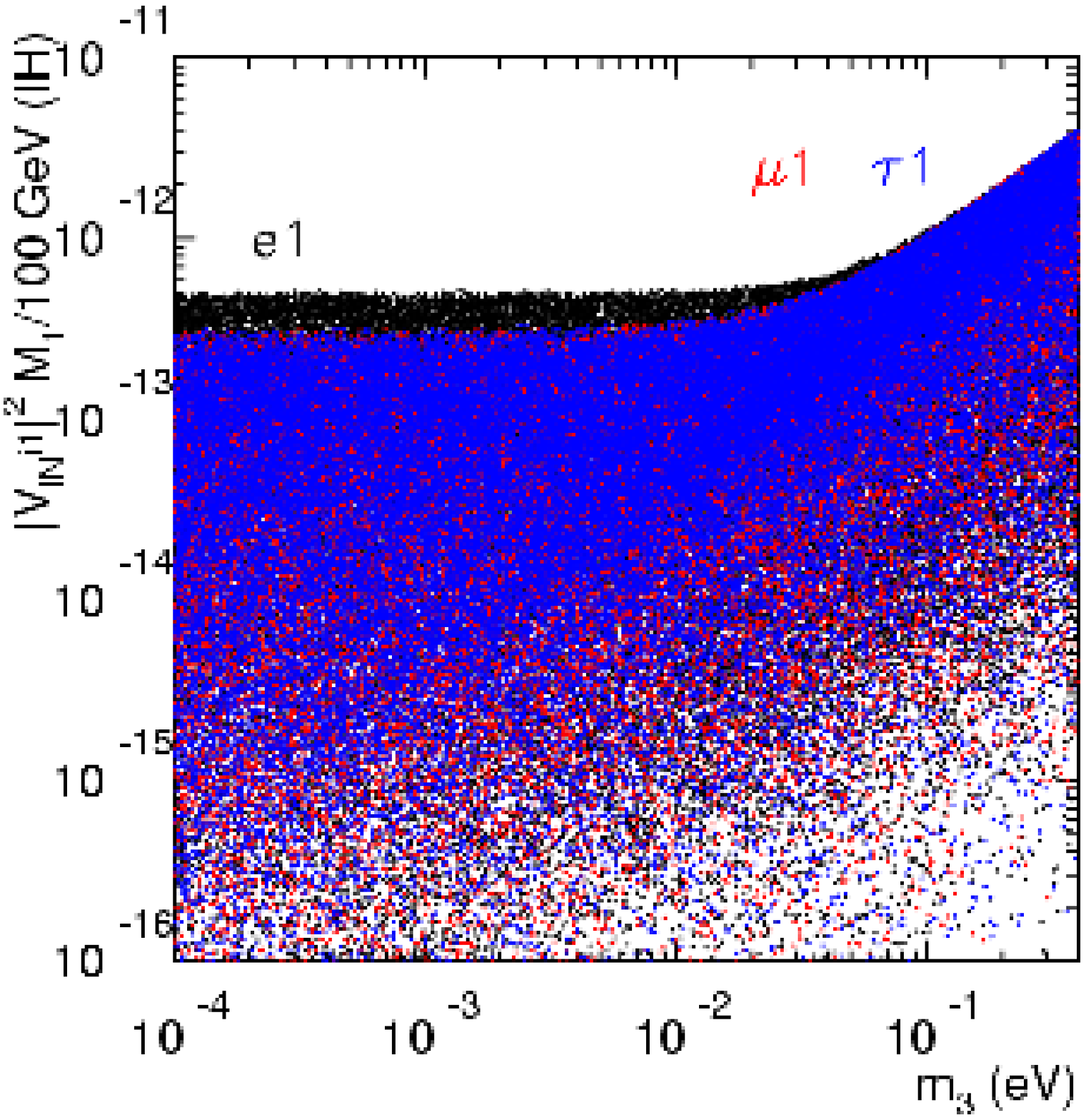}
\end{tabular}
\end{center}
\caption{$|V_{lN}^{i1}|^2 M_1/100~{\rm GeV}$ vs. the lightest
neutrino mass for NH and IH, when $0<w_{ij}<2\pi$ for Case II
without any phases in $\Omega$. } \label{yl1m1all}
\end{figure}

The allowed range for the normalized coupling,
$|V_{lN}^{i1}|^2M_1/100~{\rm GeV}$ is shown in Fig.~\ref{yl1m1all}
as a function of the light neutrino mass in each spectrum. The results for the NH and IH
cases are displayed in the left and
right panels, respectively. The ranges of
$|V_{lN}^{i2}|^2M_2/100~{\rm GeV}$ and $|V_{lN}^{i3}|^2M_3/100~{\rm
GeV}$ are almost the same and will not be shown separately. Unlike for Case I, by just looking at the
absolute values $|V_{lN}|^2$ along, it is difficult to distinguish
neutrino mass hierarchies. But there are allowed
regions for $|V_{lN}^{e1}|^2$ for the NH and IH which do not
overlap. In this particular situation, there is still a hope to
distinguish different neutrino mass hierarchies.

\subsection{Other Constraints}

In Type III seesaw model, the heavy triplet lepton masses are free
parameters. The current constraint on heavy charged lepton masses comes
from the direct search at collider~\cite{PDG}, $M_E\gtrsim 100~{\rm
GeV}$, which we will use as a lower bound and take $M_{E,N}$ to be
larger than Higgs boson mass $M_H (> 114~\rm GeV)$. Because the
charged and neutral heavy leptons have different mass matrices, they
are in general different. However the splits are small compared with
the common mass term $M_{\Sigma}$. We will take $M_N=M_E$ as the common
triplet mass in our later discussions.

Other constraints come from electroweak precision measurements, LFV
processes, and meson rare
decays~\cite{PDG,Aguila,fcnc,flavor,leptonEWPD,Han:2006ip}, we have checked
that these limits do not provide strong constraints for heavy
triplet lepton decays and productions.

\section{Heavy Triplet Lepton Decays}
In this section we study the main features of the heavy triplet lepton decays taking into account the constraints on $|V_{lN}|^2$ from the neutrino
mass and mixing data as discussed in the previous section. From Eq.~\ref{VL} one therefore anticipates that $E$ and $N$
decays could be different. We
explore this in more details in the following.

\subsection{Main Features of Heavy Triplet Lepton Decays}
\begin{figure}[tb]
\begin{center}
\begin{tabular}{cc}
\includegraphics[scale=1,width=8cm]{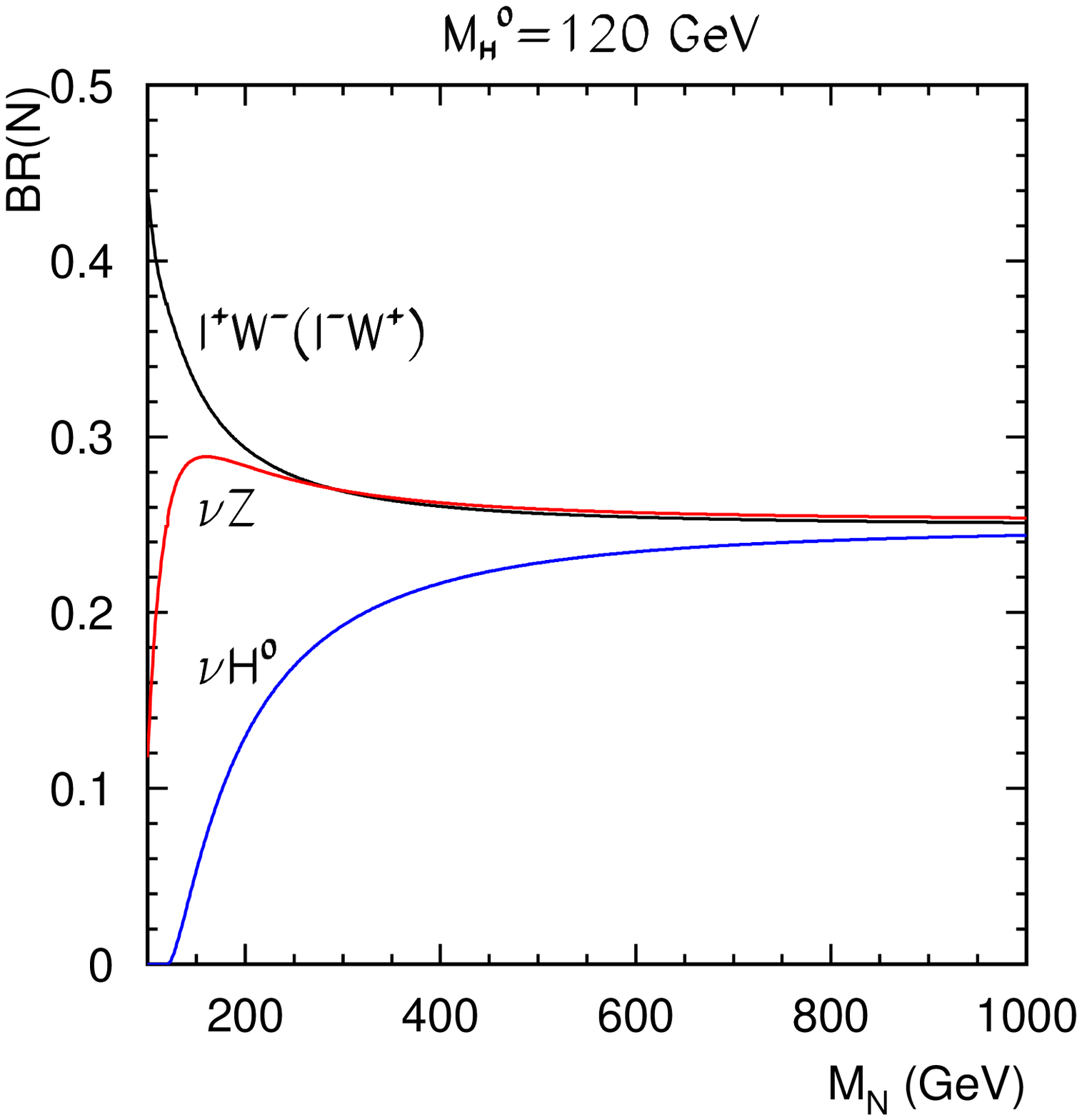}
\includegraphics[scale=1,width=8cm]{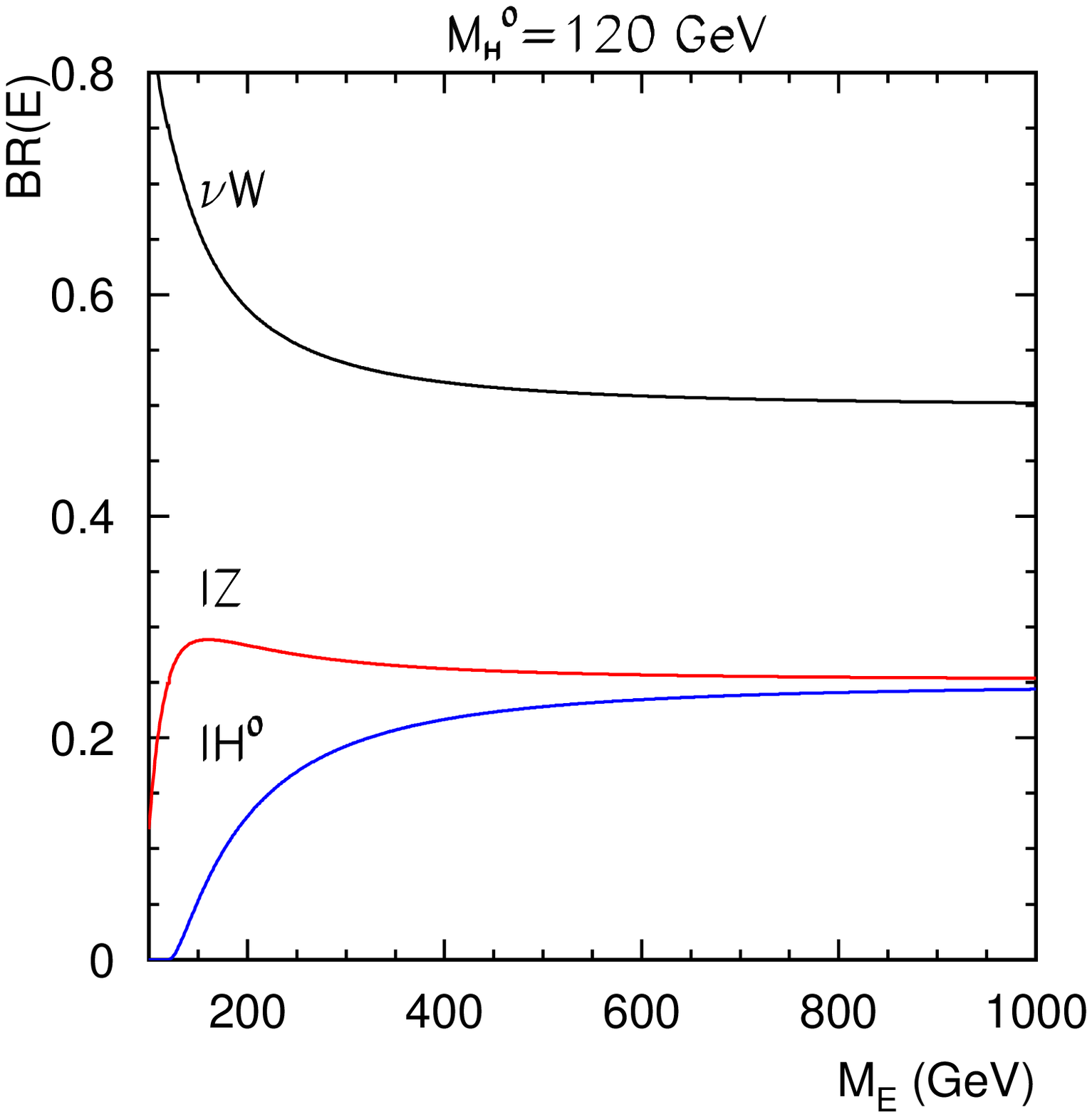}
\end{tabular}
\end{center}
\caption{Branching fractions of $N$ and $E^\pm$. }
\label{br}
\end{figure}

The partial widths for $N$ and $E$ decays are given by
~\cite{Han:2006ip,Aguila}
\begin{eqnarray}
&&\Gamma(N_i\to \ell^-W^+)= \Gamma(N_i\to \ell^+W^-)={g^2\over
64\pi}|V_{lN}^{\ell i}|^2{M_{N_i}^3\over M_W^2}\left(1-{M_W^2\over
M_{N_i}^2}\right)\left(1+{M_W^2\over M_{N_i}^2}-2{M_W^4\over
M_{N_i}^4}\right)\label{N1},\nonumber\\
&&\sum^3_{m=1}\Gamma(N_i\to \nu_mZ)={g^2\over 64\pi
c_W^2}\sum_{\ell=e}^\tau|V_{lN}^{\ell i}|^2{M_{N_i}^3\over M_Z^2}\left(1-{M_Z^2\over
M_{N_i}^2}\right)\left(1+{M_Z^2\over M_{N_i}^2}-2{M_Z^4\over
M_{N_i}^4}\right)\label{N2},\nonumber\\
&&\sum^3_{m=1}\Gamma(N_i\to \nu_mH^0)={g^2\over
64\pi}\sum_{\ell=e}^\tau|V_{lN}^{\ell i}|^2{M_{N_i}^3\over
M_W^2}\left(1-{M_H^2\over
M_{N_i}^2}\right)^2\label{N3},\\
&&\sum^3_{m=1}\Gamma(E_i^+\to \bar{\nu}_mW^+)={g^2\over
32\pi}\sum_{\ell=e}^\tau|V_{lN}^{\ell i}|^2{M_{E_i}^3\over M_W^2}\left(1-{M_W^2\over
M_{E_i}^2}\right)\left(1+{M_W^2\over M_{E_i}^2}-2{M_W^4\over
M_{E_i}^4}\right)\label{E1},\nonumber\\
&&\Gamma(E_i^+\to \ell^+Z)={g^2\over 64\pi c_W^2}|V_{lN}^{\ell i}|^2{M_{E_i}^3\over
M_Z^2}\left(1-{M_Z^2\over M_{E_i}^2}\right)\left(1+{M_Z^2\over
M_{E_i}^2}-2{M_Z^4\over
M_{E_i}^4}\right)\label{E2},\nonumber\\
&&\Gamma(E_i^+\to \ell^+H^0)= {g^2\over 64\pi}|V_{lN}^{\ell i}|^2{M_{E_i}^3\over
M_W^2}\left(1-{M_H^2\over M_{E_i}^2}\right)^2.\label{E3}\nonumber
\end{eqnarray}
In the above, we have used the relation~\cite{Han:2006ip}
\begin{eqnarray}
\sum^3_{m=1}|(V_{PMNS}^\dagger V_{lN})^{mi}|^2 =
\sum_{\ell=e}^\tau |V_{lN}^{\ell i}|^2.
\end{eqnarray}
One can see that all $E$ and $N$ decay partial widths are
proportional to $|V_{lN}|^2$ and the branching ratios of the
cleanest channels have relationship $BR(N\to \ell^\pm
W^\mp)=BR(E^\pm \to \ell^\pm Z)$ for large triplet mass. In
Fig.~\ref{br} we show the branching fractions for the decays of $N$
(left) and $E$ (right) versus their masses with $M_{H^0}=120~{\rm
GeV}$ for Cases I and II, in which the lepton and neutrino flavors
in final state are summed. Since all partial widths are proportional
to $\sum_{\ell=e}^\tau|V_{lN}^{\ell i}|^2$, there are no other
free parameters for the branching ratios displayed in the figure.

\subsection{Heavy Triplet Lepton Decays And Neutrino Mass Spectra}
We now present our results for decay branching ratios in detail for Cases I and II described before.

\subsubsection{Case I}
\begin{figure}[tb]
\begin{center}
\begin{tabular}{cc}
\includegraphics[scale=1,width=8cm]{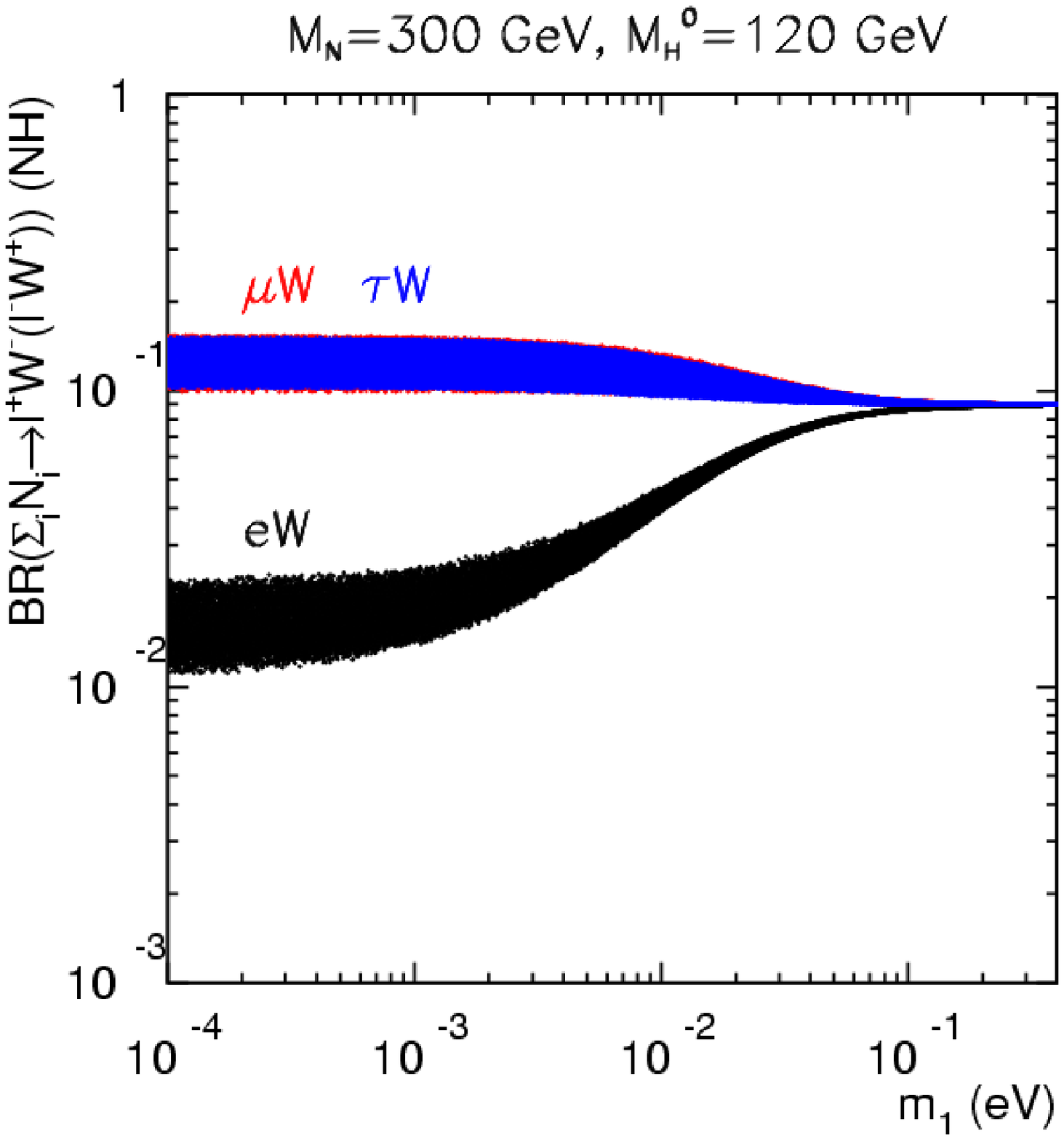}
\includegraphics[scale=1,width=8cm]{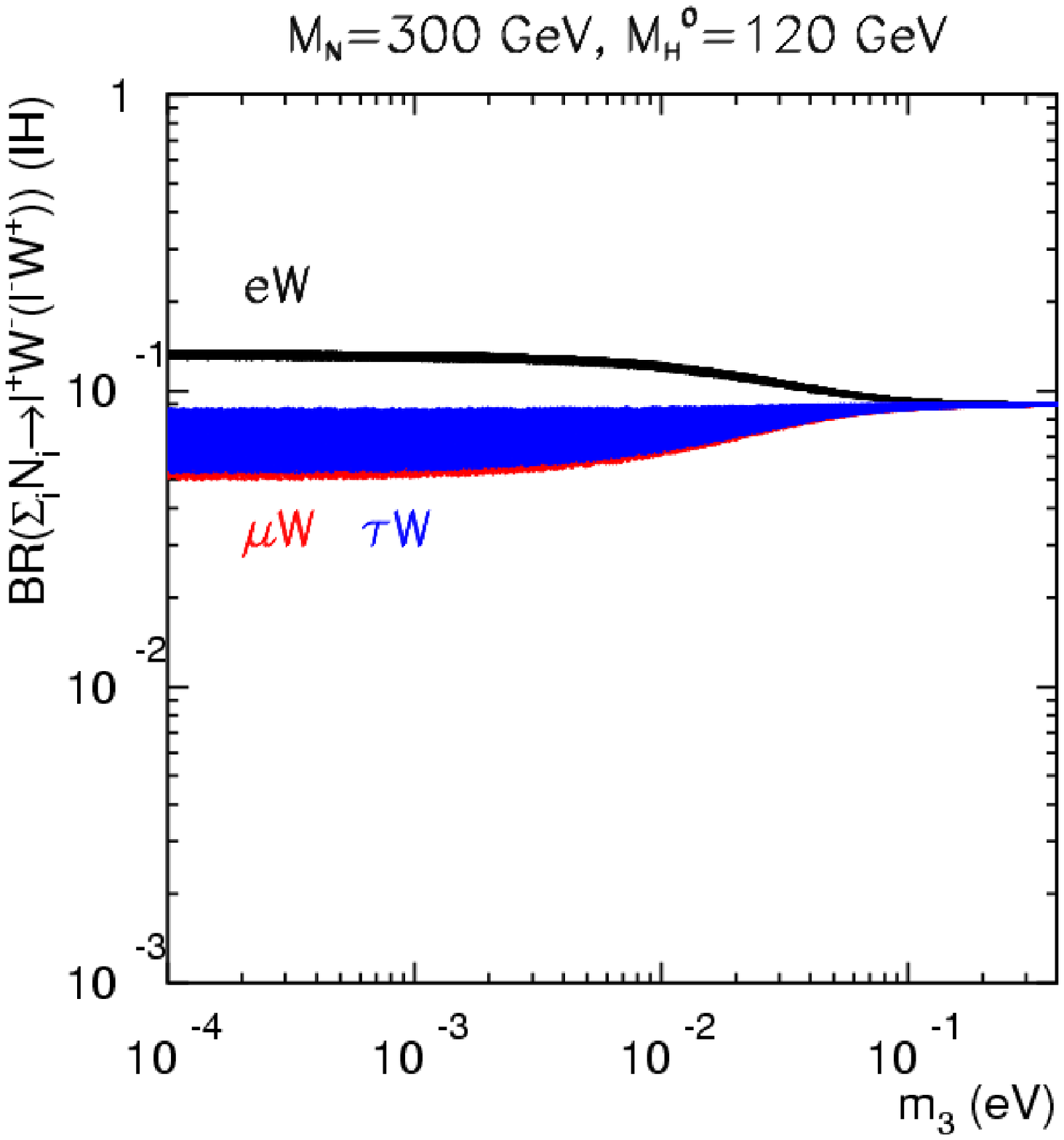}
\end{tabular}
\end{center}
\caption{The branching fractions of $\sum_{i=1,2,3}N_i\to \ell^\pm
W^\mp \ (\ell=e,\mu,\tau)$ for NH and IH versus lightest neutrino
mass when $M_N=300~{\rm GeV}, M_{H^0}=120~{\rm GeV}$ for Case I
without any phases. } \label{nbr}
\end{figure}

In Fig.~\ref{nbr} we show the impact of the neutrino masses and
mixing angles on the branching fractions summing all $N_i$
decaying into $e,\mu,\tau$ lepton plus $W$ boson respectively, with
the left panels for NH and the right panels for IH. The branching
fraction can differ by one order of magnitude in NH case with
$BR(\mu^\pm W^\mp),BR(\tau^\pm W^\mp)\gg BR(e^\pm W^\mp)$, and about
a few times of magnitude in the IH spectrum with $BR(e^\pm
W^\mp)>BR(\mu^\pm W^\mp),BR(\tau^\pm W^\mp)$. As expected that
all the channels are quite similar when the neutrino spectrum is
quasi-degenerate, $m_1\approx m_2\approx m_3\geq 0.1$ eV.

\subsubsection{Case II}

\begin{figure}[tb]
\begin{center}
\begin{tabular}{cc}
\includegraphics[scale=1,width=8cm]{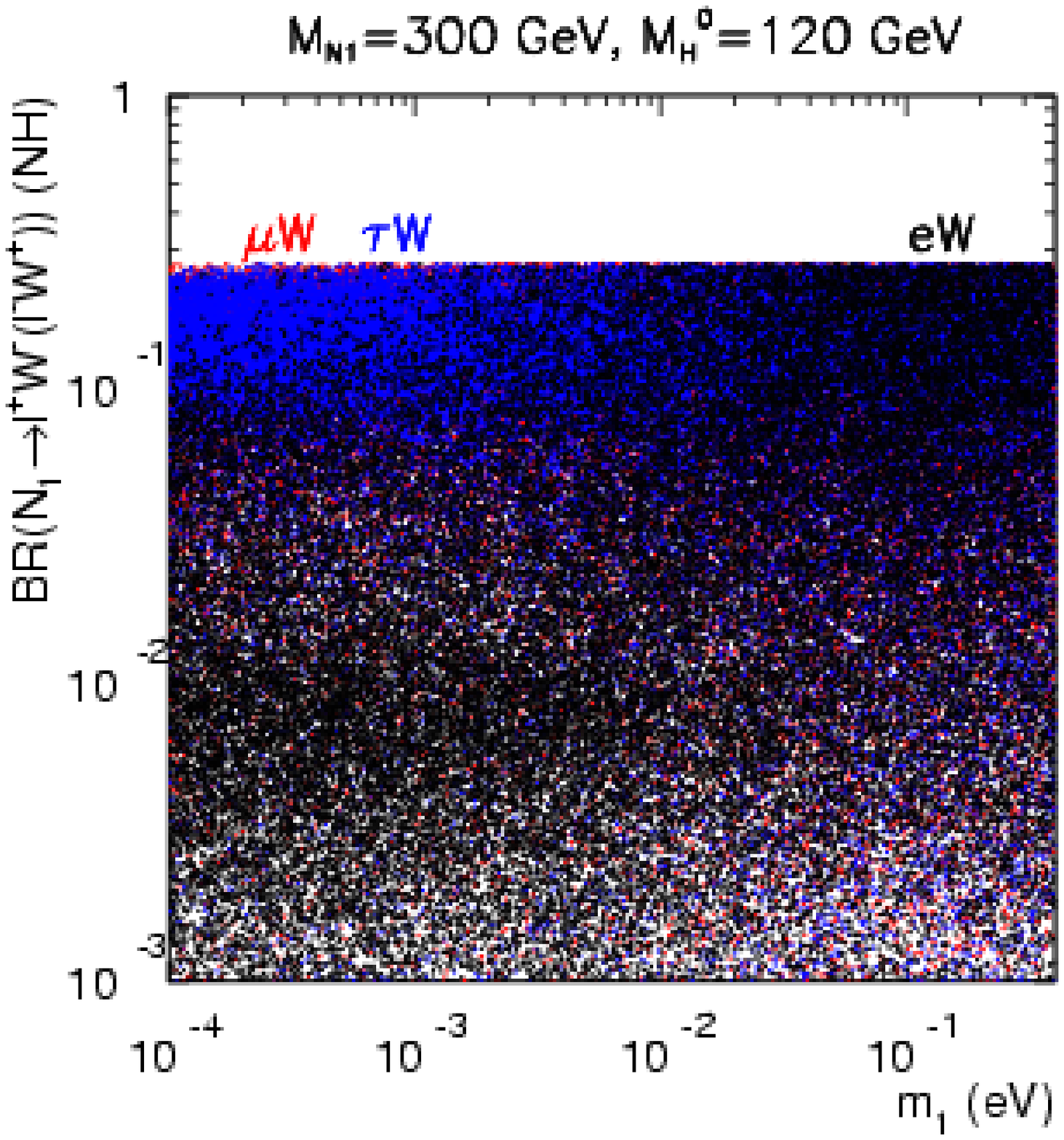}
\includegraphics[scale=1,width=8cm]{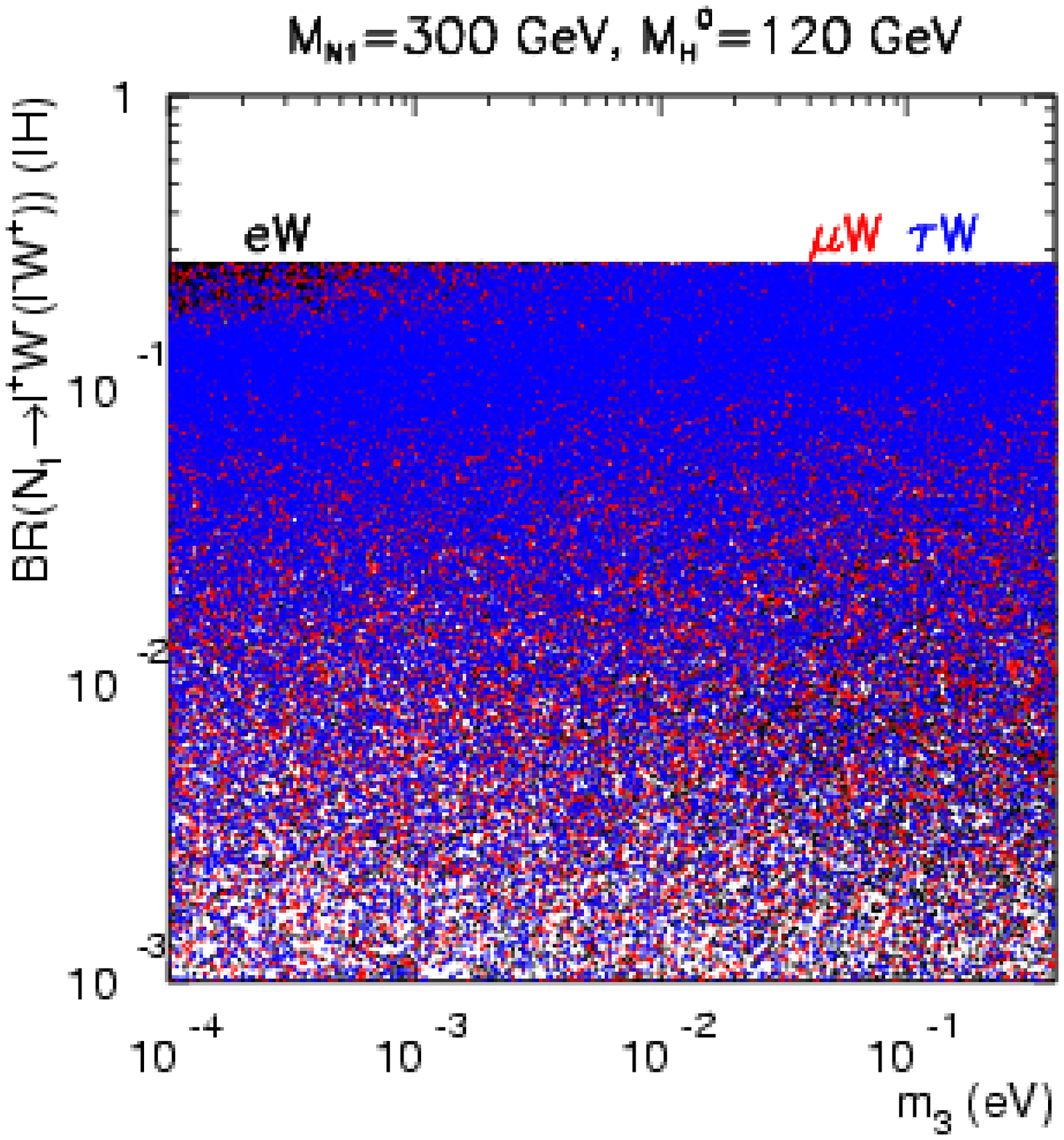}
\end{tabular}
\end{center}
\caption{Branching fractions of process $N_1\to \ell^\pm W^\mp,
\ell=e,\mu,\tau$ vs. the lightest neutrino mass for NH and IH, when
$M_1=300~{\rm GeV}$, $M_{H^0}=120~{\rm GeV}$ and $0<w_{ij}<2\pi$ for
Case II without any phases in $\Omega$. } \label{br1all}
\end{figure}

We show the branching fractions of processes $N_1\to \ell^\pm W^\mp
\ (\ell=e,\mu,\tau)$ as functions of the lightest neutrino mass for
NH and IH, when $M_1=300~{\rm GeV}$, in Fig.~\ref{br1all}. The
behaviors of $N_2$ and $N_3$ decays are almost the same.
These do not
seem to provide discrimination power between the two mass spectra.

\begin{figure}[tb]
\begin{center}
\begin{tabular}{cc}
\includegraphics[scale=1,width=8cm]{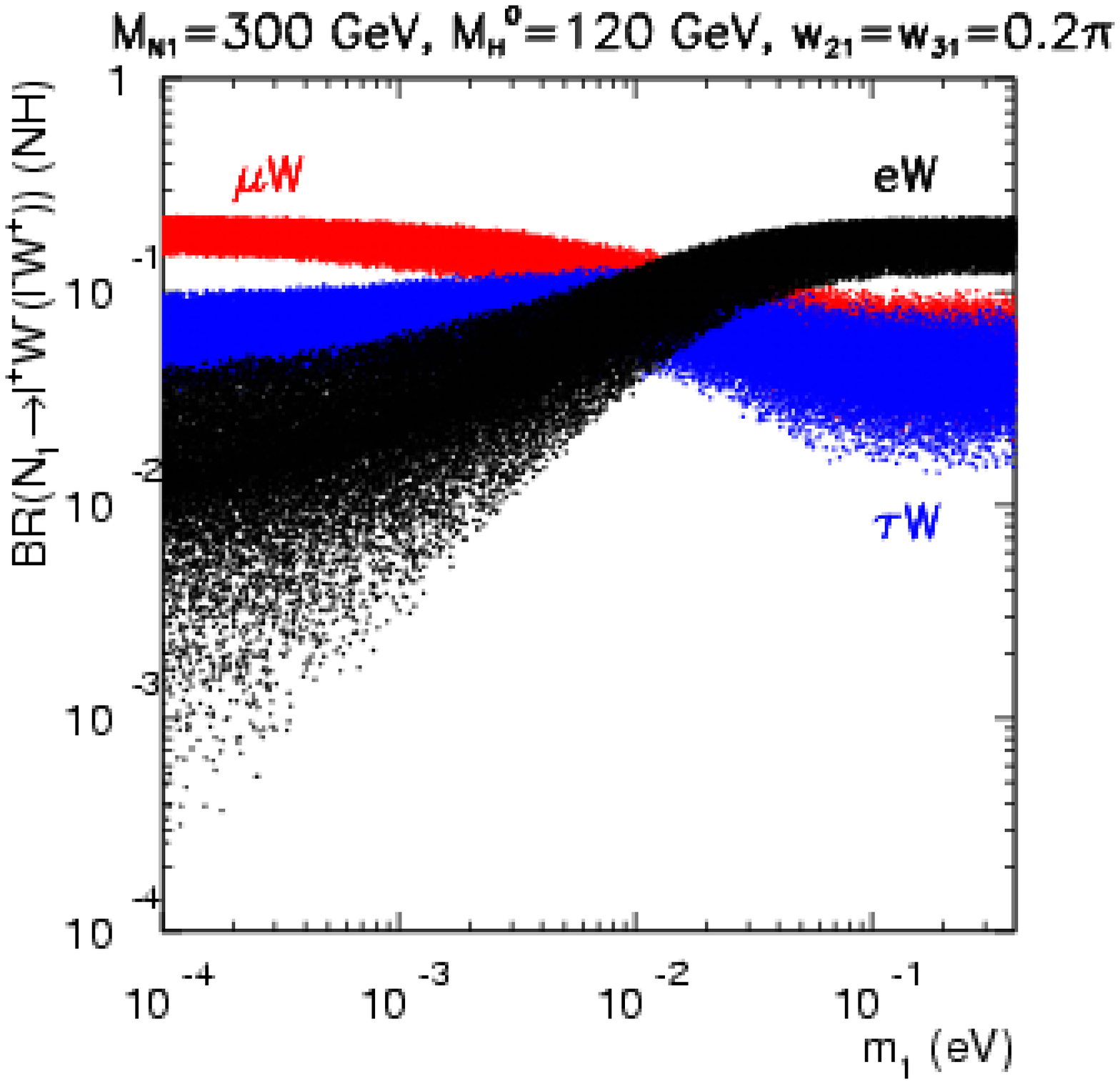}
\includegraphics[scale=1,width=8cm]{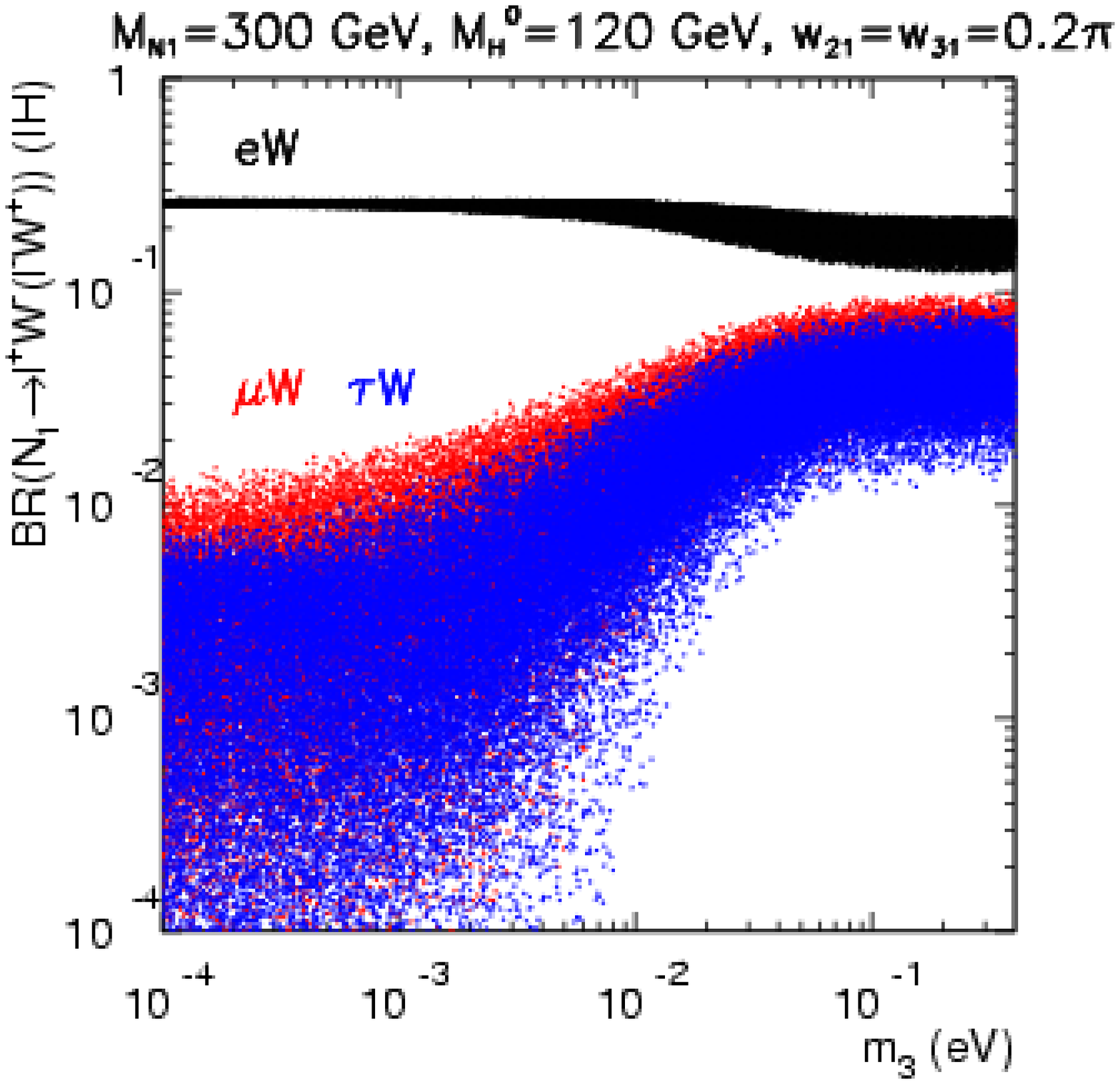}
\end{tabular}
\end{center}
\caption{Branching fractions of process $N_1\to \ell^\pm W^\mp,
\ell=e,\mu,\tau$ vs. the lightest neutrino mass for NH and IH, with
$M_1=300~{\rm GeV}$, and $w_{21}=w_{31}=0.2\pi$ for Case II without
any phases in $\Omega$. } \label{n1brcheck}
\end{figure}

It is important to note that nature would choose only one specific
form of $\Omega$, not a random selection. To illustrate in fact that
detailed analysis can still distinguish different neutrino mass
spectra in each special case, we show in Fig.~\ref{n1brcheck} the
branching fractions of $N_1$ as functions of the lightest neutrino
mass for NH and IH, respectively, with $M_1 = 300$ GeV, and $w_{21}
= w_{31}= 0.2 \pi$. Note that $N_1$ decay does not depend on
$w_{32}$. We see that the branching fractions for NH and IH cases
can be substantially different, $BR(\mu^\pm W^\mp)>BR(\tau^\pm
W^\mp)>BR(e^\pm W^\mp)$ in NH and $BR(e^\pm W^\mp)\gg BR(\mu^\pm
W^\mp),BR(\tau^\pm W^\mp)$ in IH.

The above analysis can only be useful
if there are independent ways that the angles $w_{ij}$ and phases
can be measured. We have not been able to find viable method to
achieve this. We therefore would like to turn the argument around that if in
the future the neutrino mass hierarchy is measured, then in
combination with the possible information on the sizes of the
elements in $V_{lN}$ from our later discussions in Sec. V.C.,
information on the model parameters $w_{ij}$ and phases may be
extracted at the LHC.

\subsection{Impact of Majorana Phases for $E$ and $N$ Decays}

So far we have assumed that the Majorana phases are all zero. The
unknown Majorana phases could modify the predictions for $E$ and $N$
decays. In this section we study the effect of non-zero Majorana
phases on $E$ and $N$ decays. We note that in general there are two
Majorana phases, $\Phi_{1,2}$, and a general analysis will be
complicated. The situation can be simplified for some special cases.
From Appendix B, it can be easily seen that in the limits that $m_1
= 0$ and $m_3 =0$, the phase $\Phi_1$ and $\Phi_2$ drop off the
expressions for $V_{lN}$, respectively. The $m_1 =0$ can happen for
normal hierarchy, and $m_3 = 0$ can happen for inverted hierarchy
neutrino mass patterns. We therefore will take these two cases for
illustrations. Note that with non-zero Majorana phases, it is not
possible to have Case I any more. Our discussion here will only
apply to Case II. The $V_{lN}$ dependence on Majorana phases for
Case II can be read off from Appendix B. The two illustration cases
are:

\paragraph{NH with one massless neutrino
$(m_1\approx 0)$.} In this case the $E$ and $N$ decay rates depend on only one Majorana phase
$\Phi_2$.

\paragraph{IH with one massless neutrino
$(m_3\approx 0)$.} In this case $E$ and $N$ decay rates depend on only one phase $\Phi_1$.

\begin{figure}[tb]
\begin{center}
\begin{tabular}{cc}
\includegraphics[scale=1,width=8cm]{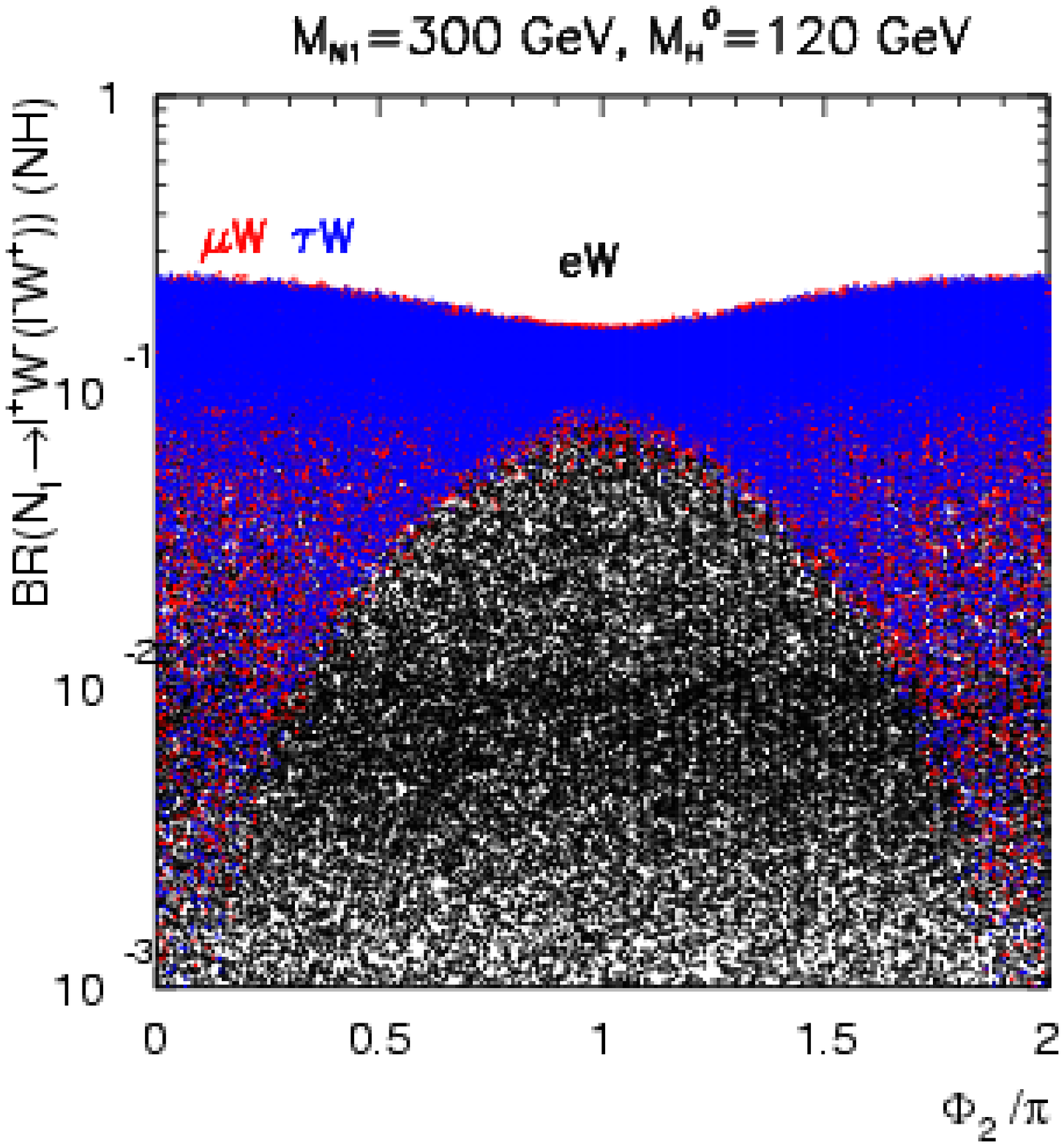}
\includegraphics[scale=1,width=8cm]{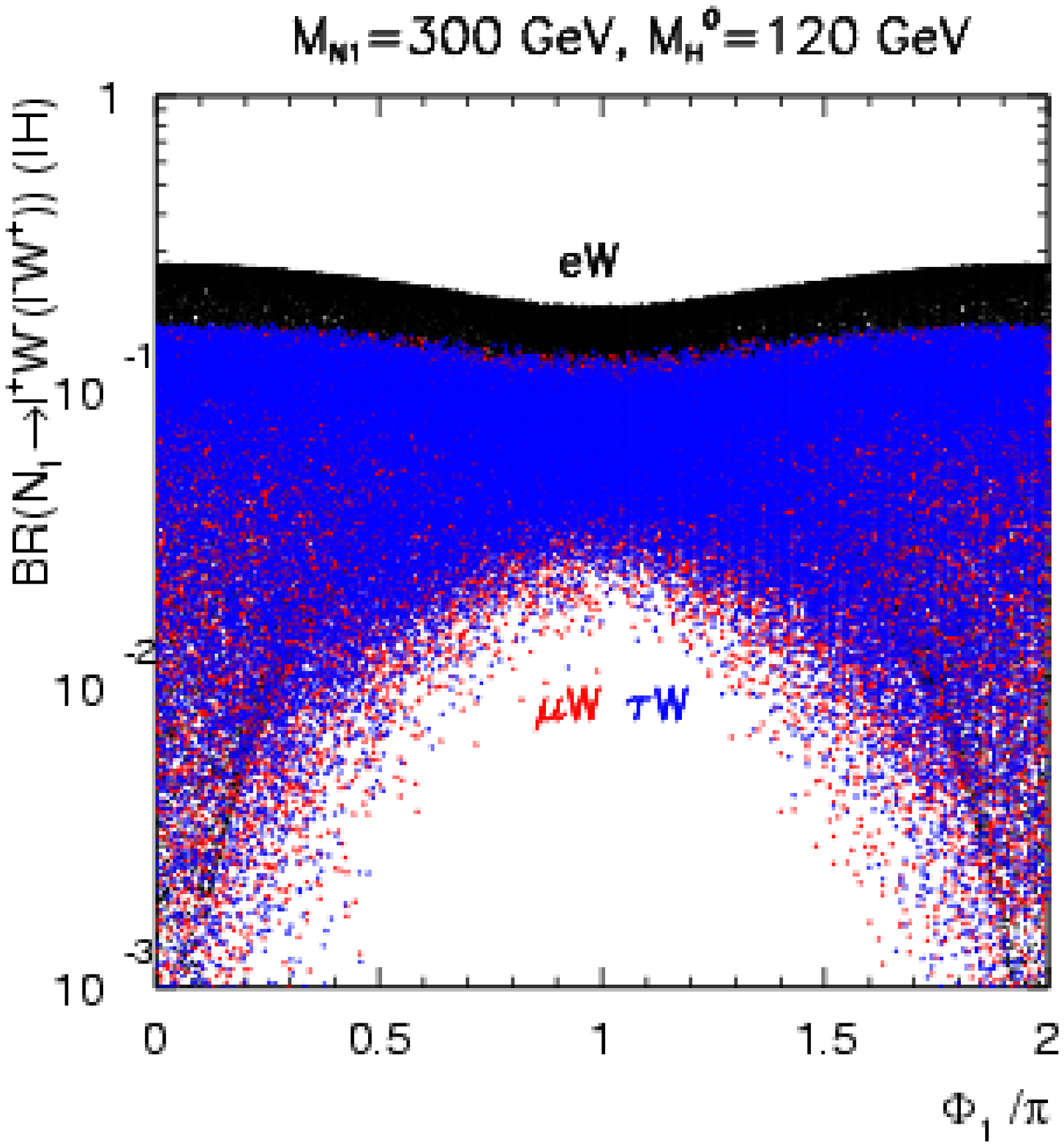}
\end{tabular}
\end{center}
\caption{The branching fractions of $N_1\to \ell^\pm W^\mp$ versus
Majorana phase $\Phi_2$ for NH and $\Phi_1$ for IH with
$M_1=300~{\rm GeV}, M_{H^0}=120~{\rm GeV}$ and $0<w_{ij}<2\pi$ for
Case II without any phases in $\Omega$. } \label{phi1}
\end{figure}

\begin{figure}[tb]
\begin{center}
\begin{tabular}{cc}
\includegraphics[scale=1,width=8cm]{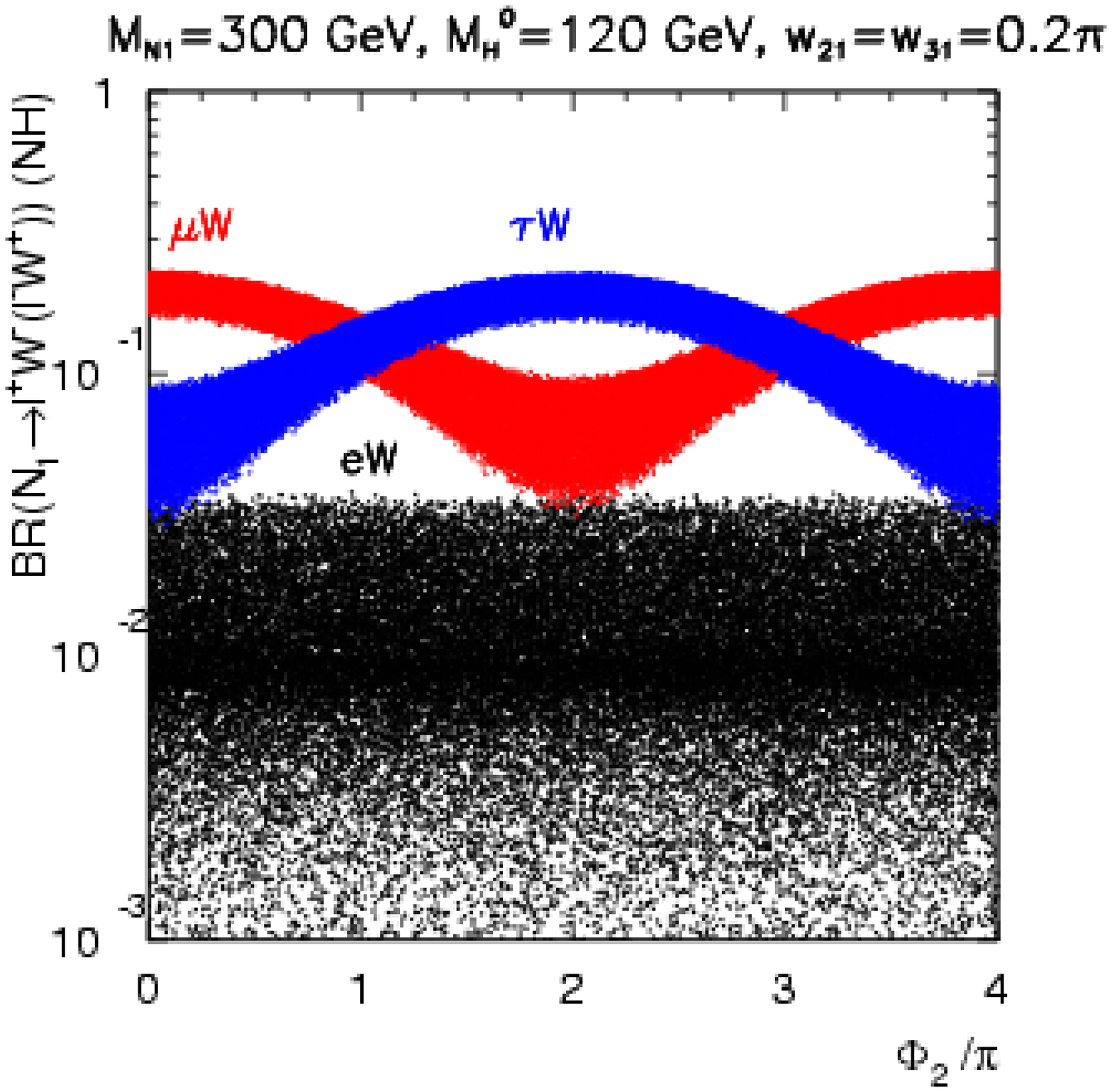}
\includegraphics[scale=1,width=8cm]{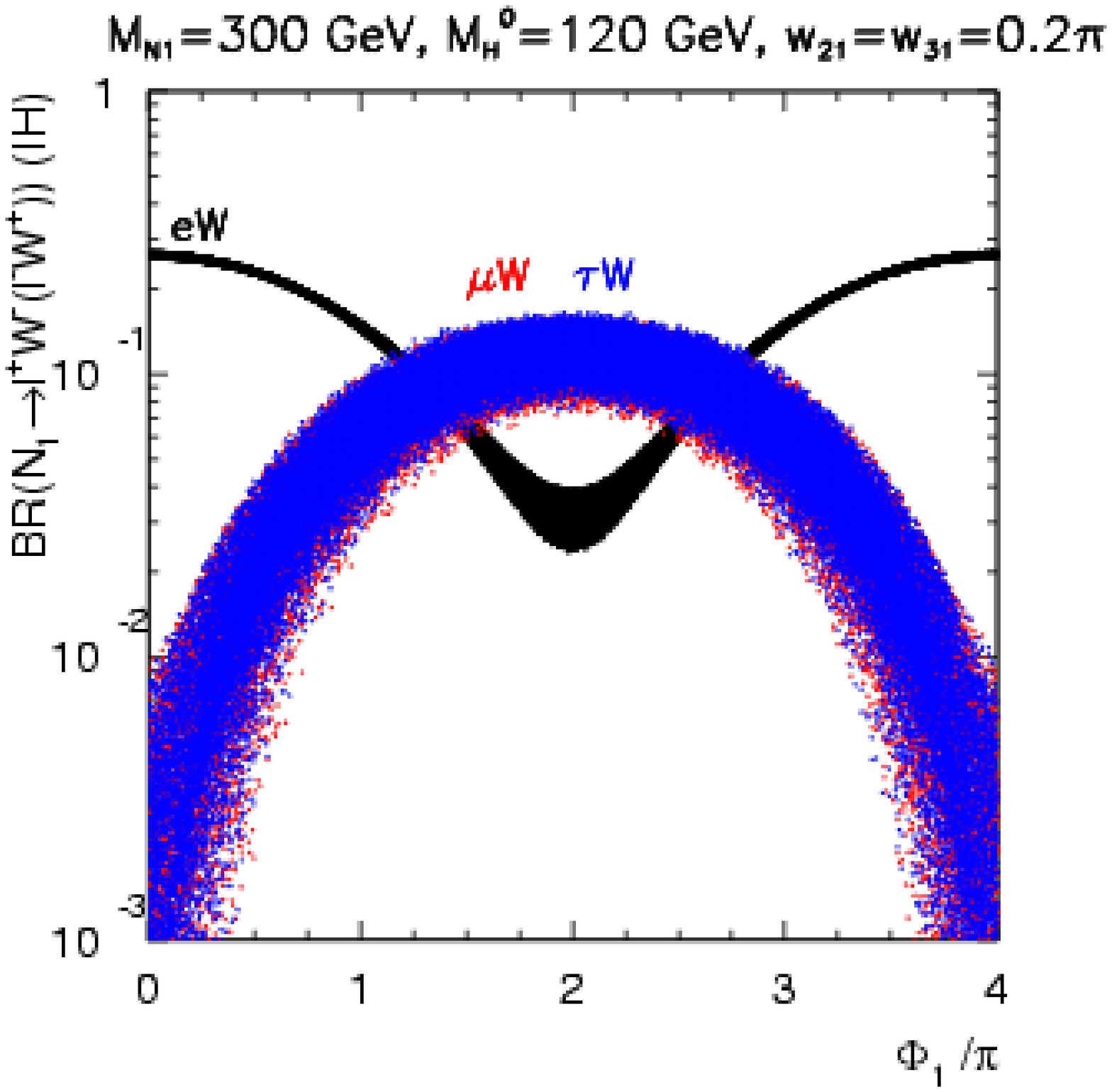}
\end{tabular}
\end{center}
\caption{The branching fractions of $N_1\to \ell^\pm W^\mp$ versus
Majorana phase $\Phi_2$ for NH and $\Phi_1$ for IH with
$M_1=300~{\rm GeV}, M_{H^0}=120~{\rm GeV}$, and $w_{21} = w_{31} =
0.2 \pi$ for Case II without any phases in $\Omega$. } \label{phi11}
\end{figure}

In Fig.~\ref{phi1} we show the dependence of $N_1$ decay rates with
Majorana phases $\Phi_2$ and $\Phi_1$ in NH and IH respectively
without any phases in $\Omega$. The dependence of $N_2$ and $N_3$
decays on Majorana phases are almost the same as that of $N_1$. For
$\Phi_2$ from $0$ to $\pi$ in NH, they are always $\mu W, \tau W$
channels that dominate. And for $\Phi_1$ from $0$ to $\pi$ in IH, $e
W$ channel always dominates. To cover the whole ranges the range of
Majorana phases needs to go from 0 to $4\pi$ according to our
definition in Eq.~\ref{pmns}. But for the fully scanned plot
Fig.~\ref{phi1} the range from 0 to $2\pi$ can reflect the complete
feature. The figures are symmetric from $2\pi$ to $4\pi$ to the ones
from 0 to $2\pi$. We see that the Majorana phases do have impact on
the branching ratios. One can extract information on Majorana phase
from different lepton-flavor final states. One can also
study more detailed correlations of the NH and IH cases with the
change of Majorana phases which we show indicative plots for the
above two cases in Fig.~\ref{phi11}. We can see that for NH when the phase
$\Phi_2=2\pi$, one obtains the maximal suppression (enhancement) for
channel $N_1\to \mu^\pm W^\mp$ ($N_1\to \tau^\pm W^\mp$) by one
order. For IH the maximal suppression and enhancement takes place
also when $\Phi_1=2\pi$. In this case the dominate channels swap
from $N_1\to e^\pm W^\mp$ when $\Phi_1=0$ to $N_1\to \mu^\pm
W^\mp,\tau^\pm W^\mp$ when $\Phi_1=2\pi$. This qualitative change
can be of useful in extracting the value of the Majorana phase
$\Phi_1$.

\subsection{Total Decay Width of Heavy Triplet Leptons}
\begin{figure}[tb]
\begin{center}
\begin{tabular}{cc}
\includegraphics[scale=1,width=7.5cm,angle=90]{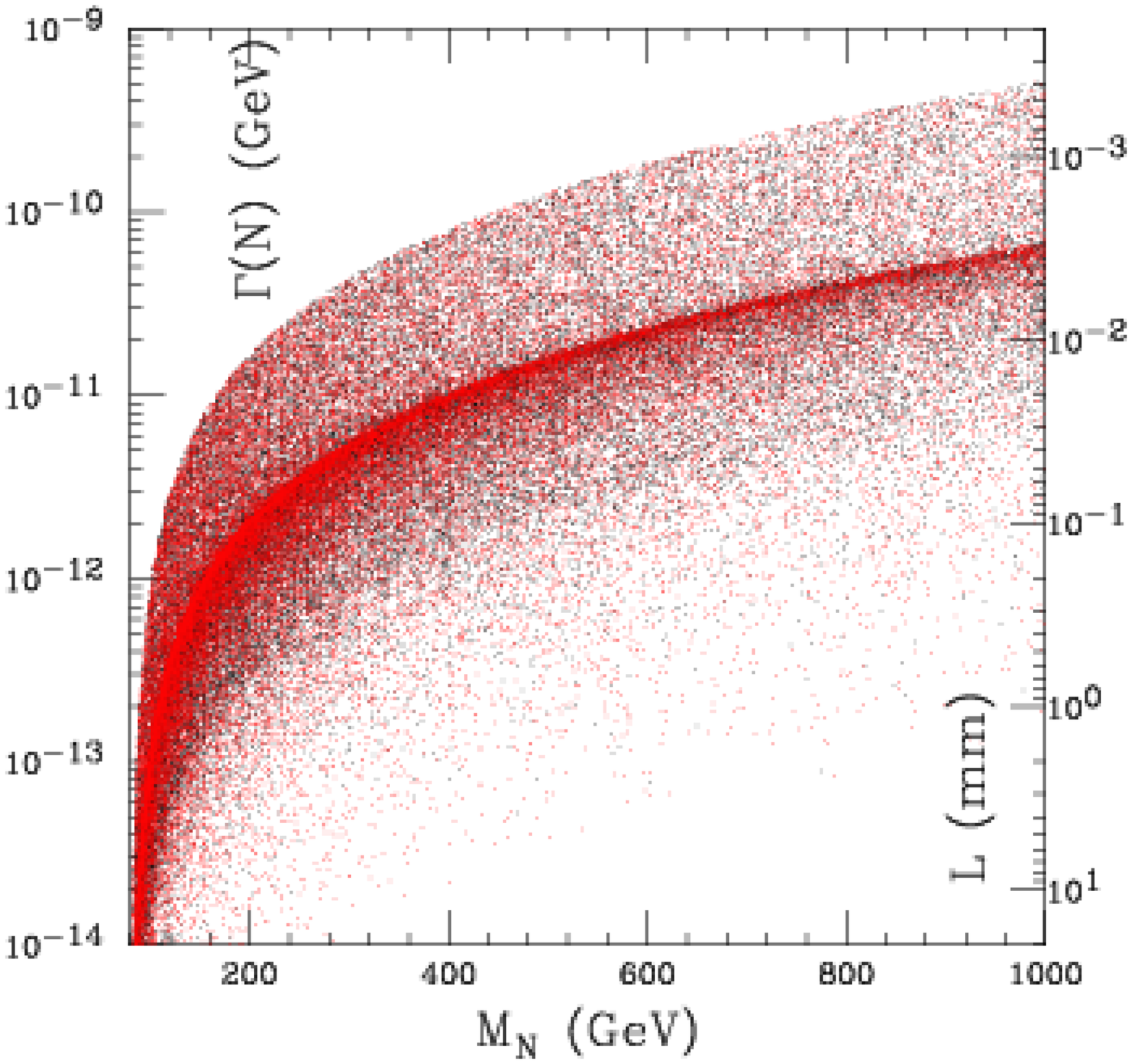}
\includegraphics[scale=1,width=7.5cm,angle=90]{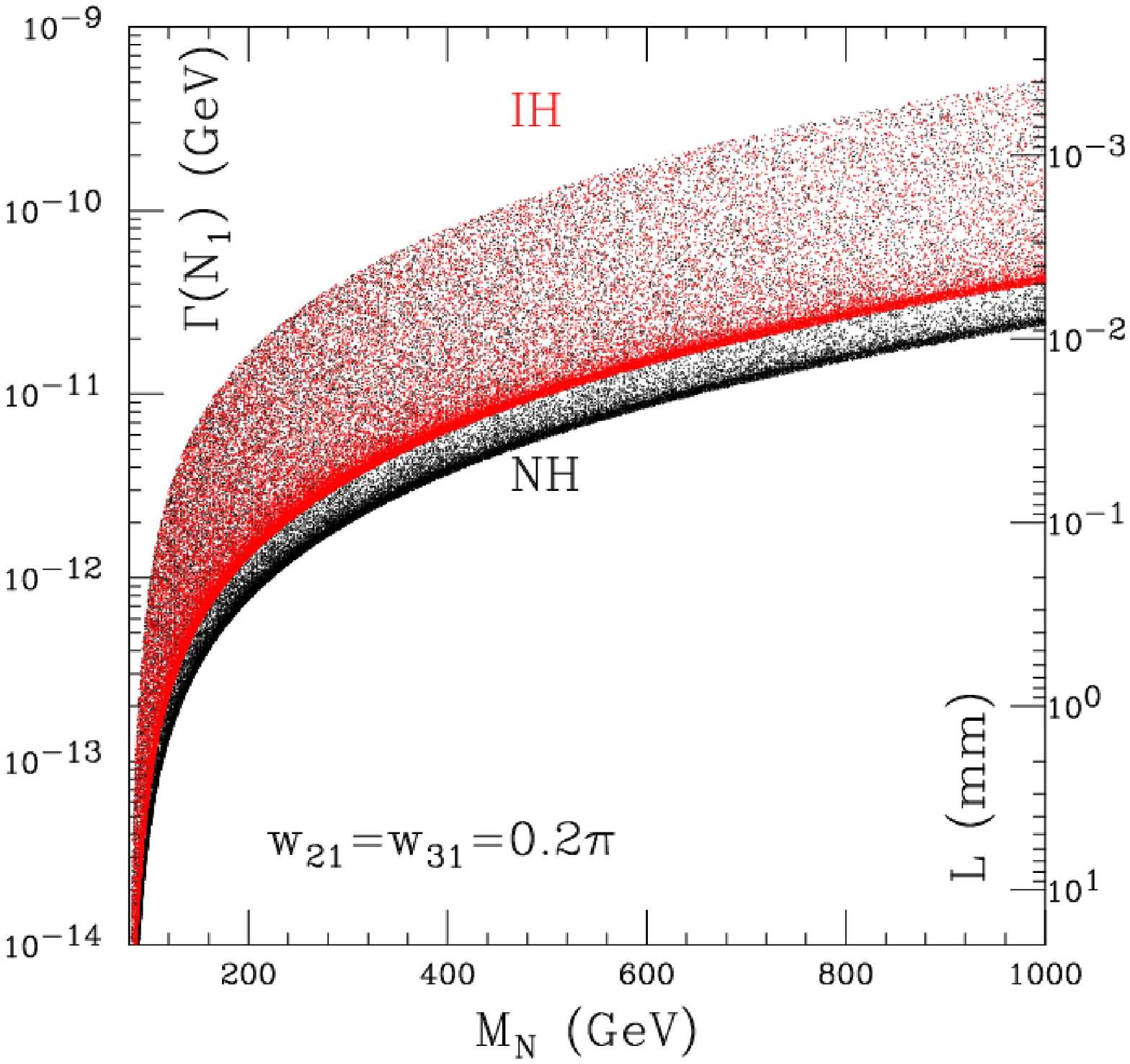}
\end{tabular}
\end{center}
\caption{The left and right panels are for the total decay widths of $N_1$ with
$M_{H^0}=120~{\rm GeV}$ for $w_{ij}$ scanned in their whole allowed ranges, and  for $w_{21} = w_{31} = 0.2\pi$, respectively. } \label{totw}
\end{figure}

To complete our study about the heavy lepton and neutrino
properties, in Fig.~\ref{totw} we plot the total width (left axis)
and decay length (right axis) for $N$ versus $M_N$ in NH and IH for
Case II. The total decay width is proportional to $M_\nu M_N^2$.
Although not considered as long-lived for large triplet mass, the
$E$ and $N$ decay could lead to a visible displaced vertex in the
detector at the LHC. This displaced vertex can be observed through
$E$ and $N$ reconstructions as first pointed out by
Ref.~\cite{production1}. Careful analysis of displaced vertex can
also provide crucial information about neutrino mass hierarchy since
the NH and IH cases have different decay widths as can be seen from
Figs.~\ref{n1brcheck}, \ref{phi11}, and also the right panel in
Fig.~\ref{totw}. Since we cannot separately determine each of the
heavy triplet lepton decays for Case I. It is not possible to give a
similar description about the decay length. But this case is just a
special case of Case II which are already implicitly included in the
results for Case II.

\section{Heavy Triplet Lepton Productions at the LHC}

\begin{figure}[tb]
\begin{center}
\begin{tabular}{cc}
\includegraphics[scale=1,width=8cm]{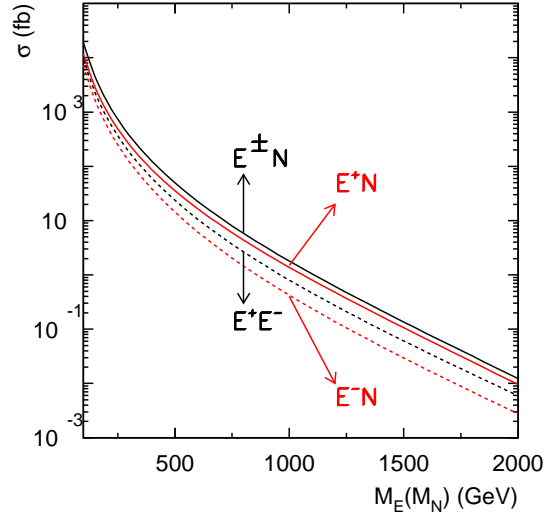}
\end{tabular}
\end{center}
\caption{Heavy lepton production total cross section at the LHC
versus its mass. The black dashed curve is for $pp\to E^+E^-$. The
black solid curve is for $pp\to E^\pm N$ via $W^\pm$ exchange,
assuming $M_E=M_N$. The two red curves are for $pp\to E^+ N$ (solid)
and $pp\to E^- N$ (dashed) respectively.} \label{t3tot}
\end{figure}

In this section we study the main production mechanisms of heavy
triplet leptons and their experimental signatures at the LHC. There
are existing previous literatures on this topic, both in theoretical
and phenomenological considerations~\cite{production1,Aguila,PDG}.
The main production channels of $E^\pm, N$ are
\begin{equation}
p p \to \gamma^\ast/Z^\ast \to E^+ E^-, \ \ \ pp\to W^\ast\to E^\pm
N\;.
\end{equation}
The relevant total production cross sections are plotted in
Fig.~\ref{t3tot}. Note that the production cross sections for $E^+ N$ and $E^-N$ are different due to the fact that the LHC is a pp machine.

We can see that up to 1.5 TeV, the cross section for each production
mode is a few times larger than $0.01 fb$. It gives the hope that
with enough integrated luminosity, saying 300 $fb^{-1}$, LHC may be
able to probe the scale up to 1.5 TeV if all three production modes
are analyzed. However, one should be more careful in carrying out
the analysis beyond the naive total cross section estimate. One has
to make appropriate cuts to reduce SM backgrounds which will also
reduce the signal rate.

The detection of $E^\pm$ and $N$ are through their decays into SM
particles. We have studied the main decay modes, $E^\pm \to
\ell^\pm Z, \nu W^\pm, \ell^\pm H^0$ and $N \to \ell^\pm W^\mp, \nu
Z, \nu H^0$ in previous sections. The signal channels for $E^\pm$
and $N$ productions can be classified according to charged usual
leptons in the final states~\cite{Aguila}: i)
6 charged leptons $\ell^\pm \ell^\pm \ell^\pm \ell^\mp \ell^\mp
\ell^\mp$; ii) 5 charged leptons $\ell^\pm \ell^\pm \ell^\mp
\ell^\mp \ell^+$; iii) 4 charged leptons $\ell^\pm \ell^\pm \ell^\pm
\ell^\mp$; iv) 4 charged leptons $\ell^+ \ell^+ \ell^- \ell^-$; v) 3
charged leptons $\ell^\pm \ell^\pm \ell^\mp$; vi) 2 charged leptons
$\ell^\pm \ell^\pm$; vii) 2 charged leptons $\ell^+\ell^- jjjj$; And
viii) 1 charged lepton $\ell^\pm jjjj$.

With the heavy triplet leptons fixed at a mass of 300 GeV, it has
been shown that signals from i) and ii) have too small cross
sections. They are not good for discovery. Signals from iii) and iv)
can provide clean measurement of the heavy triplet masses. Signals
from v) and vi) have excellent potential for the discovery with
relatively high signal rate and small background. Signals from vii)
and viii) have large cross sections, but large
background~\cite{Aguila}. We will not carry out a full comprehensive
study of all possible final states, but concentrate on two types of
particular final states, belonging to iv) and vi), which represent
two ideal signals for $E^+ E^-$ and $E^\pm N$ productions.

These two types of signals are: 1) $pp \to E^+E^- \to \ell^+ Z (\to
\ell^+ \ell^-) \ell^- Z(\to qq)$, and 2) $pp \to E^\pm N \to
\ell^\pm Z (\to q \bar q) \ell^\pm W^\mp (\to q \bar q')$. We note
that all the particles in the final states in these two processes
can be measured and the masses of $E$ and $N$ can, in principle, be
reconstructed. Therefore these processes are easier to control
compared with those with multi-neutrinos in the final states.
Replacing $Z$ by $H^0$ can also result in the same final states.
Since Higgs boson mass is not known yet, we will not consider them.
This type of events can be eliminated with reconstruction of $Z$
mass.

Our approach beyond the existing studies is to make concrete
predictions of the $E^\pm$ and $N$ signals in connection with the
neutrino oscillation parameters through Eq.~\ref{VL}, and allow
the heavy triplet lepton masses to vary. We also include $\tau$
events reconstruction which turns out to provide some interesting
information.

We now discuss the observability at the LHC in detail. In
sub-sec.~A and B, we are mainly concerned with the kinematical
features for the signal and backgrounds. We will take the decay
branching fractions of $E$ and $N$ to be $100\%$ to the
corresponding channels under discussions. In sub-sec. C, we will
devote ourselves to the determination for the branching fractions.
We also assume that the three heavy triplet leptons are hierarchical
and consider the lightest one of them. We will comment on degenerate
case as described in Case I previously at the end.

\subsection{$E^+E^-$ pair production}

In this case, one of the cleanest ways to make sure signals are from $E^+ E^-$ production is to use the decay mode
\begin{eqnarray}
E^\pm \to \ell^\pm Z \ \ (\ell=e,\mu,\tau)
\end{eqnarray}
and analyze
\begin{eqnarray}
E^+E^-\to \ell^+Z(\to \ell^+\ell^-)\ell^-Z(\to q\bar{q}).
\end{eqnarray}
We will explore the signal observability according to the different lepton flavors.

\subsubsection{$E^+E^-\to \ell^+\ell^+\ell^-\ell^-jj \ \ (\ell=e,\mu)$}

For this case, the leading irreducible SM backgrounds to this channel are
\begin{eqnarray}
&&\ell^+\ell^-Z(\to \ell^+\ell^-)jj\to \ell^+\ell^+\ell^-\ell^-jj,
\nonumber \\
&&t(\to b W^+(\to \ell^+\nu))\bar{t}(\to \bar{b} W^-(\to
\ell^-\bar{\nu}))Z(\to \ell^+\ell^-)\to
\ell^+\ell^+\ell^-\ell^-jj+\cancel{E}_T.
\end{eqnarray}
We generate the SM backgrounds using MadGraph.

Although the background rates are very large to begin with, the kinematics is quite different between the signal and the backgrounds.
We employ the following basic acceptance cuts for the event
selection~\cite{cms}
\begin{eqnarray}
&&p_T(\ell)\geq15~{\rm GeV}, \ |\eta(\ell)|<2.5,\nonumber\\
&&p_T(j)\geq25~{\rm GeV}, \ |\eta(j)|<3.0, \nonumber\\
&&\Delta R_{jj},\ \Delta R_{j\ell},\ \Delta R_{\ell\ell}\geq 0.4. \label{cut1}
\end{eqnarray}
where $p_T$ is the transverse momentum, $\eta$ is the
pseudo-rapidity and $\Delta R$ is the separation between events for any of the pairs $\ell$ and $\ell$, $\ell$ and $j$, and, $j$ and $j$.

To simulate the detector effects on the energy-momentum
measurements, we smear the electromagnetic energy and jet energy by
a Gaussian distribution whose width is parameterized as~\cite{cms}
\begin{eqnarray}
{ \Delta E\over E} &=& {a_{cal} \over \sqrt{E/{\rm GeV}} } \oplus
b_{cal}, \quad a_{cal}=10\%,\  b_{cal}=0.7\% ,
\label{ecal}\nonumber\\
{ \Delta E\over E} &=& {a_{had} \over \sqrt{E/{\rm GeV}} } \oplus
b_{had}, \quad a_{had}=50\%,\  b_{had}=3\%.
\end{eqnarray}

\begin{figure}[tb]
\begin{center}
\begin{tabular}{cc}
\includegraphics[scale=1,width=8cm]{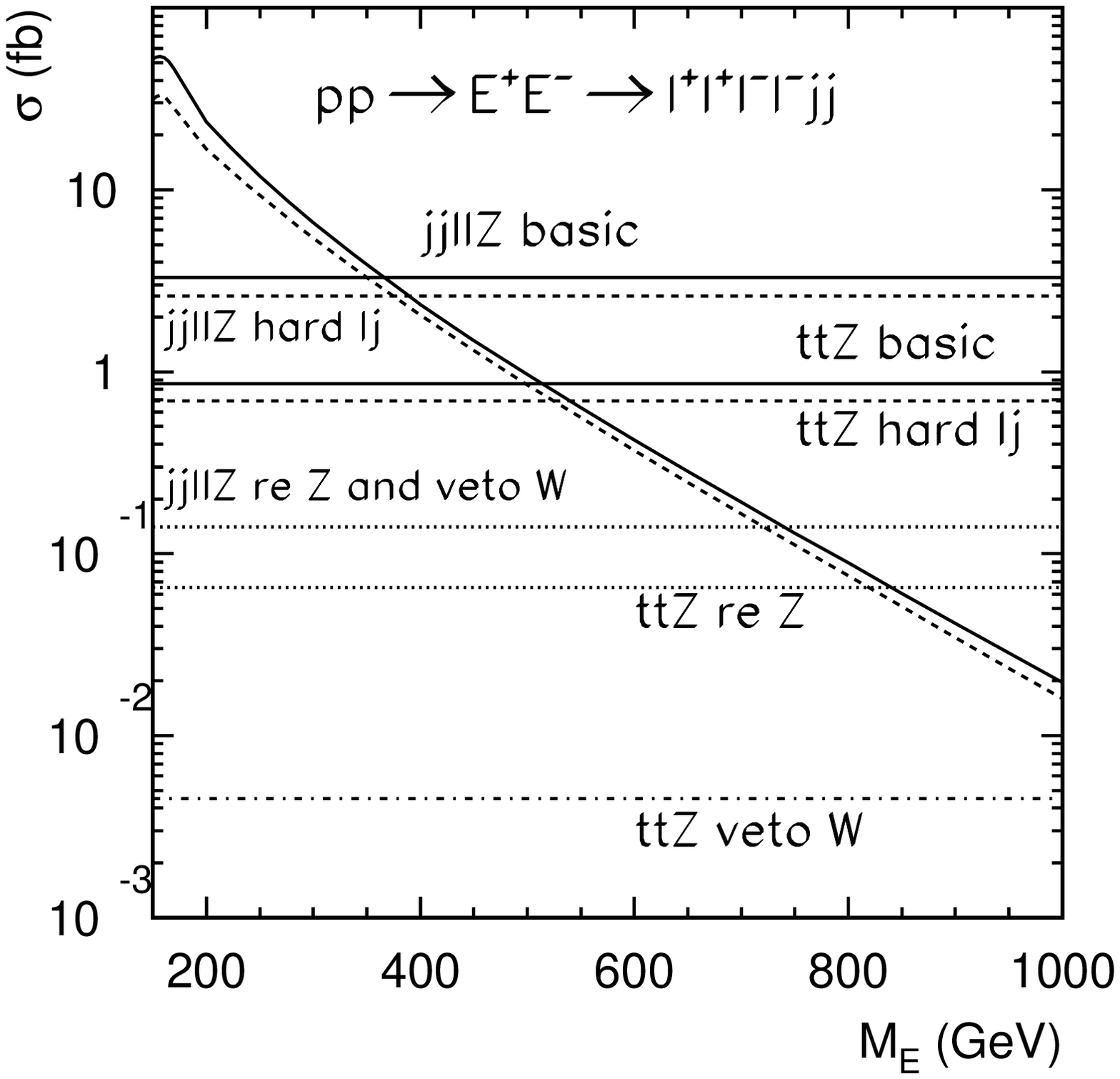}
\end{tabular}
\end{center}
\caption{Production cross section of $E^+E^-$ with basic cuts
(solid) and hard final states cuts (dashed). Branching fractions for
$E$ decay are not included in this plot. For comparison, the
background processes are also included with the sequential cuts as
indicated.} \label{t3pair}
\end{figure}

With further judicious cuts, the background can be reduced more.
We outline the characteristics and propose some judicious cuts as follows.
\begin{itemize}
\item For a few hundred GeV $E$, the leptons from heavy triplet lepton decays are very energetic. We therefore tight up the
kinematical cuts with
\begin{eqnarray}
p_T^{max}(\ell)>M_E/4, \ \ p_T^{max}(j)>50~{\rm GeV}.
\end{eqnarray}

\item To reconstruct $Z$ boson, we select among four possibilities of opposite sign lepton pair $\ell^+\ell^-$ and take advantage of the feature $M_{\ell_1^+\ell_2^-}=M_{jj}$. In practice, we take their invariant masses close to $M_Z$ with $|M_{\ell^+_1\ell^-_2,jj}-M_Z|<15~{\rm GeV}$.
\item To reconstruct heavy lepton $E$, we take advantage of the feature that two $E$'s have equal mass $M_{\ell_1^+\ell_2^-\ell_3^\pm}=M_{jj\ell_4^\mp}$. In practice, we take $|M_{\ell_1^+\ell_2^-\ell_3^\pm}-M_{jj\ell_4^\mp}|<M_E/25$. This helps for the background reduction, in particular for $\ell^+\ell^-Zjj$.
\item To remove the $t\bar{t}Z$ background, we veto the events with large missing energy from $W$ decay $\cancel{E}_T<20~{\rm GeV}$.
\end{itemize}
The production cross section of $E^+E^-$ signal with basic cuts (solid curve) and all of the cuts (dotted curve) above are plotted in Fig.~\ref{t3pair}. For comparison, the background processes of $\ell^+\ell^-Zjj$ and $t\bar{t}Z$ are also included with the sequential cuts as indicated. The backgrounds are suppressed substantially.

\begin{figure}[tb]
\begin{center}
\begin{tabular}{cc}
\includegraphics[scale=1,width=8cm]{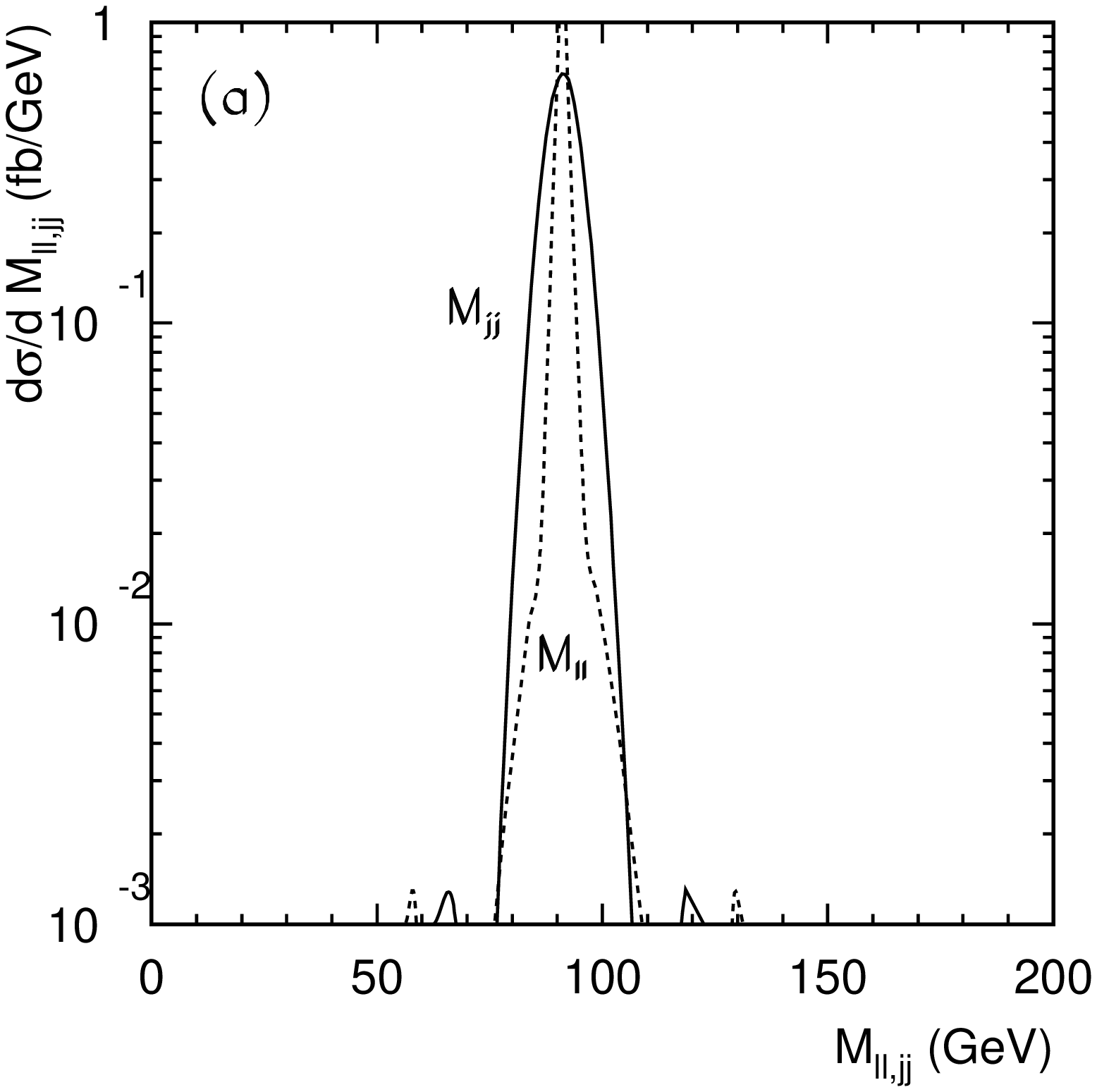}
\includegraphics[scale=1,width=8cm]{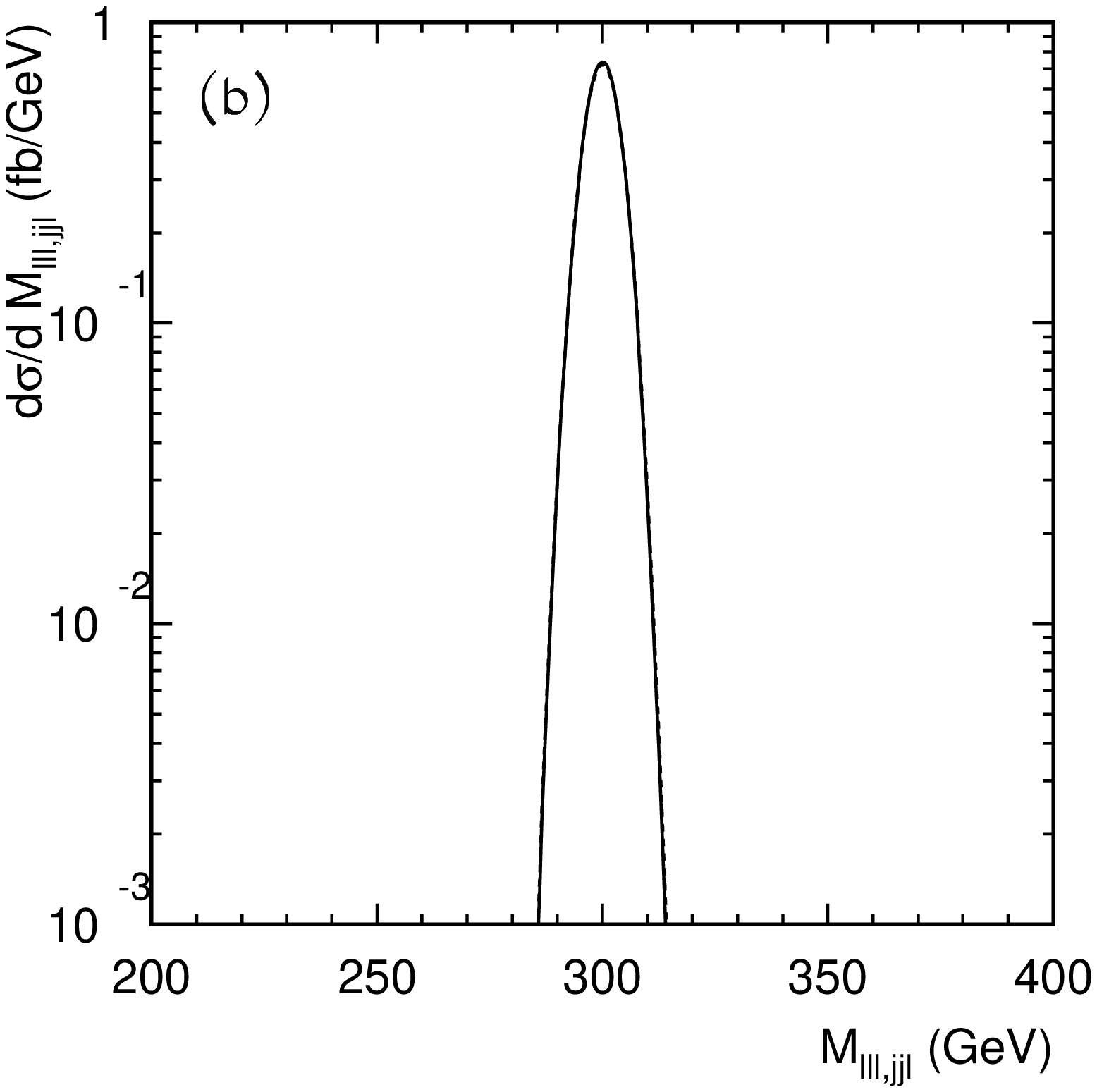}
\end{tabular}
\end{center}
\caption{Reconstructed invariant mass of $M(\ell_1^+\ell_2^-), M(jj)$ (a) and $M(\ell_1^+\ell_2^-\ell_3^\pm),M(jj\ell_4^\mp)$ (b) for $\ell^+\ell^+\ell^-\ell^-jj$ production, with a $E$ mass of $300$ GeV. }
\label{pairre}
\end{figure}

Finally, when we perform the signal significance analysis, we look for the resonance in the mass distribution of $\ell_1^+ \ell_2^- \ell_3^\pm$ and $jj\ell_4^\mp$. The invariant masses of them are plotted in Fig.~\ref{pairre} for 300 GeV $E$ pair production. If we look at a mass window of $|M_{\ell_1^+\ell_2^-\ell_3^\pm,jj\ell_4^\mp}-M_E|<M_E/20$, the backgrounds will be at a negligible level.

As a remark, we would like to comment on the other potentially
large, but reducible backgrounds, like $t\bar{t}jj$. The
$t\bar{t}jj$ production rate is very high, leading to the
$\ell^+\ell^- X$ final state with about $40~{\rm pb}$. Demanding
another isolated lepton presumably from the $b$ quarks and with the
basic cuts, the background rate will be reduced by about three to
four orders of magnitude. The stringent lepton isolation cut for
multiple charged leptons can substantially remove the $b$-quark
cascade decays. With the additional $M_{jj}, M_{\ell^+\ell^-},
M_{\ell^+\ell^-\ell^\pm}, M_{jj\ell^\mp}$ cuts, the backgrounds
should be under control.

\subsubsection{$E^+E^-\to \tau^\pm\ell^\mp\ell^+\ell^-jj,\tau^+\tau^-\ell^+\ell^-jj$}

The $\tau$ lepton final state from $E$ decay can provide additional information.
Its identification and reconstruction are different from $e,\mu$ final states because a $\tau$ decays promptly
and there will always be missing neutrinos in $\tau$ decay products.

\begin{figure}[tb]
\begin{center}
\begin{tabular}{cc}
\includegraphics[scale=1,width=8cm]{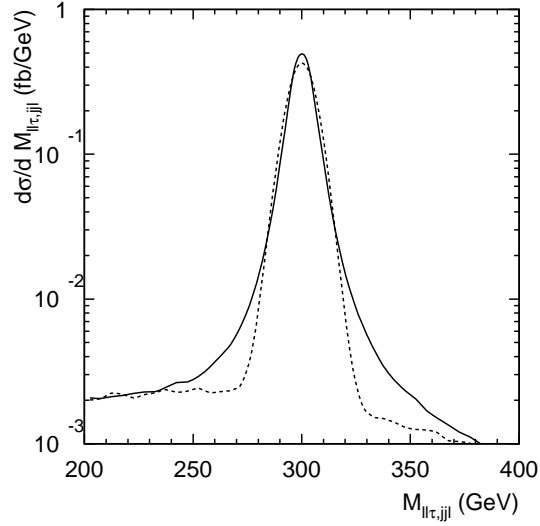}
\end{tabular}
\end{center}
\caption{Reconstructed invariant mass of
$M(\ell_1^+\ell_2^-\tau^\pm) ({\rm solid}),M(jj\ell^\mp) ({\rm
dashed})$ for $\tau^\pm\ell^\mp\ell^-\ell^-jj$ production, with a
$E$ mass of $300$ GeV. } \label{pairere1t}
\end{figure}

In order to reconstruct the events with $\tau$'s we note that all
the $\tau$'s are very energetic from the decay of a few
hundred ${\rm GeV}$ heavy lepton $E$. The missing momentum will be
along the direction of the charged track. We thus assume the momentum of  the missing
neutrinos to be reconstructed by
\begin{eqnarray}
\overrightarrow{p}({\rm invisible}) =\kappa\overrightarrow{p}({\rm track}).
\end{eqnarray}
Identifying $\overrightarrow{p_T}({\rm invisible})$ with the measured $\cancel{E}_T$,
we thus obtain the $\tau$ momentum by
\begin{eqnarray}
\nonumber
\overrightarrow{p}_T(\tau)=\overrightarrow{p}_T(\ell) +\overrightarrow{\cancel{E}}_T,\quad
p_L^{}(\tau)=p_L^{}(\ell) + { \cancel{E}_T \over p_T^{}(\ell) } p_L^{}(\ell).
\end{eqnarray}
The $E$ pair kinematics is, thus, fully reconstructed. The
reconstructed invariant masses of $M(\ell_1^+\ell_2^-\tau^\pm)$ and
$M(jj\ell^\mp)$ are plotted in Fig.~\ref{pairere1t}. We see that
$M(\ell_1^+\ell_2^-\tau^\pm)$ distribution (solid curve) is slightly
broader as anticipated. We always can find the rather narrow mass
distribution. The $jj\ell^\mp$ system right here (dashed curve) serves as the most distinctive kinematical feature for
the signal identification. Of course the single $\tau$ lepton could
also be produced with the hadronic decay $Z$ boson from the same
parent $E$. The feature in this case is the same.

For $\tau^+\tau^- \ell^+\ell^- jj$ events with two $\tau$'s, we generalize the momenta reconstruction to
\begin{eqnarray}
\overrightarrow{p}({\rm invisible})=\kappa_1\overrightarrow{p}({\rm track}_1)+\kappa_2\overrightarrow{p}({\rm track}_2).
\end{eqnarray}
The proportionality constants $\kappa_1$ and $\kappa_2$ can be determined
from the missing energy measurement as long as the two charge tracks are
linearly independent. In practice when we wish to identify the events with $\tau$'s, we require a minimal
missing transverse energy
\begin{equation}
\cancel{E}_T > 20\ {\rm GeV}.
\end{equation}
This will effectively separate them from the $\ell^+\ell^+\ell^-\ell^- jj$ events.

\begin{figure}[tb]
\begin{center}
\begin{tabular}{cc}
\includegraphics[scale=1,width=8cm]{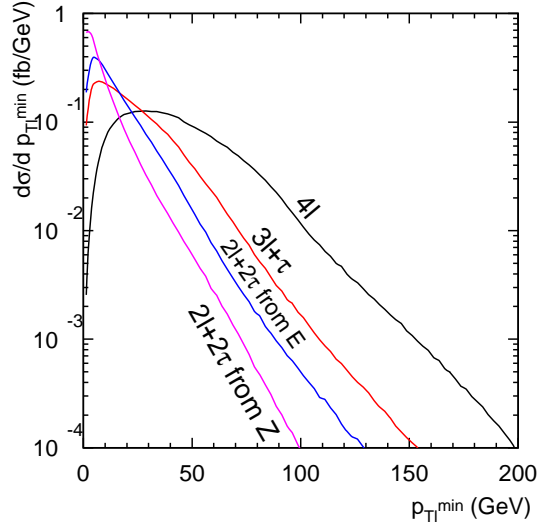}
\end{tabular}
\end{center}
\caption{$p_T$ distribution of the softer lepton from the heavy lepton $E$ and $\tau$ decays in the events of
$\ell^+\ell^+\ell^-\ell^- jj$,  $\tau^\pm \ell^\mp\ell^+\ell^- jj$ and $\tau^+\tau^- \ell^+\ell^- jj$, for a $E$ mass of $300$ GeV. }
\label{pairptl}
\end{figure}

Another important difference between the leptons from the primary
$E$ decay and from the $\tau$ decay is that the latter is much
softer. In Fig.~\ref{pairptl} we show the $p_T$ distribution of the
softer lepton from the heavy lepton and $\tau$ decays in the events
of $\ell^+\ell^+\ell^-\ell^- jj$,  $\tau^\pm \ell^\mp\ell^+\ell^-
jj$ and $\tau^+\tau^- \ell^+\ell^- jj$.

\begin{figure}[tb]
\begin{center}
\begin{tabular}{cc}
\includegraphics[scale=1,width=8cm]{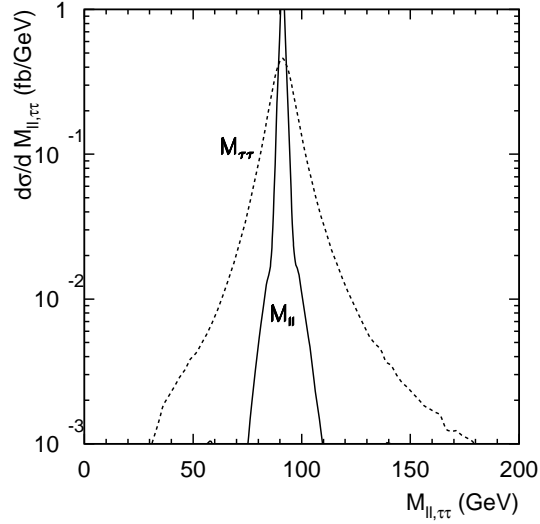}
\end{tabular}
\end{center}
\caption{Reconstructed invariant mass of $M(\ell^+\ell^-),M(\tau^+\tau^-)$ for
$\tau^+\tau^-\ell^+\ell^-jj$ production, with a $E$ mass of $300$ GeV. }
\label{pairzre2t}
\end{figure}

\begin{figure}[tb]
\begin{center}
\begin{tabular}{cc}
\includegraphics[scale=1,width=8cm]{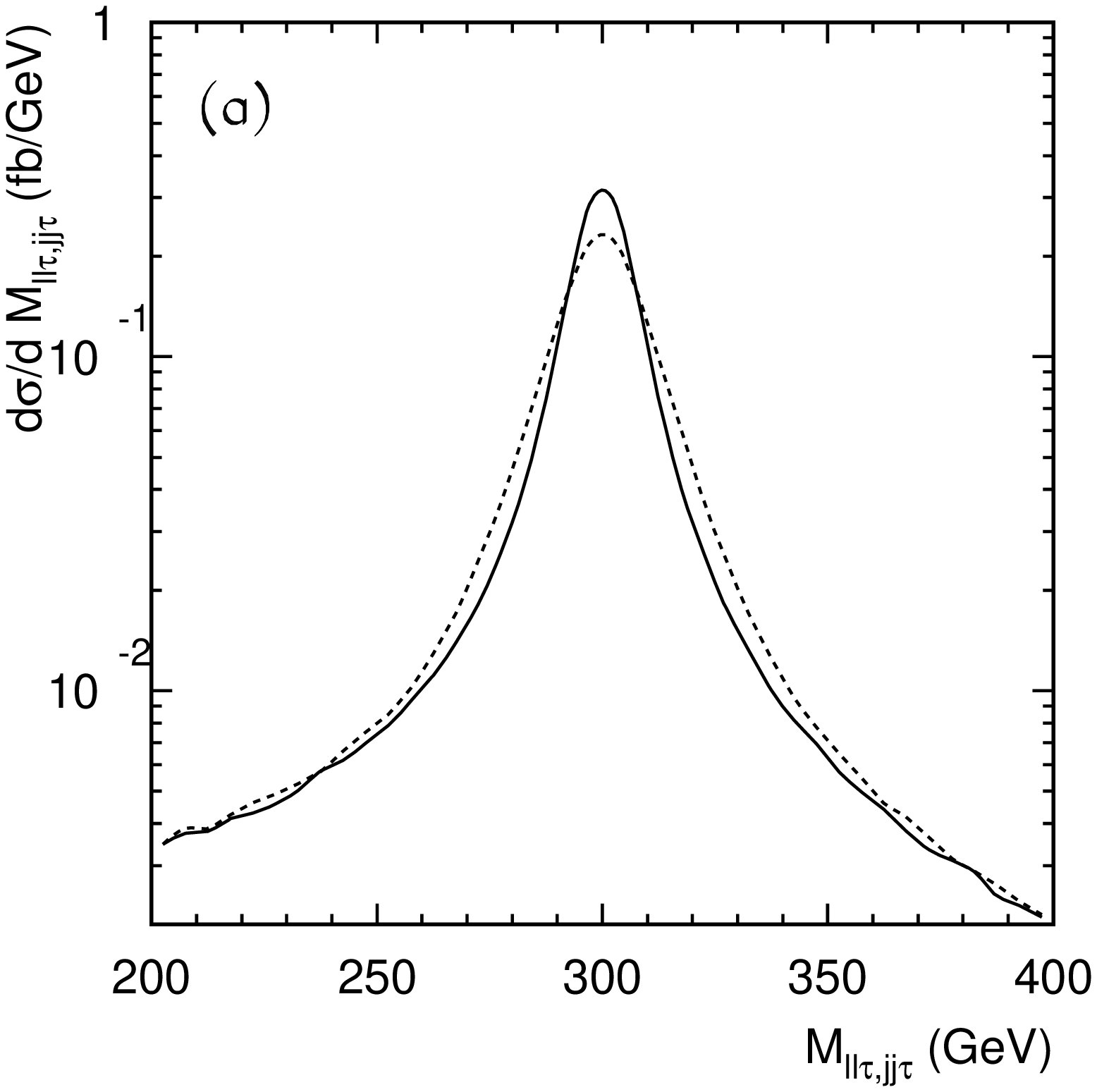}
\includegraphics[scale=1,width=8cm]{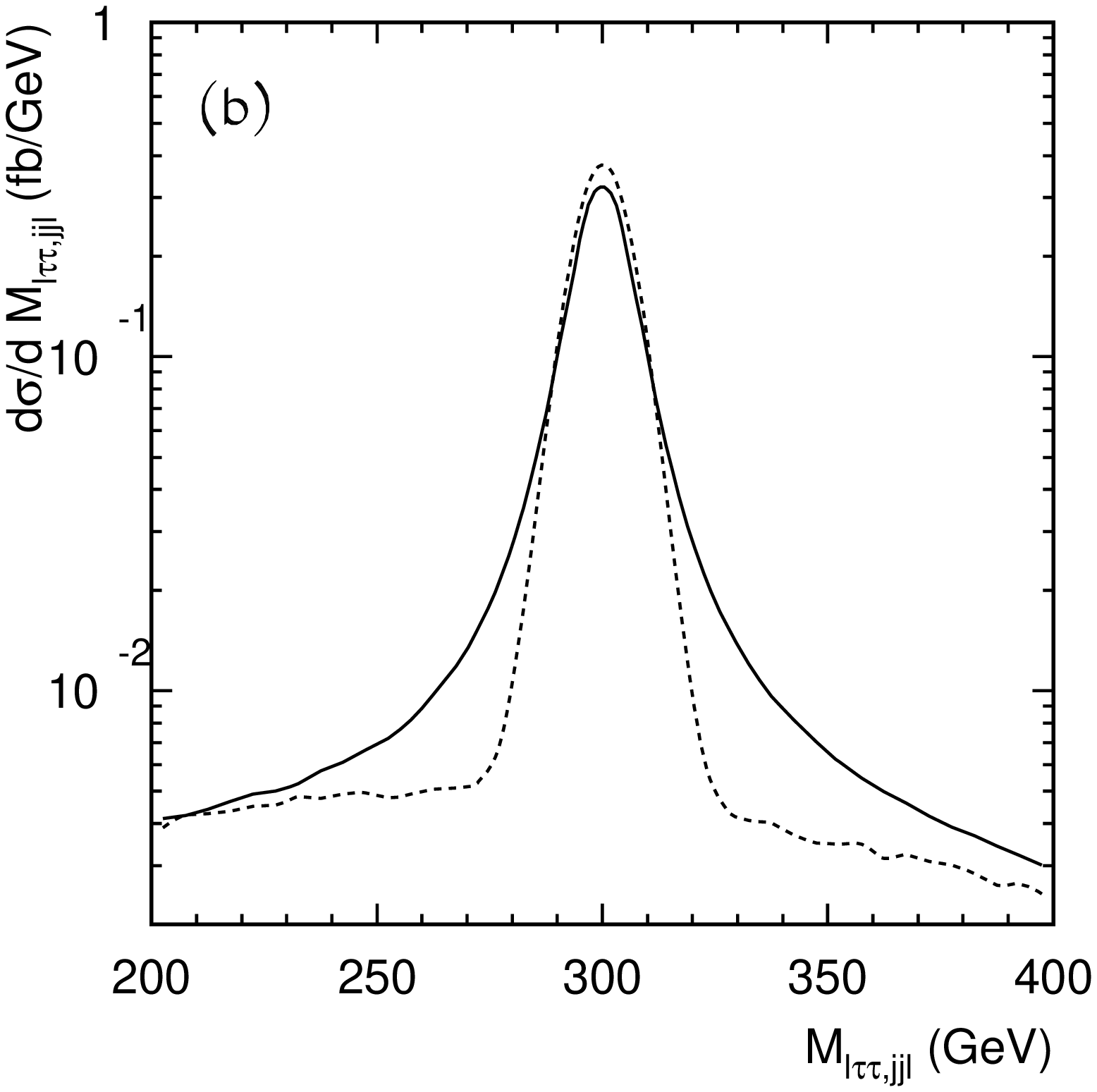}
\end{tabular}
\end{center}
\caption{Reconstructed invariant mass of $M(\tau^\pm\ell^+\ell^-)$
(solid), $M(\tau^\mp jj)$ (dashed) (a) and $M(\ell^\pm
\tau^+\tau^-)$ (solid), $M(\ell^\mp jj)$ (dashed) (b) for
$\tau^+\tau^-\ell^+\ell^-jj$ production, with a $E$ mass of $300$
GeV. } \label{pairere2t}
\end{figure}

Note that the two $\tau$'s could be produced
from either two heavy leptons or $Z$ boson. It is easy to
distinguish the two cases. Firstly it is the two hard $e,\mu$
leptons that reconstruct $Z$ boson when two $\tau$'s are from $E$
decay, but if two $\tau$'s reconstruct $Z$ boson the invariant mass
distribution must be much broader. We plot the reconstructed
invariant mass distributions of $M(\ell^+\ell^-)$ and
$M(\tau^+\tau^-)$ in Fig.~\ref{pairzre2t}. On the other hand, for
$E$ reconstruction, the invariant masses of $\tau^\pm\ell^+\ell^-$
and $\tau^\mp jj$ are almost the same. But that of $\ell^\pm
\tau^+\tau^-$ is broader than $\ell^\mp jj$, see
Fig.~\ref{pairere2t}.

\subsection{$E^\pm N$ associated production}
In this case the cleanest decay modes of $E^\pm$ and $N$ are
\begin{eqnarray}
E^\pm \to \ell^\pm Z, \ \ N\to \ell^\pm W^\mp \ \ (\ell=e,\mu,\tau)
\end{eqnarray}
and the signals for $E^\pm N$ associated production are
\begin{eqnarray}
&&E^\pm N\to \ell^\pm Z(\to q\bar{q})\ell^\pm W^\mp(\to q\bar{q}').
\end{eqnarray}


We employ the same basic acceptance cuts and smearing parameters as in the previous section. The leading irreducible SM background to this channel is
\begin{eqnarray}
&&t\bar{t}W^\pm\to \ell^\pm\ell^\pm jjjj+\cancel{E}_T.
\end{eqnarray}
The QCD processes $jjjjW^\pm W^\pm, jjW^\pm W^\pm W^\mp$ are much
smaller. This is estimated based on the fact that QCD induced $jjW^\pm
W^\pm\rightarrow jj\ell^\pm\ell^\pm\cancel{E}_T$ is about 15 fb.
With an additional $\alpha^2_s$ and 6 body phase space or one more
$W$ suppression, they are much smaller than $t\bar{t}W^\pm$. Other
EW backgrounds $WWWW,WWWZ$ are also negligible.

\begin{figure}[tb]
\begin{center}
\begin{tabular}{cc}
\includegraphics[scale=1,width=8cm]{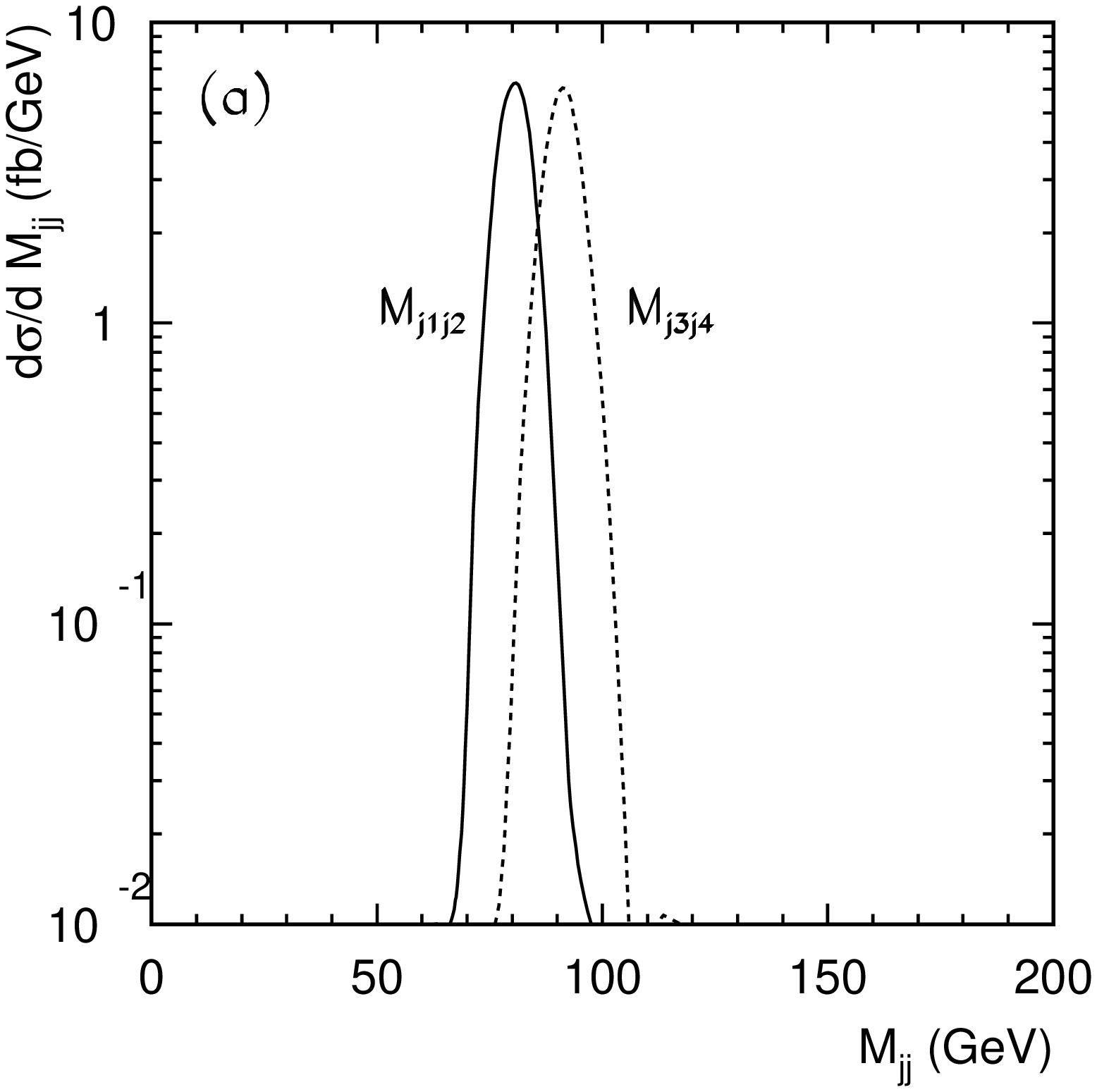}
\includegraphics[scale=1,width=8cm]{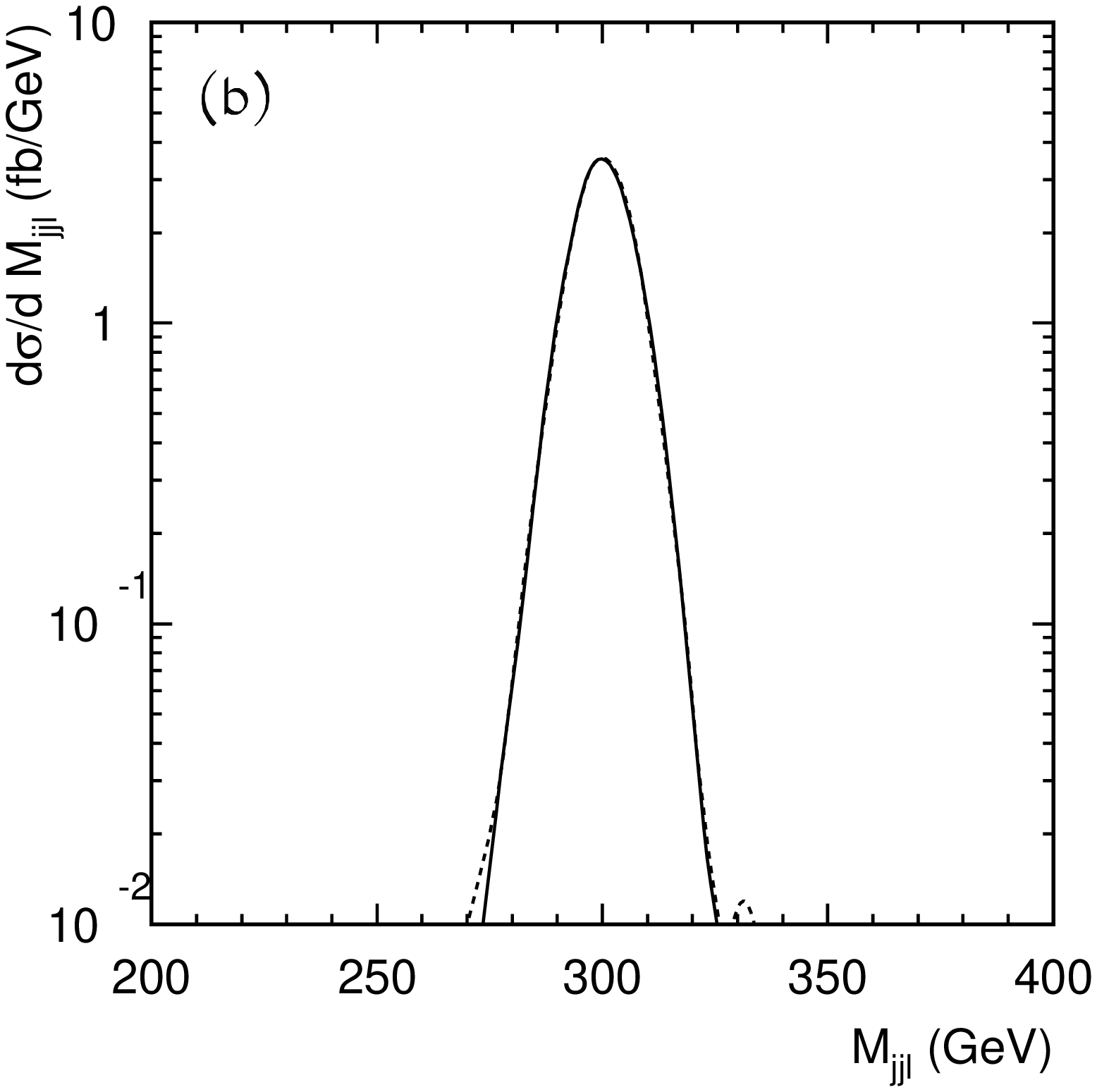}
\end{tabular}
\end{center}
\caption{Reconstructed invariant mass of $M(j_1j_2),M(j_3j_4)$ (a) and $M(\ell_1 j_1j_2),M(\ell_2j_3j_4)$ (b) for $\ell^\pm\ell^\pm jjjj$ production, with a $E$ mass of $300$ GeV. }
\label{aso1re}
\end{figure}

\begin{figure}[tb]
\begin{center}
\begin{tabular}{cc}
\includegraphics[scale=1,width=8cm]{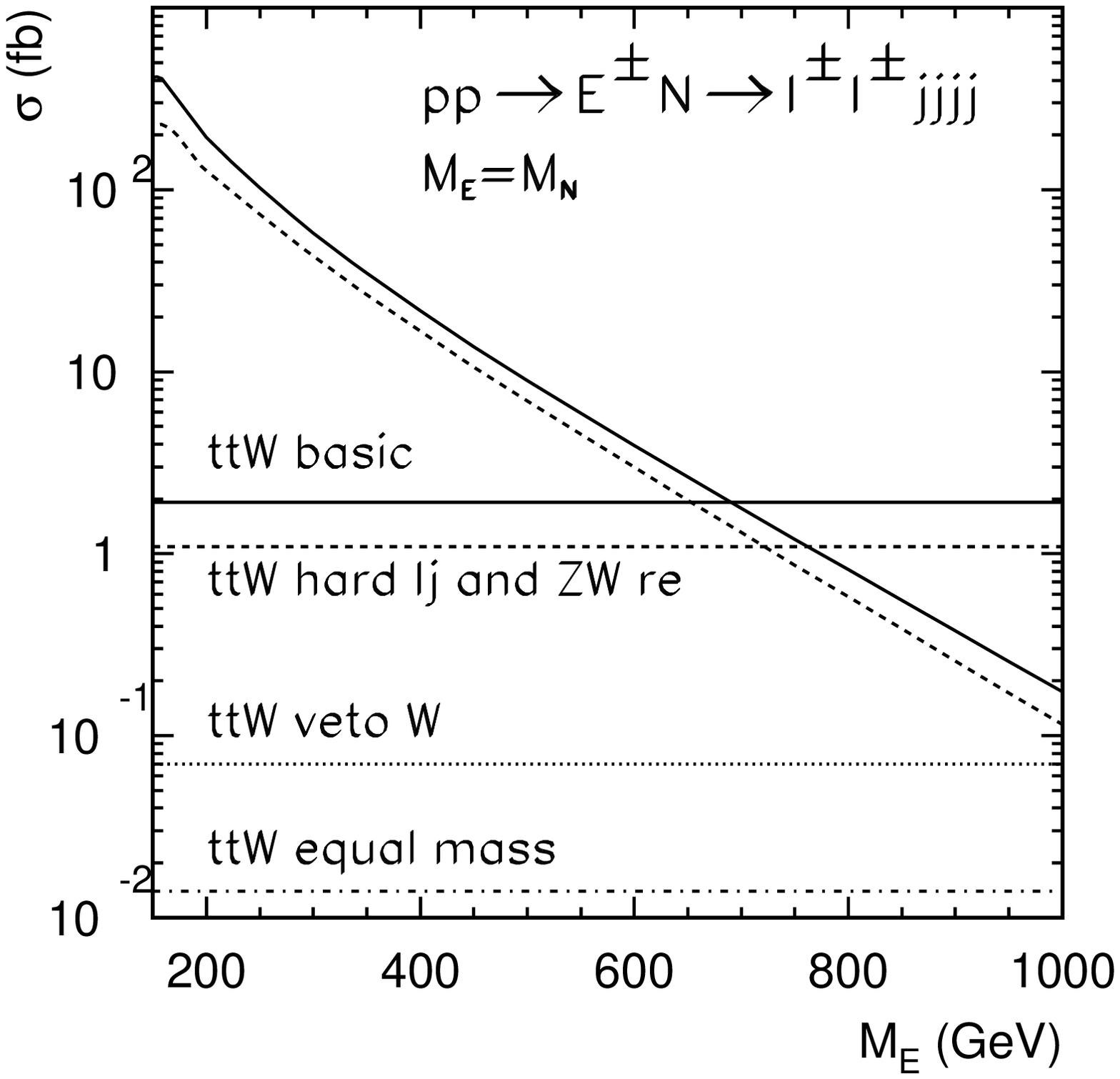}
\end{tabular}
\end{center}
\caption{Total cross section for $E^\pm N$ production after the
basic cuts (solid) and all cuts (dashed) and the leading
$t\bar{t}W^\pm$ background after all cuts. } \label{aso1}
\end{figure}

We again apply further judicious cuts as following to reduce the background,
\begin{itemize}
\item We set additional cuts for hard leptons and jets
\begin{eqnarray}
p_T^{max}(\ell)>M_E/4, \ \ p_T^{max}(j)>50~{\rm GeV}.
\end{eqnarray}
\item To reconstruct $Z$ and $W$ boson masses, we select two pair jets among the four ones and take their invariant masses closest to $M_Z$ and $M_W$ with $|M_{j_1j_2}-M_Z(M_{j_3j_4}-M_W)|<15~{\rm GeV}$. Their invariant masses are plotted in Fig.~\ref{aso1re} (a).
\item To reconstruct heavy lepton $E$ and $N$, we take advantage of the feature that they have equal mass $M_{\ell_1 j_1j_2}=M_{\ell_2j_3j_4}$. In practice, we take $|M_{\ell_1 j_1j_2}-M_{\ell_2j_3j_4}|<M_E/25$. This helps for the background reduction.
\item To remove the $t\bar{t}Z$ background, we veto the events with large missing energy from $W$ decay $\cancel{E}_T<20~{\rm GeV}$.
\end{itemize}
Next, when we perform the signal significance analysis, we look for
the resonance in the mass distribution of $\ell_1 j_1j_2$ and
$\ell_2j_3j_4$. The invariant masses of them are plotted in
Fig.~\ref{aso1re} (b) for 300 GeV $E$ pair production. If we look at
a mass window of $|M_{\ell_1 j_1j_2,\ell_2j_3j_4}-M_E|<M_E/20$, the
backgrounds will be at a negligible level. The production cross
section of $E^\pm N$ signal with basic cuts (solid curve) and all of
the cuts (dotted curve) above are plotted in Fig.~\ref{aso1}. For
comparison, the background processes of $t\bar{t}W^\pm$ is also
included with the sequential cuts as indicated.

Finally, we would like to give a comment for another excellent
signal for $E^\pm N$ production
\begin{eqnarray}
E^\pm N\to \ell^\pm Z(\to \ell^+\ell^-)\ell^\pm W^\mp(\to q\bar{q}')
\end{eqnarray}
This signal has almost no standard model background. However, its
production rate is about 10 times lower than that we consider above.
We will not study it here further.

\subsection{Measuring Branching Fractions}

\begin{figure}[tb]
\begin{center}
\begin{tabular}{cc}
\includegraphics[scale=1,width=8cm]{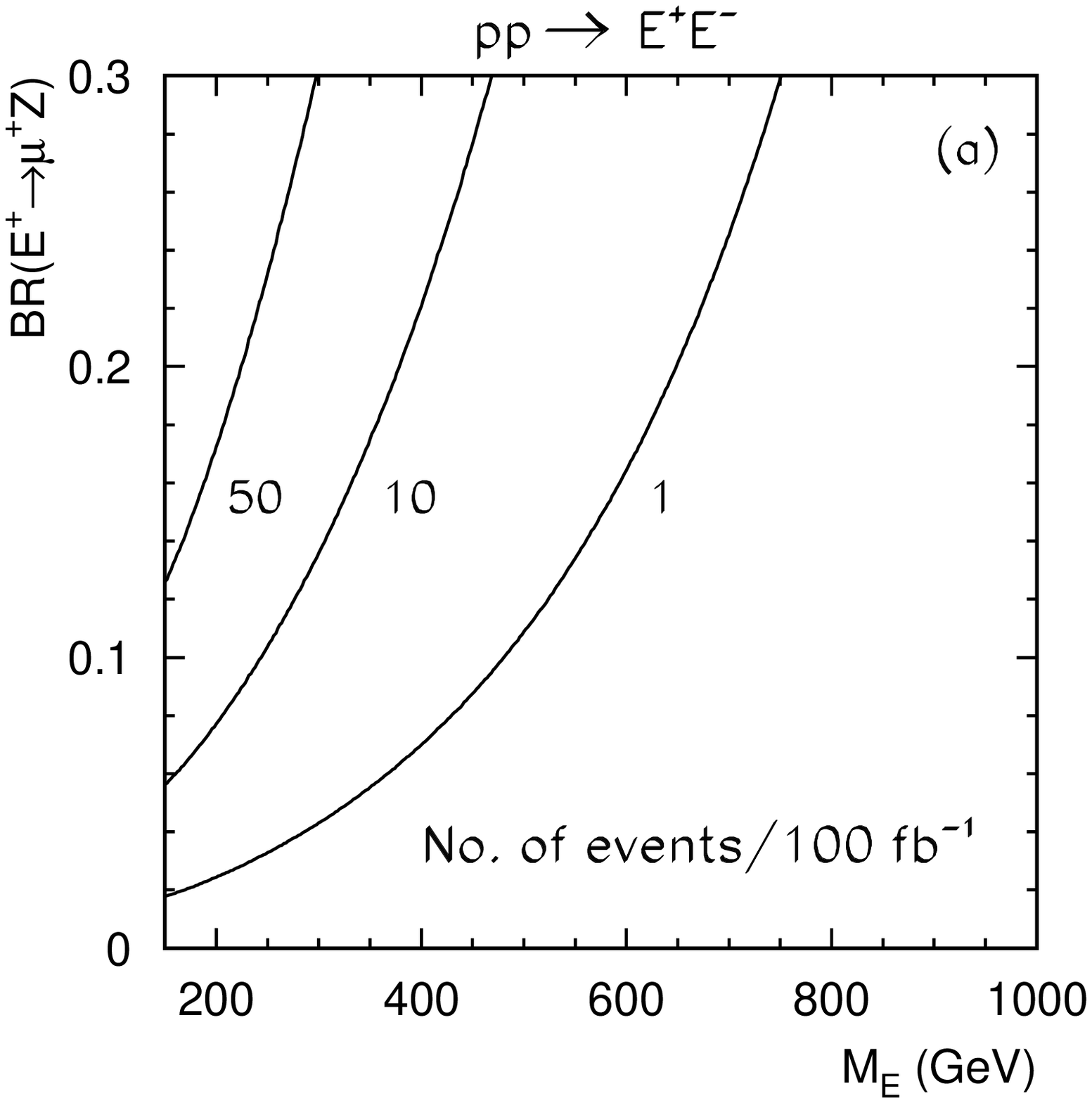}
\includegraphics[scale=1,width=8cm]{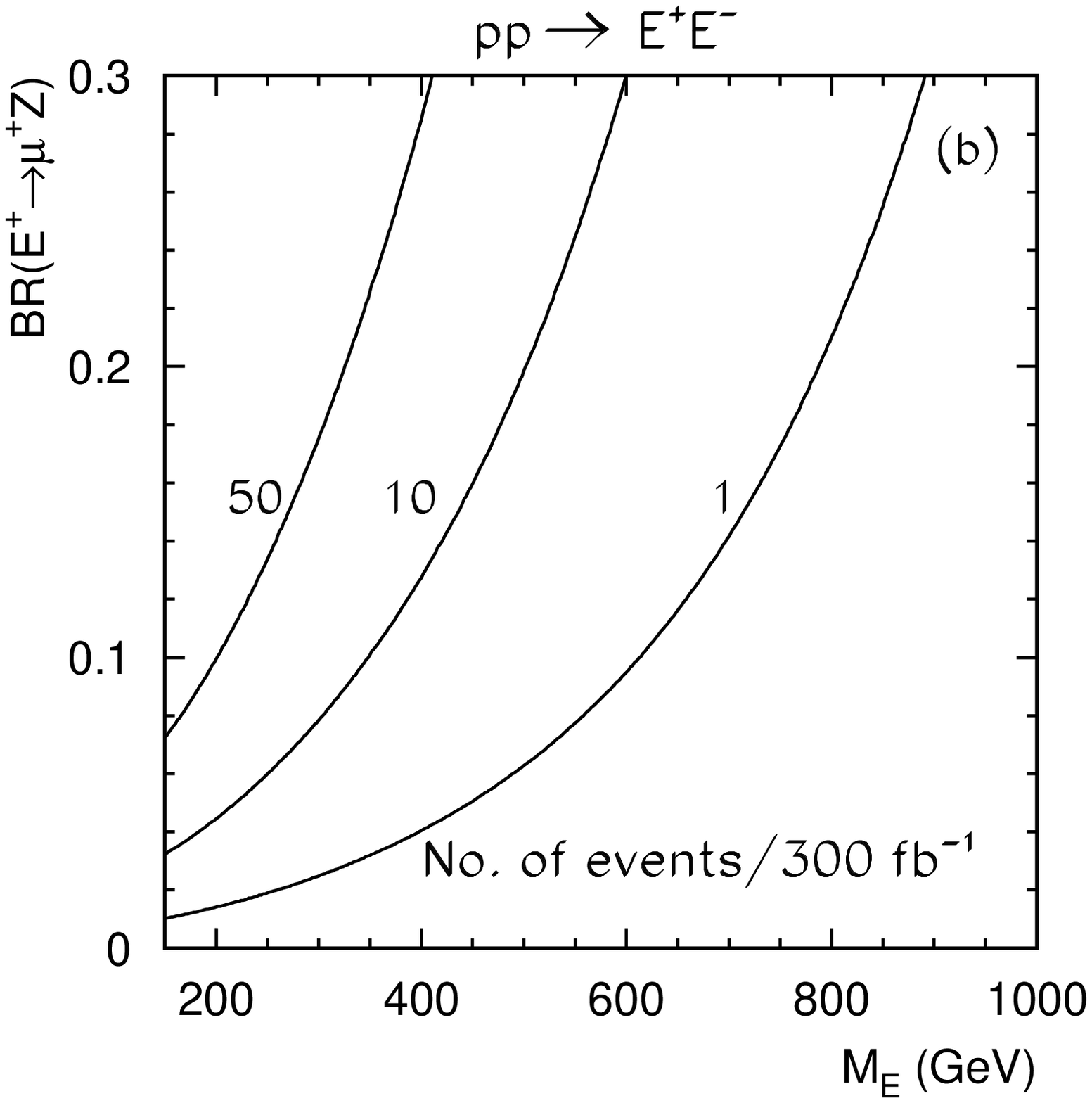}
\end{tabular}
\end{center}
\caption{Event contours in the BR-$M_E$ plane at the LHC with
integrated luminosity $100~{\rm fb}^{-1}$ (a) and $300~{\rm
fb}^{-1}$ (b) for $E^+E^-\to \mu^+Z\mu^-Z\to \mu^+\mu^+\mu^-\mu^-
jj$, including all the judicious cuts described in the early
section. } \label{eve1}
\end{figure}

So far, we have only studied the characteristic features of the
signal and backgrounds for the leading channels with the
decay branching fractions of $E$ and $N$ to be $100\%$. For
illustration, we consider first the cleanest channel, $E^+E^-\to
\mu^+Z \ \mu^- Z$. The number of events is written as
\begin{eqnarray}
N=L\times \sigma(pp\to E^+E^-)\times {\rm BR}^2(E^+\to \mu^+Z),\label{eventpair}
\end{eqnarray}
where $L$ is the integrated luminosity. Given a sufficient number of
events $N$, the mass of $E$ is determined by the invariant mass of
three leptons and one lepton and two jets. We thus predict the
corresponding production rate $\sigma(pp\to E^+E^-)$ for this given
mass. The only unknown in the Eq.~\ref{eventpair} is the decay
branching fraction. We present the event contours in the BR$-M_{E}$
plane in Figs.~\ref{eve1} (a) and (b) for $100~{\rm
fb}^{-1}$ and $300~\rm fb^{-1}$ luminosities including all the
judicious cuts described earlier, with which the backgrounds are
insignificant. We see that with the estimated branching fraction for
$\mu+Z$, one can reach the coverage of about $M_{E}\lesssim 0.8$ TeV for
$100~{\rm fb}^{-1}$ luminosity and $M_{E}\lesssim 0.9$ TeV for
$300~{\rm fb}^{-1}$ luminosity.

\begin{figure}[tb]
\begin{center}
\begin{tabular}{cc}
\includegraphics[scale=1,width=8cm]{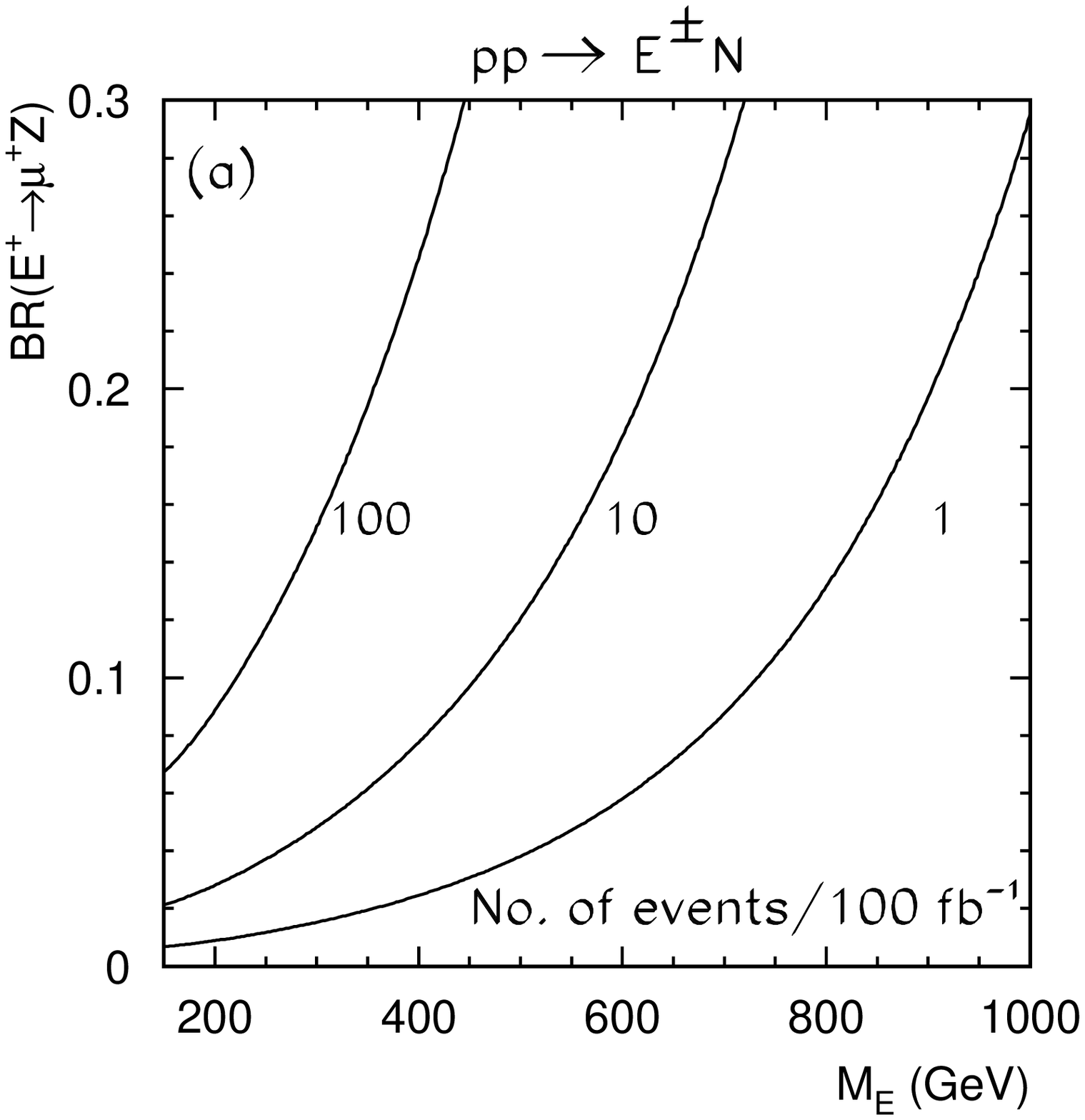}
\includegraphics[scale=1,width=8cm]{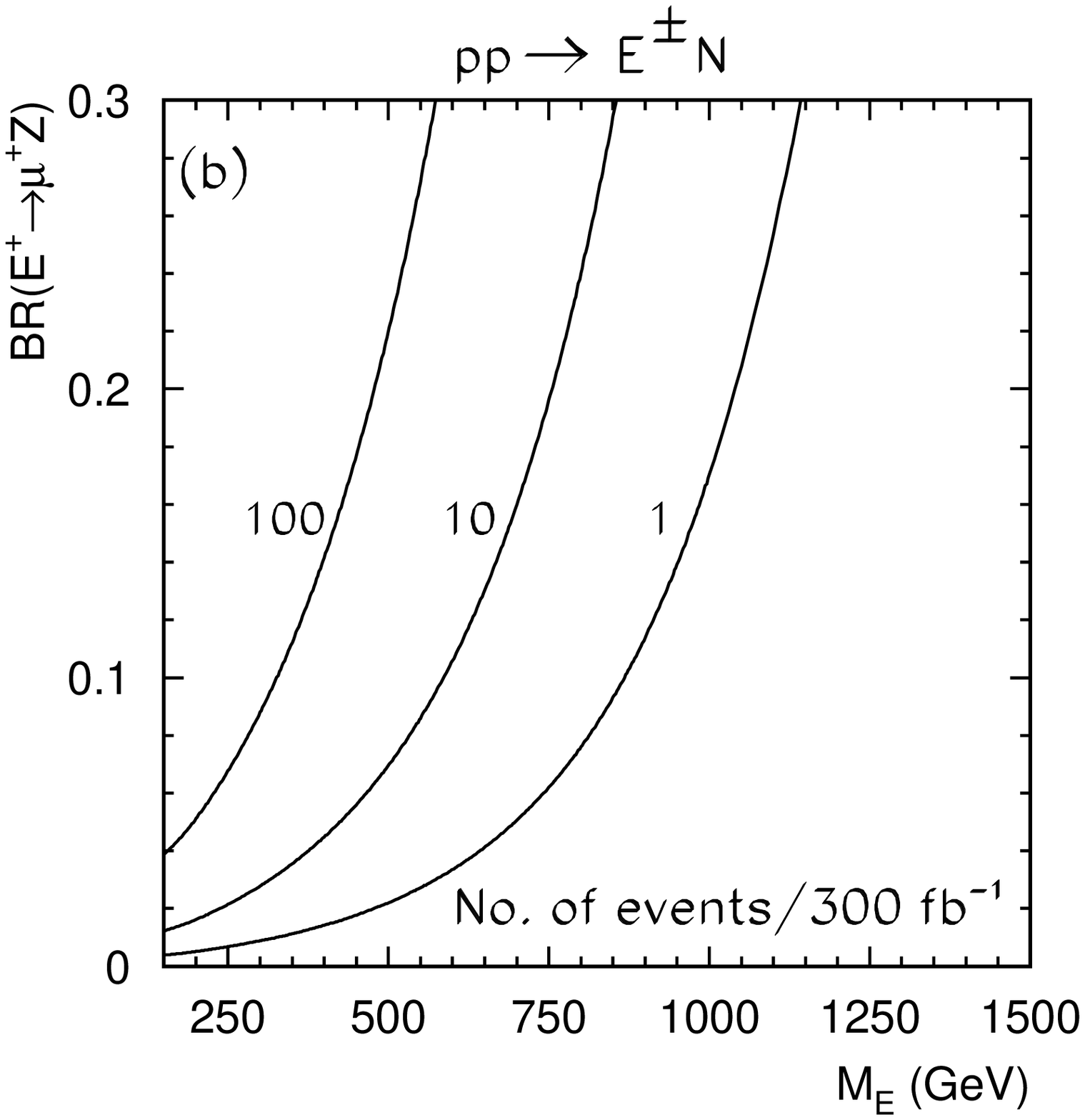}
\end{tabular}
\end{center}
\caption{Event contours in the BR-$M_E$ plane at the LHC with
integrated luminosity $100~{\rm fb}^{-1}$ (a) and $300~{\rm
fb}^{-1}$ (b) for $E^\pm N\to \mu^\pm Z \mu^\pm W^\mp\to \mu^\pm
\mu^\pm jjjj$, including all the judicious cuts described in the
early section. } \label{eve2}
\end{figure}

\begin{figure}[tb]
\begin{center}
\begin{tabular}{cc}
\includegraphics[scale=1,width=8cm]{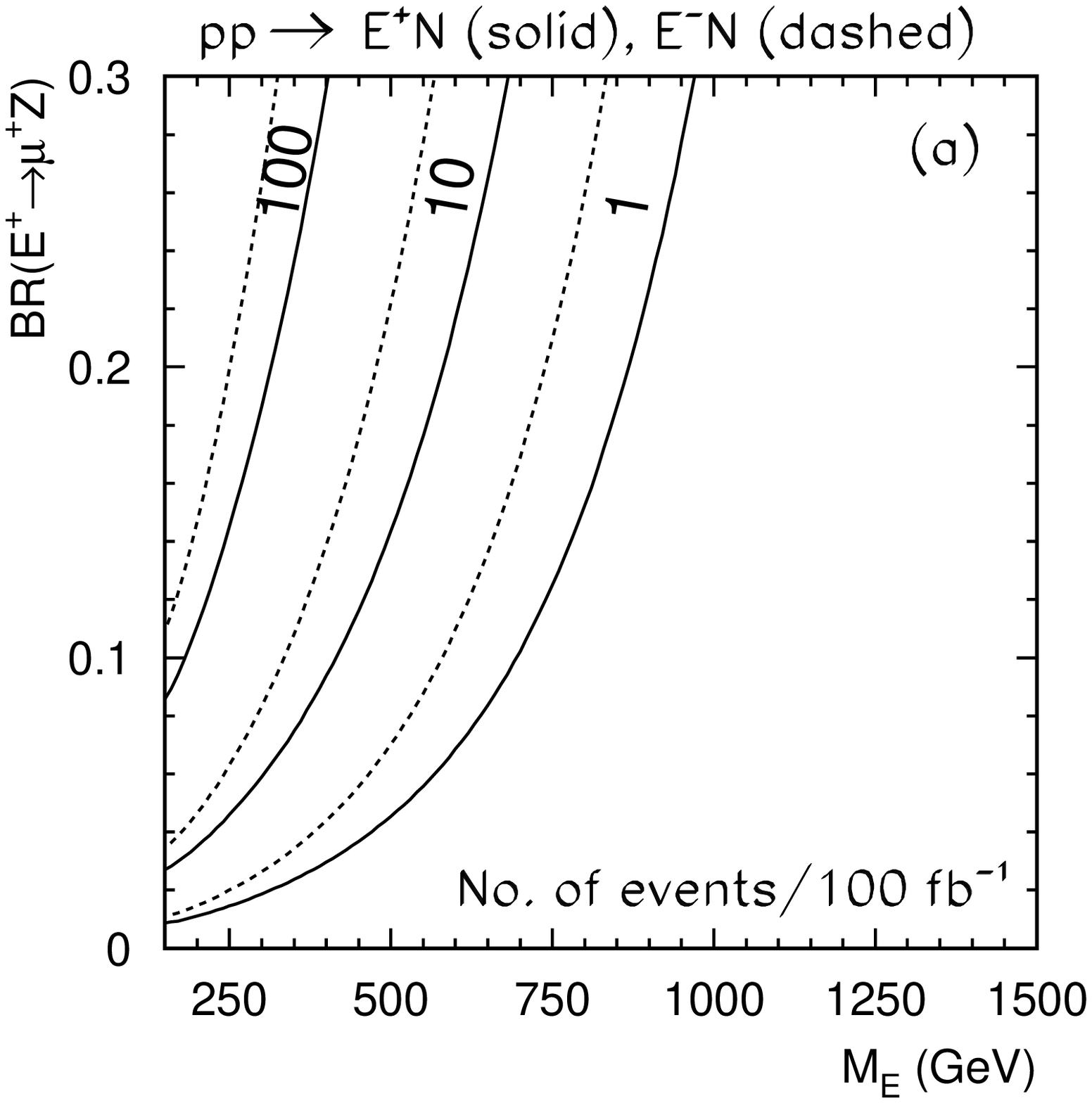}
\includegraphics[scale=1,width=8cm]{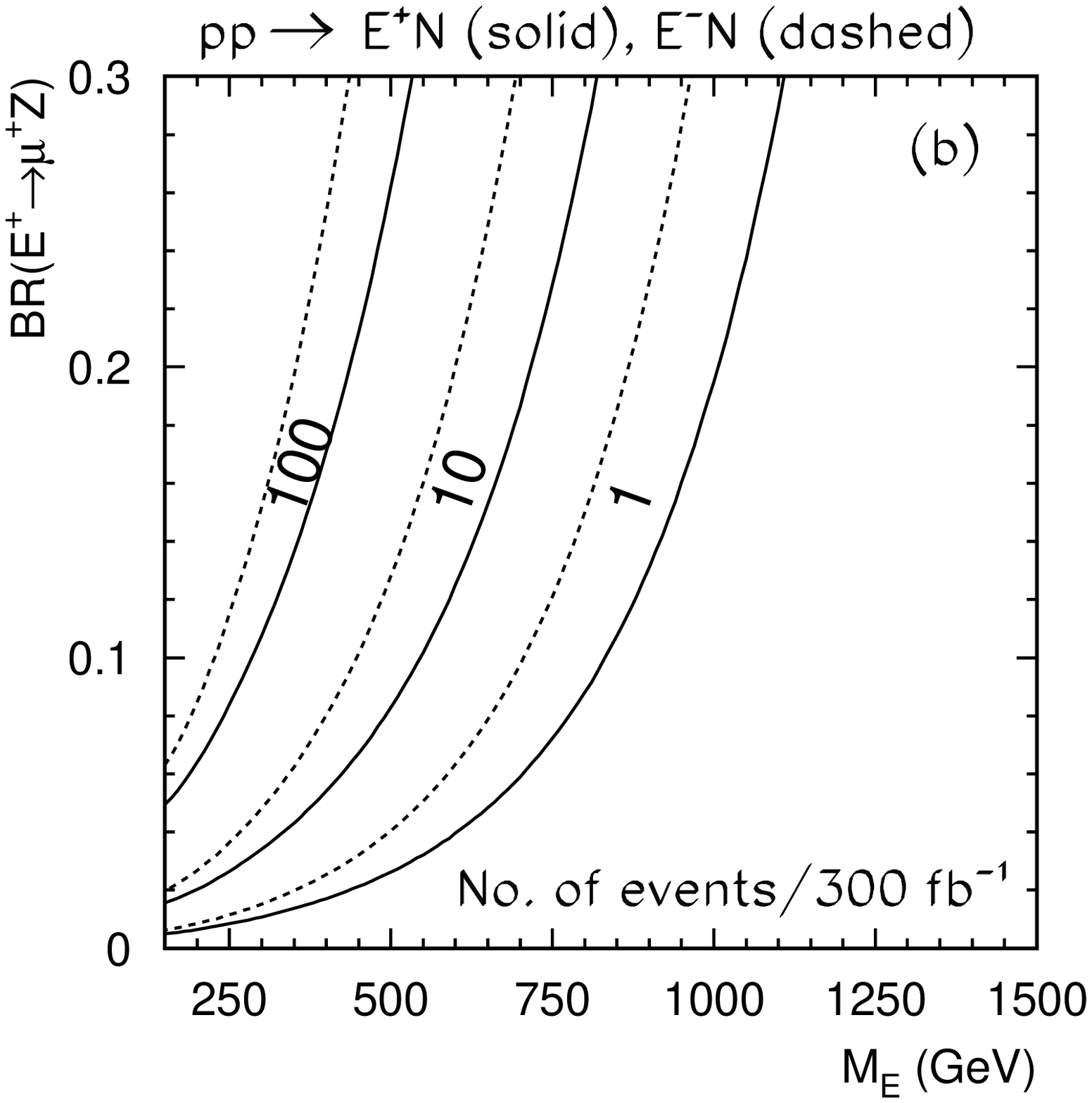}
\end{tabular}
\end{center}
\caption{Event contours in the BR-$M_E$ plane at the LHC with
integrated luminosity $100~{\rm fb}^{-1}$ (a) and $300~{\rm
fb}^{-1}$ (b) for $E^+ N\to \mu^+ Z \mu^+ W^-\to \mu^+ \mu^+ jjjj$
(solid line) and $E^- N\to \mu^- Z \mu^- W^+\to \mu^- \mu^- jjjj$
(dashed line), including all the judicious cuts described in the
early section. } \label{eve3}
\end{figure}

The associated $E^\pm N$ production has larger cross section
compared with $E^+E^-$ production and can provide even better
signals. In Figs.~\ref{eve2} (a) and (b), we show the event contours
in the BR-$M_E$ plane, for $100~{\rm fb}^{-1}$ and $300~{\rm
fb}^{-1}$ luminosities including all the judicious cuts described
earlier. Note again the events number for $E^+N$ and $E^-N$ cases
are different due to the LHC being a pp machine. In Fig.~\ref{eve3}
(a) and (b) we show the event contour for $E^+ N$ and $E^- N$
separately. One can see that the LHC has tremendous sensitivity to
probe the channel $E^\pm\to \mu^\pm Z$ or $N\to \mu^\pm W^\mp$ in
this production mechanism. One can reach the coverage of about
$M_{E}\lesssim 1$ TeV for $100~{\rm fb}^{-1}$ luminosity and
$M_{E}\lesssim 1.2$ TeV for $300~{\rm fb}^{-1}$ luminosity.

We comment that for Case I, where the three heavy triplet charged
and neutral leptons are degenerate, one can not distinguish among
the three heavy particles and the detection signals will add up.
This will enhance the event number by roughly a factor of 3.

As discussed earlier that in Case I, the $E$ and $N$ decay branching
fractions and the light neutrino mass matrix are directly
correlated, therefore measuring the BR's of different flavor
combinations becomes crucial in understanding the neutrino mass
hierarchy pattern and thus the mass generation mechanism. We find
the following for Case I when all phases are neglected,

\beq {\rm BR}(E^+E^-(E^\pm N) \to \ell\ell ZZ(\ell\ell ZW)) \approx
\left\{
\begin{array}{ll}
\displaystyle (23\%)^2 \quad &{\rm for\ NH:}\ (\mu^\pm+\tau^\pm)
(\mu^\pm+\tau^\pm) ZZ(ZW),  \\ [1mm] \displaystyle (13\%)^2 \quad
&{\rm for\ IH:}\  e^\pm e^\pm ZZ(ZW),  \\ [1mm] \displaystyle
(17\%)^2 \quad &{\rm for\ QD:}\ (e^\pm+\mu^\pm+\tau^\pm)
(e^\pm+\mu^\pm+\tau^\pm) ZZ(ZW),
\end{array}
\right. \label{BRN} \eeq
supporting statement above. These predictions are the consequence
from the neutrino oscillation experiments and are subject to be
tested at the LHC to confirm the seesaw theory. However for the more
general situation Case II when heavy neutrinos are not degenerate,
no such information can be extracted since the correlation between
the BR and light neutrino mass patterns is not strong.

We would like to comment that even for the complicated Case II,
interesting information about the model can still be extracted. As
pointed out earlier that using information on the neutrino mass
pattern from other experiments and the sizes of elements in $V_{lN}$
from analysis here, one may be able to obtain more information about
the model parameters such as the angles $w_{ij}$ and the Majorana
phases.

\section{Summary}
We have studied the properties of heavy $SU(2)_L$ triplet lepton in
Type III seesaw model and also their signatures at the LHC for small
mixing solution between light and heavy leptons. The small mixing
solution is characterized by the fact that in the limit that the
light neutrino masses go to zero, the mixing also goes to zero. The
smallness of light neutrino masses then leads to the fact that the
total decay widths of heavy leptons are small. With such small decay
widths, although not considered as long-lived for large triplet
mass, the heavy lepton decays could lead to a visible displaced
vertex in the detector at the LHC. This displaced vertex can be
observed through $E$ and $N$ reconstructions. We summarize our main
results with small mixing in the following:

\begin{itemize}

\item To a good approximation, the couplings of light charged lepton and heavy triplet leptons $V_{lN}$ to $Z$, $W$ and $H^0$ bosons in Eq.~\ref{SSS}
can be expressed with measurable neutrino mass and mixing through
Eq.~\ref{VL} with three unknown complex parameters $w_{ij}$ in a
$3\times 3$ orthogonal matrix given in Eq.~\ref{vlno}. This allowed
us to study the correlation between the decays of heavy triplet
leptons, light neutrino masses and mixing and model parameters. With
real $w_{ij}$, the mixing between light and heavy is small which
leads to displaced vertex at the LHC for heavy leptons decays if
produced.

\item Using the relation in Eq.~\ref{VL}, we have tried to study possible correlation in neutrino mass hierarchy and heavy lepton decays with real $w_{ij}$.
We find that only in certain limited cases, for example Case I
studied in this paper, the correlation is strong. The study of heavy
lepton productions and decays at the LHC may help to determine the
neutrino mass hierarchy. For the more general situation Case II when
heavy neutrinos are not degenerate, no such information can be
extracted because the correlation is weak. However, even for this
case interesting information about the model can still be extracted.
If in the future the neutrino mass pattern is determined from other
experiments and the sizes of elements in $V_{lN}$ from analysis of
heavy lepton productions and decays at the LHC, one may be able to
obtain more information about the model parameters such as the
angles $w_{ij}$ and the Majorana phases $\Phi_i$.



\item We have studied production and detection of heavy triplet leptons at LHC with
judicious cuts to reduce SM background to see how large the seesaw
scale can be reach at the LHC. The associated production $E^\pm N$
is crucial to identify the quantum numbers of the triplet leptons
and to distinguish between the neutrino mass hierarchies. Even with
only the cleanest channels $\mu^\pm \mu^\pm+\rm jets$, the signal
observability can reach about $M_{E}\lesssim 1$ TeV for $100~{\rm
fb}^{-1}$ luminosity and $M_{E}\lesssim 1.2$ TeV for $300~{\rm
fb}^{-1}$ luminosity.

\item Although the rate of pair production $E^+E^-$ is
smaller than $E^\pm N$, we demonstrated that besides the clean
4-lepton channels from $e, \mu$, the $\tau$ final state can be
effectively reconstructed as well. Even with only the cleanest
channels $\mu^+ \mu^+ \mu^- \mu^-+\rm jets$, the signal
observability can reach $M_{E}\lesssim 0.8$ TeV for $100~{\rm
fb}^{-1}$ luminosity and $M_{E}\lesssim 0.9$ TeV for $300~{\rm
fb}^{-1}$ luminosity.

\end{itemize}

If Nature does use low scale, as low as 1 TeV, to facilitate seesaw mechanism, there will be a lot of surprises to come soon after LHC
will be in full operation. We urge our experimentalists to carry out searches for low scale seesaw effects.

\newpage

\subsection*{Acknowledgment}
XGH was supported in part by the NSC and NCTS. We acknowledge Tao
Han for providing his Fortran codes HANLIB for our calculations. T.
Li would like to thank Tao Han and Pavel Fileviez P{\'e}rez for
helpful discussions.

\appendix

\section{Derivation of Equation \ref{VVL}}
To derive Eq.~\ref{VVL}, we need to have some detailed relation of the block matrices in
the unitary matrices $U_{0}$ and $U_{L,R}$. For $U_0$, we have
\begin{eqnarray}
&&U_{0\nu\nu}U_{0\nu\nu}^\dagger+U_{0\nu\Sigma}U_{0\nu\Sigma}^\dagger
=U_{0\Sigma\nu}U_{0\Sigma\nu}^\dagger+U_{0\Sigma\Sigma}U_{0\Sigma\Sigma}^\dagger=1,\nonumber\\
&&U_{0\nu\nu}^\dagger U_{0\nu\nu}+U_{0\Sigma\nu}^\dagger
U_{0\Sigma\nu}
=U_{0\nu\Sigma}^\dagger U_{0\nu\Sigma}+U_{0\Sigma\Sigma}^\dagger U_{0\Sigma\Sigma}=1,\\
&&U_{0\nu\nu} U_{0\Sigma\nu}^\dagger+U_{0\nu\Sigma}
U_{0\Sigma\Sigma}^\dagger =U_{0\nu\nu}^\dagger
U_{0\nu\Sigma}+U_{0\Sigma\nu}^\dagger U_{0\Sigma\Sigma}=0.\nonumber
\end{eqnarray}

From neutrino mass matrix diagonalization, we have
\begin{eqnarray}
&&U_0^\dagger \left(
  \begin{array}{cc}
    0 & Y_\Sigma^\dagger v/\sqrt{2} \\
    Y_\Sigma^\ast v/\sqrt{2} & M_\Sigma^\ast \\
  \end{array}
\right) U_0^\ast
=
\left(
  \begin{array}{cc}
    m_\nu^{diag} & 0 \\
    0 & M_{\Sigma N}^{diag} \\
  \end{array}
\right)\;,
\end{eqnarray}
and
\begin{eqnarray}
&&U_{0\Sigma\nu}^\dagger Y_\Sigma^\ast
v/\sqrt{2}=m_\nu^{diag} U_{0\nu\nu}^T, \ \ Y_\Sigma^\dagger
vU_{0\Sigma\Sigma}^\ast/\sqrt{2} =U_{0\nu\Sigma}M_{\Sigma N}^{diag},\nonumber\\
&&U^\dagger_{0\nu\nu}Y^\dagger_\Sigma
v/\sqrt{2}+U^\dagger_{0\Sigma\nu}M_\Sigma^\ast=m_\nu^{diag}U_{0\Sigma
\nu}^T, \ \ Y_\Sigma^\ast
vU^\ast_{0\nu\Sigma}/\sqrt{2}+M_\Sigma^\ast U^\ast_{0\Sigma\Sigma}=U_{0\Sigma
\Sigma}M_{\Sigma N}^{diag}.
\end{eqnarray}
For $U_{L,R}$, we have
\begin{eqnarray}
&&U_{L,Rll}U_{L,Rll}^\dagger+U_{L,Rl\Psi}U_{L,Rl\Psi}^\dagger
=U_{L,R\Psi l}U_{L,R\Psi l}^\dagger+U_{L,R\Psi\Psi}U_{L,R\Psi\Psi}^\dagger=1,\nonumber\\
&&U_{L,Rll}^\dagger U_{L,Rll}+U_{L,R\Psi l}^\dagger U_{L,R\Psi l}
=U_{L,Rl\Psi}^\dagger U_{L,Rl\Psi}+U_{L,R\Psi\Psi}^\dagger U_{L,R\Psi\Psi}=1,\\
&&U_{L,Rll} U_{L,R\Psi l}^\dagger+U_{L,Rl\Psi}
U_{L,R\Psi\Psi}^\dagger =U_{L,Rll}^\dagger U_{L,Rl\Psi}+U_{L,R\Psi
l}^\dagger U_{L,R\Psi\Psi}=0.\nonumber
\end{eqnarray}

From charged lepton mass matrix diagonalization, we have
\begin{eqnarray}
&&U_L^\dagger \left(
  \begin{array}{cc}
    m_l^\dagger & Y_\Sigma^\dagger v \\
    0 & M_\Sigma^\dagger\\
  \end{array}
\right) U_R = \left(
  \begin{array}{cc}
    m_l^{diag} & 0 \\
    0 & M_{\Sigma C}^{diag} \\
  \end{array}
\right)
\end{eqnarray}
and
\begin{eqnarray}
&&U_{Lll}^\dagger m_l^\dagger=m_l^{diag}U_{Rll}^\dagger, \ \
M_\Sigma^\dagger U_{R\Psi\Psi}=U_{L\Psi\Psi}M_{\Sigma C}^{diag},\nonumber\\
&&U_{R\Psi l}^\dagger M_\Sigma=m_l^{diag}U_{L\Psi l}^\dagger, \ \
m_l U_{Ll\Psi}=U_{Rl\Psi}M_{\Sigma C}^{diag},\nonumber\\
&&U_{Rll}^\dagger m_l+U_{R\Psi l}^\dagger Y_\Sigma
v=m_l^{diag}U_{Lll}^\dagger, \ \ Y_\Sigma vU_{Ll\Psi}+M_\Sigma
U_{L\Psi\Psi}=U_{R\Psi\Psi}M_{\Sigma C}^{diag},\\
&&U_{Lll}^\dagger Y_\Sigma^\dagger v+U_{L\Psi l}^\dagger
M_\Sigma^\dagger=m_l^{diag} U_{R\Psi l}^\dagger, \ \ m_l^\dagger
U_{Rl\Psi}+Y_\Sigma^\dagger v U_{R\Psi\Psi}=U_{Ll\Psi}M_{\Sigma
C}^{diag}\;.\nonumber
\end{eqnarray}
Combining the above relations and the definition of $V^L_{l\Sigma}$, and using the approximation $M_{\Sigma N} = M_{\Sigma C} = M_\Sigma$, we obtain
\begin{eqnarray}
V_{l\Sigma}^L&=&U_{Lll}^\dagger
U_{0\nu\Sigma}+\sqrt{2}U_{L\Psi l}^\dagger U_{0\Sigma\Sigma}\;,
\end{eqnarray}
which leads to
\begin{eqnarray}
V_{l\Sigma}^L&=&V_{PMNS}U_{0\nu\nu}^\dagger U_{0\nu\Sigma}+\sqrt{2}U_{L\Psi
l}^\dagger U_{0\Sigma\Sigma}+V_{l\Sigma}^L U_{0\nu \Sigma}^\dagger
U_{0\nu \Sigma}\;.
\end{eqnarray}
Then we have
\begin{eqnarray}
V_{l\Sigma}^\ast M_{\Sigma N}^{diag}U_{0\nu\Sigma}^\dagger
U_{Lll}&=& -V_{PMNS}^\ast m_\nu^{diag} U_{0\nu\nu}^\dagger
U_{Lll}+\sqrt{2}U_{L\Psi l}^T Y_\Sigma v\sqrt{2}U_{Lll}\;.\label{ee}
\end{eqnarray}

From the definition of $V_{l\Sigma}^L$ we can also get
\begin{eqnarray}
V_{l\Sigma}^L&=&V_{PMNS}U_{0\Sigma\nu}^\dagger U_{0\Sigma\Sigma}+U_{Lll}^\dagger
U_{0\nu\Sigma}+V_{l\Sigma}^L U_{0\Sigma \Sigma}^\dagger U_{0\Sigma
\Sigma}
\end{eqnarray}
which leads to
\begin{eqnarray}
V_{l\Sigma}^{L\ast} M_{\Sigma N}^{diag}U_{0\Sigma\Sigma}^\dagger
U_{L\Psi l}&=&-V_{PMNS}^\ast
m_\nu^{diag}U_{0\Sigma\nu}^\dagger U_{L\Psi
l}+V_{l\Sigma}^{L\ast}U_{0\Sigma\Sigma}^TM_\Sigma^T U_{L\Psi
l}\nonumber \\
&+&V^\ast_{PMNS}U_{0\Sigma\nu}^TM_\Sigma U_{L\Psi l}
+U_{Lll}^TY_\Sigma^Tv U_{L\Psi l}/\sqrt{2}\;.\label{eee}
\end{eqnarray}

Combining Eqs. \ref{ee} and \ref{eee}, we finally obtain
\begin{eqnarray}
V_{l\Sigma}^{L\ast} M_{\Sigma N}^{diag} V_{l\Sigma}^{L\dagger}&=&
-V_{PMNS}^\ast m_\nu^{diag} V_{PMNS}^\dagger
+V_{l\Sigma}^{L\ast}U_{0\Sigma\Sigma}^TM_\Sigma^T U_{L\Psi
l}\sqrt{2}+V^\ast_{PMNS}U_{0\Sigma\nu}^TM_\Sigma U_{L\Psi
l}\sqrt{2}\nonumber
\\
&+&U_{Lll}^TY_\Sigma^TvU_{L\Psi l}+U_{L\Psi
l}^TY_\Sigma v U_{Lll}\nonumber \\
&=&-V_{PMNS}^\ast m_\nu^{diag}
V_{PMNS}^\dagger+m_l^{diag}U^T_{R\Psi l}U_{L\Psi l}+U_{L\Psi
l}^TU_{R\Psi l}m_l^{diag}\;.
\end{eqnarray}

\section{Explicit Expressions of $V_{lN}$ for Case II}
From Eq.~\ref{VL} we can write $V_{lN}$ explicitly as
\begin{eqnarray}
V_{l N}&=& i V_{PMNS} \ (m_\nu^{diag})^{1/2} \ \Omega \
(M_N^{diag})^{-1/2}\;, \label{b1}
\end{eqnarray}
where $\Omega$ is matrix which satisfies $\Omega \Omega^T =1$. It can be parameterized as
\begin{equation}
\Omega (w_{21}, w_{31}, w_{32})= R_{12} (w_{21}) R_{13}(w_{31})
R_{23}(w_{32}),
\end{equation}
with
\begin{equation}
R_{12}=\left(
\begin{array}{ccc}
    cw_{21} & - sw_{21}  & 0
    \\
    sw_{21} & cw_{21} & 0
    \\
    0 & 0 & 1
  \end{array}
\right), \;\;
R_{13}=\left(
\begin{array}{ccc}
    cw_{31} &  0 & -sw_{31}
    \\
    0 & 1 & 0
    \\
    sw_{31} & 0 & cw_{31}
  \end{array}
\right),\;\;
R_{23}=\left(
\begin{array}{ccc}
    1 & 0 & 0 \\
    0 & cw_{{32}} & - sw_{{32}}
    \\
    0 & sw_{{32}} & cw_{{32}}
  \end{array}
\right),\nonumber
\end{equation}
where $sw_{ij} = \sin(w_{ij})$ and $\cos(w_{ij})$.

The couplings
$V_{lN}$ for different charged lepton and heavy neutrino flavors are
\begin{eqnarray}
-iV_{lN}^{e1}\sqrt{M_1}&=&\sqrt{m_2}c_{13}s_{12}sw_{21}cw_{31}
+\sqrt{m_1}c_{12}c_{13}cw_{21}cw_{31} e^{i\Phi_1/2}
+\sqrt{m_3}s_{13}sw_{31}e^{i(\Phi_2/2-\delta)}\;, \nonumber\\
-iV_{lN}^{\mu
1}\sqrt{M_1}&=&\sqrt{m_2}(c_{12}c_{23}-s_{12}s_{13}s_{23}e^{i
\delta})sw_{21}cw_{31}\nonumber \\
&+&\sqrt{m_1}(-s_{12}c_{23}-c_{12}s_{13}s_{23}e^{i
\delta})cw_{21}cw_{31} e^{i\Phi_1/2}
+\sqrt{m_3}c_{13}s_{23}sw_{31}e^{i\Phi_2/2}\;,
\nonumber\\
-iV_{lN}^{\tau
1}\sqrt{M_1}&=&\sqrt{m_2}(-c_{12}s_{23}-s_{12}s_{13}c_{23}e^{i
\delta})sw_{21}cw_{31}\nonumber \\
&+&\sqrt{m_1}(s_{12}s_{23}-c_{12}s_{13}c_{23}e^{i
\delta})cw_{21}cw_{31}e^{i\Phi_1/2}
+\sqrt{m_3}c_{13}c_{23}sw_{31}e^{i\Phi_2/2}\;,
\end{eqnarray}

\begin{eqnarray}
-iV_{lN}^{e2}\sqrt{M_2}&=&\sqrt{m_2}c_{13}s_{12}(-sw_{21}sw_{31}sw_{32}+cw_{21}cw_{32})\;,\nonumber\\
&+&\sqrt{m_1}c_{12}c_{13}(-sw_{31}sw_{32}cw_{21}-sw_{21}cw_{32})e^{i\Phi_1/2}
+\sqrt{m_3}s_{13}sw_{32}cw_{31}e^{i(\Phi_2/2-\delta)}\;,\nonumber\\
-iV_{lN}^{\mu
2}\sqrt{M_2}&=&\sqrt{m_2}(c_{12}c_{23}-s_{12}s_{13}s_{23}e^{i\delta})
(-sw_{21}sw_{31}sw_{32}+cw_{21}cw_{32})\;,\nonumber\\
&+&\sqrt{m_1}(-s_{12}c_{23}-c_{12}s_{13}s_{23}e^{i\delta})
(-sw_{32}sw_{31}cw_{21}-sw_{21}cw_{32})e^{i\Phi_1/2}\nonumber\\
&+&\sqrt{m_3}c_{13}s_{23}sw_{32} cw_{31} e^{i\Phi_2/2}\;,\nonumber\\
-iV_{lN}^{\tau
2}\sqrt{M_2}&=&\sqrt{m_2}(-c_{12}s_{23}-s_{12}s_{13}c_{23}e^{i\delta})
(-sw_{21}sw_{31}sw_{32}+cw_{21}cw_{32})\nonumber\\
&+&\sqrt{m_1}(s_{12}s_{23}-c_{12}s_{13}c_{23}e^{i\delta})
(-sw_{32}sw_{31}cw_{21}-sw_{21}cw_{32})e^{i\Phi_1/2}\nonumber\\
&+&\sqrt{m_3}c_{13}c_{23}sw_{32}cw_{31} e^{i\Phi_2/2}\;,
\end{eqnarray}

\begin{eqnarray}
-iV_{lN}^{e3}\sqrt{M_3}&=&\sqrt{m_2}c_{13}s_{12}(-sw_{32}cw_{21}-sw_{21}sw_{31}cw_{32})\;,\nonumber\\
&+&\sqrt{m_1}c_{12}c_{13}(sw_{21}sw_{32}-sw_{31}cw_{21}cw_{32})e^{i\Phi_1/2}\nonumber\\
&+&
\sqrt{m_3}s_{13}cw_{31}cw_{32}e^{i(\Phi_2/2-\delta)}\;,\nonumber\\
-iV_{lN}^{\mu
3}\sqrt{M_3}&=&\sqrt{m_2}(c_{12}c_{23}-s_{12}s_{13}s_{23}e^{i\delta})
(-sw_{32}cw_{21}-sw_{21}sw_{31}cw_{32})\nonumber\\
&+&\sqrt{m_1}(-s_{12}c_{23}-c_{12}s_{13}s_{23}e^{i\delta})
(sw_{32}sw_{21}-sw_{31}cw_{21}cw_{32})e^{i\Phi_1/2}\nonumber\\
&+&\sqrt{m_3}c_{13}s_{23}cw_{31}cw_{32}e^{i\Phi_2/2}\;,\nonumber\\
-iV_{lN}^{\tau
3}\sqrt{M_3}&=&\sqrt{m_2}(-c_{12}s_{23}-s_{12}s_{13}c_{23}e^{i\delta})
(-sw_{32}cw_{21}-sw_{21}sw_{31}cw_{32})\nonumber\\
&+&\sqrt{m_1}(s_{12}s_{23}-c_{12}s_{13}c_{23}e^{i\delta})
(sw_{32}sw_{21}-sw_{31}cw_{21}cw_{32})e^{i\Phi_1/2}\nonumber\\
&+&\sqrt{m_3}c_{13}c_{23}cw_{31}cw_{32}e^{i\Phi_2/2}\;.
\end{eqnarray}

Note that $\Omega$ is only required to satisfy $\Omega \Omega^T =
1$, the angles $w_{ij}$ can take complex values. In principle, the
elements in $\Omega$ is unbounded. For example taking $w_{ij}$ to be
imaginary and arbitrarily large will lead to large light and heavy
neutrino mixing. Since we are only interested in small mixing with
element in $V_{lN}$ of order $\sqrt{m_\nu/M_N}$, we will consider in
the main text that the element in $\Omega$ to be real numbers by
restricting the ranges of $w_{ij}$ to be $0 \leq w_{ij} \leq 2\pi$.
In this case the above general solution belongs to the small mixing
solution. In the limit the light neutrino masses go to zero, all
elements in $V_{\l N}$ is guaranteed go to zero. Also these elements
are of order $(m_\nu /M_{\nu_R})^{1/2}$.


\end{document}